\pdfoutput=1
\documentclass[12pt,a4paper]{article}

\usepackage{ifthen} 
\newboolean{pdflatex}
\setboolean{pdflatex}{true} 

\newboolean{articletitles}
\setboolean{articletitles}{true} 

\newboolean{uprightparticles}
\setboolean{uprightparticles}{false} 

\usepackage{longtable} 
\usepackage{rotating}
\usepackage{amsmath}
\usepackage{float}
\usepackage{booktabs}

\def\paperauthors{LHCb collaboration} 
\def\paperasciititle{Precision measurement of forward Z boson production in proton-proton collisions at sqrt(s) = 13 TeV} 
\def\papertitle{Precision measurement of forward {\Z~boson} production in proton-proton collisions\\ at $\sqs = 13\tev$} 
\def\paperkeywords{{High Energy Physics}, {LHCb}} 
\def\papercopyright{\the\year\ CERN for the benefit of the LHCb collaboration} 
\def\paperlicence{CC BY 4.0 licence}
\def\paperlicenceurl{https://creativecommons.org/licenses/by/4.0/}

\def\Zmm     {\decay{\Z}{\mumu}} 
\def\Ztt     {\decay{\Z}{\tautau}}
\def\zpt     {\ensuremath{p_{\mathrm{T}}^{Z}}\xspace}
\def\zy      {\ensuremath{y^{\mathrm{Z}}}\xspace}
\def\ww      {W^{+}W^{-}}
\def\wz      {W^{\pm}\Z}
\def\zz      {\Z\Z}
\def\phistar {\ensuremath{\phi_{\eta}^{*}}\xspace}
\def\totxsec {\ensuremath{196.4 \pm 0.2 \pm 1.6 \pm 3.9}\xspace}
\def\runtwolumi {\ensuremath{5.1 \pm 0.1}\xspace}

\usepackage[top=1in, bottom=1.25in, left=1in, right=1in]{geometry}

\usepackage{microtype}
\usepackage{lineno}  
\usepackage{xspace} 
\usepackage{caption} 

\usepackage{graphicx}  
\usepackage{color}
\usepackage{colortbl}
\graphicspath{{./figs/}} 

\usepackage{amsmath} 
\usepackage{amssymb}
\usepackage{amsfonts}
\usepackage{upgreek} 

\newcommand*\patchAmsMathEnvironmentForLineno[1]{%
\expandafter\let\csname old#1\expandafter\endcsname\csname #1\endcsname
\expandafter\let\csname oldend#1\expandafter\endcsname\csname
end#1\endcsname
 \renewenvironment{#1}%
   {\linenomath\csname old#1\endcsname}%
   {\csname oldend#1\endcsname\endlinenomath}%
}
\newcommand*\patchBothAmsMathEnvironmentsForLineno[1]{%
  \patchAmsMathEnvironmentForLineno{#1}%
  \patchAmsMathEnvironmentForLineno{#1*}%
}
\AtBeginDocument{%
\patchBothAmsMathEnvironmentsForLineno{equation}%
\patchBothAmsMathEnvironmentsForLineno{align}%
\patchBothAmsMathEnvironmentsForLineno{flalign}%
\patchBothAmsMathEnvironmentsForLineno{alignat}%
\patchBothAmsMathEnvironmentsForLineno{gather}%
\patchBothAmsMathEnvironmentsForLineno{multline}%
\patchBothAmsMathEnvironmentsForLineno{eqnarray}%
}

\usepackage{hyperxmp}

\usepackage[pdftex,
            pdfauthor={\paperauthors},
            pdftitle={\paperasciititle},
            pdfkeywords={\paperkeywords},
            pdfcopyright={Copyright (C) \papercopyright},
            pdflicenseurl={\paperlicenceurl}]{hyperref}

\usepackage[colorinlistoftodos,textsize=scriptsize]{todonotes}

\usepackage[bottom,flushmargin,hang,multiple]{footmisc}

\usepackage[all]{hypcap} 

\usepackage{xspace} 
\usepackage{upgreek}

\def\lhcb   {\mbox{LHCb}\xspace}

\def\dzero  {\mbox{D0}\xspace}

\def\MagUp {\mbox{\em Mag\kern -0.05em Up}\xspace}

\ifthenelse{\boolean{uprightparticles}}%
{

 \def\Pmu         {\ensuremath{\upmu}\xspace}

 \def\Ptau        {\ensuremath{\uptau}\xspace}

 \def\PDelta      {\ensuremath{\Delta}\xspace}                 
 \def\PXi         {\ensuremath{\Xi}\xspace}                 
 \def\PLambda     {\ensuremath{\Lambda}\xspace}                 
 \def\PSigma      {\ensuremath{\Sigma}\xspace}                 
 \def\POmega      {\ensuremath{\Omega}\xspace}                 
 \def\PUpsilon    {\ensuremath{\Upsilon}\xspace}

 \def\PB      {\ensuremath{\mathrm{B}}\xspace}                 
                  
 \def\PD      {\ensuremath{\mathrm{D}}\xspace}

 \def\PK      {\ensuremath{\mathrm{K}}\xspace}

 \def\PW      {\ensuremath{\mathrm{W}}\xspace}

 \def\PZ      {\ensuremath{\mathrm{Z}}\xspace}                 
                  
 \def\Pb      {\ensuremath{\mathrm{b}}\xspace}                 
 \def\Pc      {\ensuremath{\mathrm{c}}\xspace}

 \def\Pi      {\ensuremath{\mathrm{i}}\xspace}

 \def\Ps      {\ensuremath{\mathrm{s}}\xspace}                 
 \def\Pt      {\ensuremath{\mathrm{t}}\xspace}

 \def\thebaroffset{0.0em}
}
{

 \def\Pmu         {\ensuremath{\mu}\xspace}

 \def\Ptau        {\ensuremath{\tau}\xspace}

 \mathchardef\PDelta="7101
 \mathchardef\PXi="7104
 \mathchardef\PLambda="7103
 \mathchardef\PSigma="7106
 \mathchardef\POmega="710A
 \mathchardef\PUpsilon="7107
                  
 \def\PB      {\ensuremath{B}\xspace}                 
                  
 \def\PD      {\ensuremath{D}\xspace}

 \def\PK      {\ensuremath{K}\xspace}

 \def\PW      {\ensuremath{W}\xspace}

 \def\PZ      {\ensuremath{Z}\xspace}                 
                  
 \def\Pb      {\ensuremath{b}\xspace}                 
 \def\Pc      {\ensuremath{c}\xspace}

 \def\Pi      {\ensuremath{i}\xspace}

 \def\Ps      {\ensuremath{s}\xspace}                 
 \def\Pt      {\ensuremath{t}\xspace}

 \def\thebaroffset{0.18em}
}
\newcommand{\offsetoverline}[2][\thebaroffset]{\kern #1\overline{\kern -#1 #2}}%

\makeatletter
\ifcase \@ptsize \relax
  \newcommand{\miniscule}{\@setfontsize\miniscule{4}{5}}
\or
  \newcommand{\miniscule}{\@setfontsize\miniscule{5}{6}}
\or
  \newcommand{\miniscule}{\@setfontsize\miniscule{5}{6}}
\fi
\makeatother

\DeclareRobustCommand{\optbar}[1]{\shortstack{{\miniscule (\rule[.5ex]{1.25em}{.18mm})}
  \\ [-.7ex] $#1$}}

\def\mumu       {{\ensuremath{\Pmu^+\Pmu^-}}\xspace}

\def\tautau     {{\ensuremath{\Ptau^+\Ptau^-}}\xspace}

\def\W      {{\ensuremath{\PW}}\xspace}

\def\Z      {{\ensuremath{\PZ}}\xspace}

\def\squark    {{\ensuremath{\Ps}}\xspace}

\def\cquark    {{\ensuremath{\Pc}}\xspace}
\def\cquarkbar {{\ensuremath{\overline \cquark}}\xspace}
\def\ccbar     {{\ensuremath{\cquark\cquarkbar}}\xspace}
\def\bquark    {{\ensuremath{\Pb}}\xspace}
\def\bquarkbar {{\ensuremath{\overline \bquark}}\xspace}
\def\bbbar     {{\ensuremath{\bquark\bquarkbar}}\xspace}
\def\tquark    {{\ensuremath{\Pt}}\xspace}
\def\tquarkbar {{\ensuremath{\overline \tquark}}\xspace}
\def\ttbar     {{\ensuremath{\tquark\tquarkbar}}\xspace}

\def\KorKbar {\kern \thebaroffset\optbar{\kern -\thebaroffset \PK}{}\xspace}

\def\D       {{\ensuremath{\PD}}\xspace}

\def\DorDbar {\kern \thebaroffset\optbar{\kern -\thebaroffset \PD}\xspace}

\def\Dp      {{\ensuremath{\D^+}}\xspace}
\def\Dm      {{\ensuremath{\D^-}}\xspace}

\def\DpDm    {\ensuremath{\Dp {\kern -0.16em \Dm}}\xspace}

\def\B       {{\ensuremath{\PB}}\xspace}

\def\BorBbar {\kern \thebaroffset\optbar{\kern -\thebaroffset \PB}\xspace}

\def\Bd      {{\ensuremath{\B^0}}\xspace}

\def\BdorBdbar {\kern \thebaroffset\optbar{\kern -\thebaroffset \Bd}\xspace}

\def\Bs      {{\ensuremath{\B^0_\squark}}\xspace}

\def\BsorBsbar {\kern \thebaroffset\optbar{\kern -\thebaroffset \Bs}\xspace}

\def\Y#1S{\ensuremath{\PUpsilon{(#1S)}}\xspace}

\def\LorLbar     {\kern \thebaroffset\optbar{\kern -\thebaroffset \PLambda}\xspace}

\newcommand{\decay}[2]{\ensuremath{#1\!\to #2}\xspace} 

\def\to                 {\ensuremath{\rightarrow}\xspace}

\newcommand{\as}{{\ensuremath{\alpha_s}}\xspace}

\def\AT#1     {\ensuremath{A_{\mathrm{T}}^{#1}}\xspace}           

\def\C#1      {\ensuremath{\mathcal{C}_{#1}}\xspace}                       
\def\Cp#1     {\ensuremath{\mathcal{C}_{#1}^{'}}\xspace}                    
\def\Ceff#1   {\ensuremath{\mathcal{C}_{#1}^{\mathrm{(eff)}}}\xspace}        
\def\Cpeff#1  {\ensuremath{\mathcal{C}_{#1}^{'\mathrm{(eff)}}}\xspace}       
\def\Ope#1    {\ensuremath{\mathcal{O}_{#1}}\xspace}                       
\def\Opep#1   {\ensuremath{\mathcal{O}_{#1}^{'}}\xspace}                    

\newcommand{\nospaceunit}[1]{\ensuremath{\text{#1}}}       
\newcommand{\aunit}[1]{\ensuremath{\text{\,#1}}}       

\newcommand{\tev}{\aunit{Te\kern -0.1em V}\xspace}
\newcommand{\gev}{\aunit{Ge\kern -0.1em V}\xspace}
\newcommand{\mev}{\aunit{Me\kern -0.1em V}\xspace}
\newcommand{\kev}{\aunit{ke\kern -0.1em V}\xspace}
\newcommand{\ev}{\aunit{e\kern -0.1em V}\xspace}
 
\newcommand{\mevc}{\ensuremath{\aunit{Me\kern -0.1em V\!/}c}\xspace}
\newcommand{\gevc}{\ensuremath{\aunit{Ge\kern -0.1em V\!/}c}\xspace}
\newcommand{\mevcc}{\ensuremath{\aunit{Me\kern -0.1em V\!/}c^2}\xspace}
\newcommand{\gevcc}{\ensuremath{\aunit{Ge\kern -0.1em V\!/}c^2}\xspace}

\def\mum  {\ensuremath{\,\upmu\nospaceunit{m}}\xspace}

\def\pb {\aunit{pb}\xspace}

\def\fb   {\ensuremath{\aunit{fb}}\xspace}
\def\invfb   {\ensuremath{\fb^{-1}}\xspace}

\newcommand{\chisq}{\ensuremath{\chi^2}\xspace}

\def\gsim{{~\raise.15em\hbox{$>$}\kern-.85em
          \lower.35em\hbox{$\sim$}~}\xspace}
\def\lsim{{~\raise.15em\hbox{$<$}\kern-.85em
          \lower.35em\hbox{$\sim$}~}\xspace}

\def\sqs   {\ensuremath{\protect\sqrt{s}}\xspace}

\def\pt         {\ensuremath{p_{\mathrm{T}}}\xspace}

\def\ptot       {\ensuremath{p}\xspace}

\newcommand{\lum} {\ensuremath{\mathcal{L}}\xspace}

\def\fewz       {\mbox{\textsc{Fewz}}\xspace}

\def\geant      {\mbox{\textsc{Geant4}}\xspace}

\def\herwig     {\mbox{\textsc{Herwig}}\xspace}

\def\photos     {\mbox{\textsc{Photos}}\xspace}
\def\powheg     {\mbox{\textsc{Powheg}}\xspace}
\def\pythia     {\mbox{\textsc{Pythia}}\xspace}

\def\resbos     {\mbox{\textsc{ResBos}}\xspace}

\def\matchbox   {\mbox{\textsc{MatchBox}}\xspace}

\def\tell1  {TELL1\xspace}
\def\ukl1   {UKL1\xspace}

\newcommand{\ie}{\mbox{\itshape i.e.}\xspace}

\usepackage{cite} 
\usepackage{mciteplus}

\usepackage{longtable} 
\begin{document}

\renewcommand{\thefootnote}{\fnsymbol{footnote}}
\setcounter{footnote}{1}


\begin{titlepage}
\pagenumbering{roman}

\vspace*{-1.5cm}
\centerline{\large EUROPEAN ORGANIZATION FOR NUCLEAR RESEARCH (CERN)}
\vspace*{1.5cm}
\noindent
\begin{tabular*}{\linewidth}{lc@{\extracolsep{\fill}}r@{\extracolsep{0pt}}}
\ifthenelse{\boolean{pdflatex}}
{\vspace*{-1.5cm}\mbox{\!\!\!\includegraphics[width=.14\textwidth]{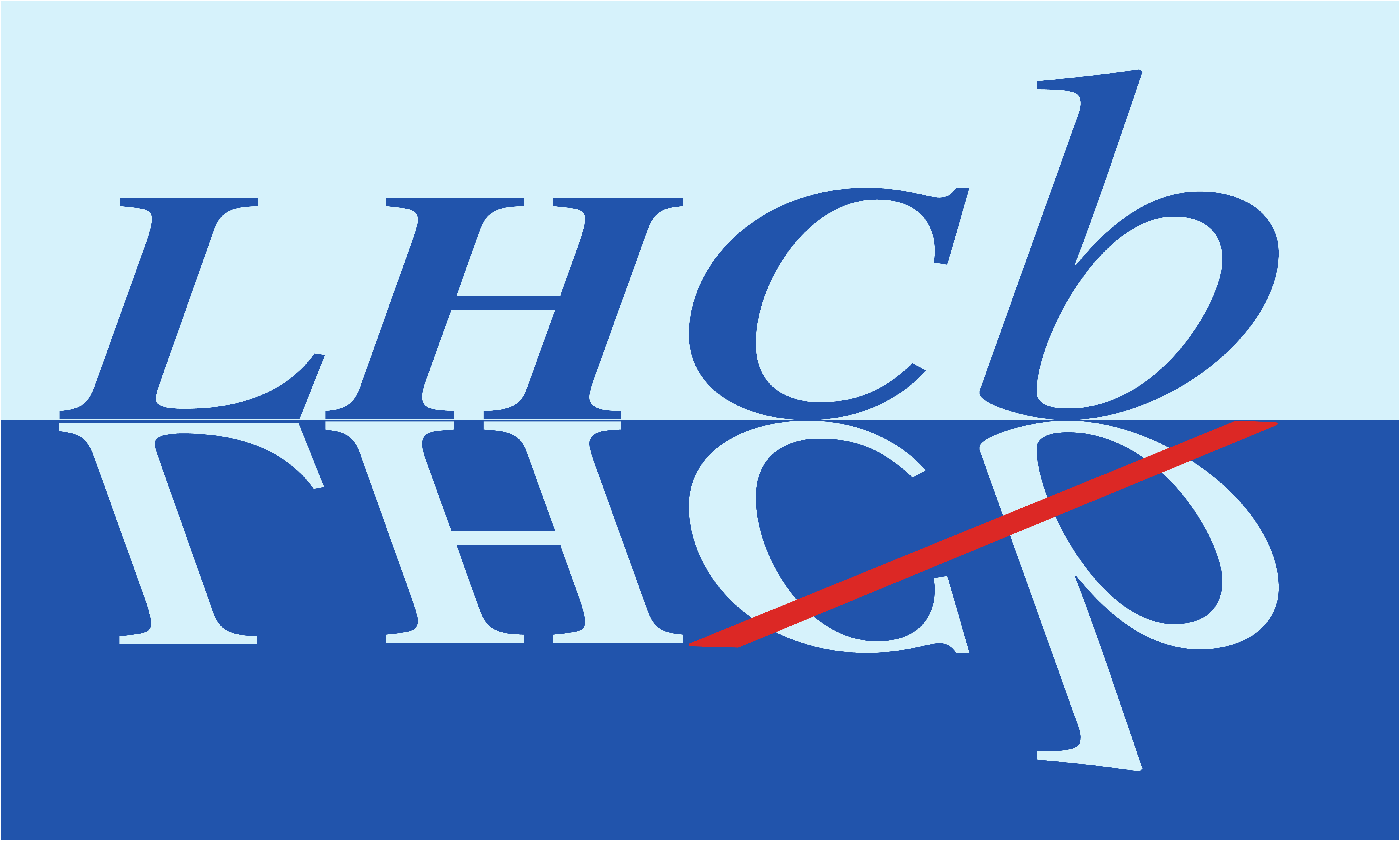}} & &}%
{\vspace*{-1.2cm}\mbox{\!\!\!\includegraphics[width=.12\textwidth]{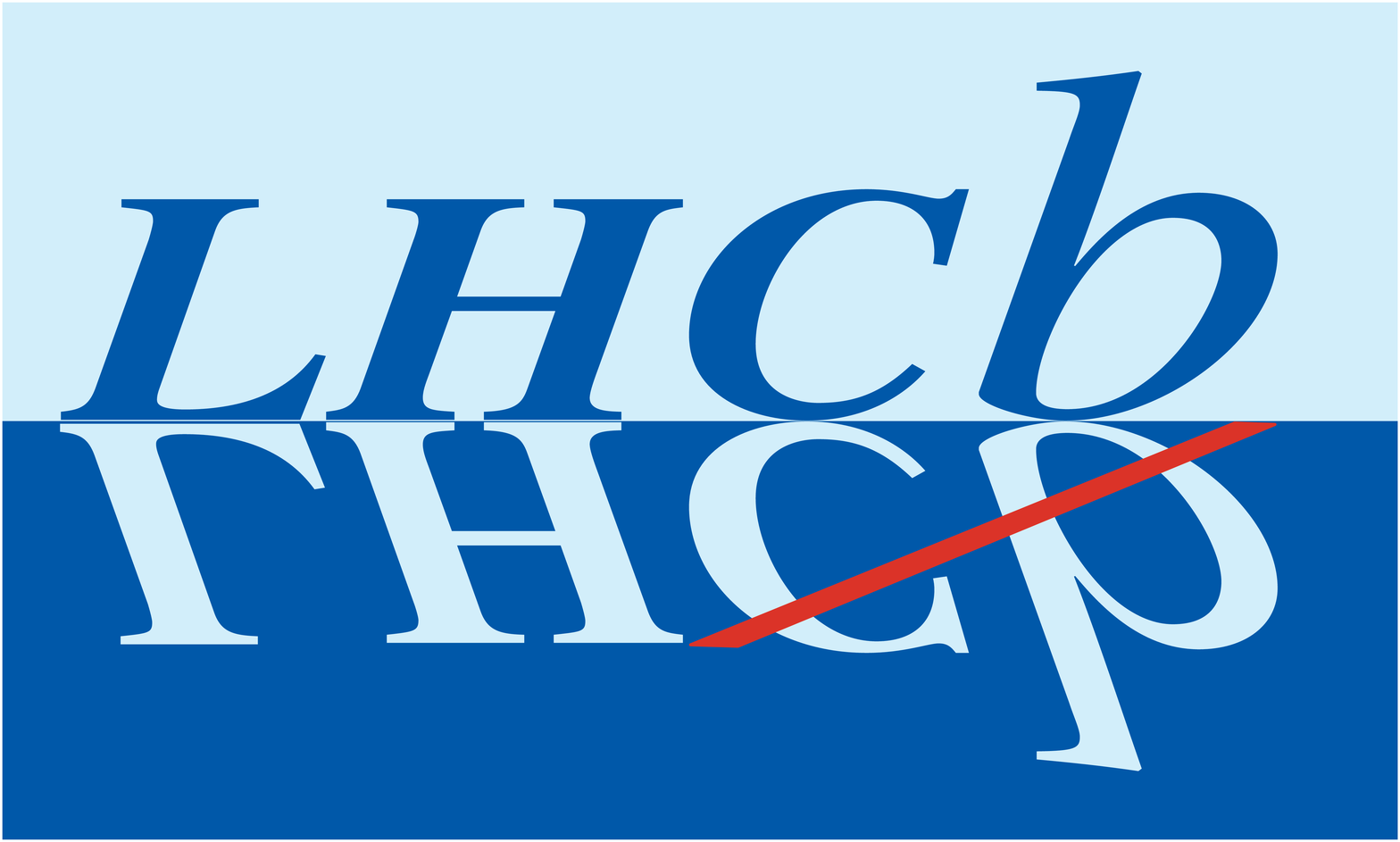}} & &}%
\\
 & & CERN-EP-2021-246 \\  
 & & LHCb-PAPER-2021-037 \\  
 & & July 05, 2022 
\end{tabular*}

\vspace*{4.0cm}

{\normalfont\bfseries\boldmath\huge
\begin{center}
  \papertitle 
\end{center}
}

\vspace*{.5cm}

\begin{center}
\paperauthors\footnote{Authors are listed at the end of this paper.}
\end{center}

\vspace{\fill}

\begin{abstract}
  \noindent
A precision measurement of the \Z boson production cross-section at $\sqs = 13\tev$ in the forward region
is presented, using $pp$ collision data collected by the \lhcb detector, 
corresponding to an integrated luminosity of 5.1\invfb. 
The production cross-section is measured using \Zmm events within the fiducial region defined as pseudorapidity $2.0<\eta<4.5$ and transverse momentum $\pt>20\gevc$ for both muons
and dimuon invariant mass $60<M_{\mu\mu}<120\gevcc$.
 The integrated cross-section is determined to be
 \begin{equation*}
    \sigma(\Zmm) = \totxsec\pb,
\end{equation*}
where the first uncertainty is statistical, the second is systematic, and the third is due to the luminosity determination.
The measured results are in agreement with theoretical predictions within uncertainties.
\end{abstract}

\vspace*{2.0cm}

\begin{center}
  Published in JHEP 07 (2022) 26
\end{center}

\vspace{\fill}

{\footnotesize 
\centerline{\copyright~\papercopyright. \href{\paperlicenceurl}{\paperlicence}.}}
\vspace*{2mm}

\end{titlepage}


\newpage
\setcounter{page}{2}
\mbox{~}


\renewcommand{\thefootnote}{\arabic{footnote}}
\setcounter{footnote}{0}

\cleardoublepage


\pagestyle{plain} 
\setcounter{page}{1}
\pagenumbering{arabic}



\section{Introduction}
\label{sec:introduction}
Precision measurements of the single \Z boson\footnote{In this article, the label \Z boson is defined to include contributions from virtual photons and interference between them.} production cross-section at the CERN Large Hadron Collider (LHC)
provide an important test of the quantum chromodynamics (QCD) and electroweak (EW) sectors of the Standard Model. 
Theoretical predictions for the \Z boson production cross-section are available up to next-to-next-to-next-to-leading order in 
perturbative QCD~\cite{Camarda:2021ict,Duhr:2021vwj}
and have comparable precision as the measured results to date.
Further validations and tests on theoretical predictions~\cite{Rijken:1994sh,Hamberg:1990np,Harlander:2002wh,vanNeerven:1991gh,Anastasiou:2003ds,Scimemi:2019cmh,Bury:2022czx} require precision measurements of the \Z boson production cross-section in different experiments.
The \lhcb collaboration has previously reported the measurement of the \W boson mass~\cite{LHCb-PAPER-2021-024},
using a data sample corresponding to
an integrated luminosity of 1.7\invfb, and sizable uncertainties from parton distribution functions (PDFs) and boson \pt modelling are seen. 
A measurement of the \Z boson production cross-section will provide information to 
 reduce these uncertainties, and the future measurements of the \W boson mass
and weak mixing angle~\cite{LHCb-PAPER-2015-039} at \lhcb could also benefit from this measurement.

The \Z boson candidates collected with the \lhcb detector are highly boosted, and produced by a parton with 
large Bjorken-$x$ and another with small $x$.
The Bjorken-$x$ is the fraction of the proton momentum carried by a parton.
Therefore, a precision measurement of \Z boson production cross-section with the \lhcb detector is particularly 
sensitive to 
PDFs, especially in the very large and small $x$ ranges.
The PDFs are constrained by the results from deep inelastic scattering and hadron collider experiments~\cite{H1:2009pze,H1:2015ubc,CDF:1996zyp,CDF:2005cgc,CDF:2010vek,CDF:2008hmn,D0:2007pcy,D0:2006eri,D0:2014kma,D0:2008nou,ATLAS:2011qdp,ATLAS:2015iiu,CMS:2013pzl,CMS:2012ivw,ATLAS:2019zci,CMS:2019raw}.
However, these measurements provide limited information for the PDFs for very large $x$ (up to $\sim 0.8$) or very small $x$ ($\sim 5\times 10^{-5}$), 
which leads to large PDF uncertainty, 
and consequently a large uncertainty in theoretical predictions of vector boson production cross-section in the forward region.
As the \lhcb detector has fully instrumented coverage in the forward region, with complementary acceptance compared to the ATLAS and CMS detectors, the collected \Z boson candidates can provide 
unique and important information for the determining the PDFs.
Previous measurements of single \W and \Z production by the \lhcb collaboration~\cite{LHCb-PAPER-2014-033,LHCb-PAPER-2015-001,LHCb-PAPER-2015-003,LHCb-PAPER-2015-049,LHCb-PAPER-2016-024,LHCb-PAPER-2016-021} 
have been included in PDF calculations~\cite{Dulat:2015mca,Harland-Lang:2014zoa,NNPDF:2014otw,Hou:2019efy}, and contribute significantly to the determination of the valence quark
PDFs at large and small values of $x$.
Furthermore, the \lhcb measurements constrain strange and charm PDFs at high $x$, including intrinsic charm~\cite{Rojo:2017xpe,Deng:2020sol}.
Recently, the SeaQuest collaboration~\cite{SeaQuest:2021zxb} reported a measurement of the Drell-Yan process, which 
is sensitive to $\bar{d}/\bar{u}$ PDF ratio. 
Tensions between the SeaQuest~\cite{SeaQuest:2021zxb} and NuSea~\cite{NuSea:2001idv} 
results in the large $x$ region are observed. 
Since both results have large contributions from nuclear effects, 
\lhcb measurements using the proton-proton ($pp$) collision data can provide complementary constraints in that $x$ region.

In this article, the integrated and differential \Z boson production cross-sections are measured at the Born level in QED using 
$pp$ collision data collected by the \lhcb detector at a centre-of-mass energy \sqs = 13\tev in 2016, 2017 and 2018,
corresponding to an integrated luminosity of $\runtwolumi\invfb$~\cite{LHCb-PAPER-2014-047}. 
The production cross-section is measured in a fiducial region that closely matches the acceptance of the \lhcb detector.
The fiducial region is defined as pseudorapidity $2.0<\eta<4.5$ and transverse momentum $\pt>20\gevc$ for both muons and 
dimuon invariant mass $60<M_{\mu\mu}<120\gevcc$.
A similar measurement using the \lhcb dielectron events is foreseen in future.
The differential cross-section is measured as a function of the \Z boson rapidity (\zy), transverse momentum (\zpt) and \phistar.
The observable $\phistar$, which was first measured by the \dzero~\cite{D0:2010qhp} experiment, probes similar physics as the \Z boson \pt, 
but is an angular variable that can be measured with better resolution by collider detectors.
It is defined as
\begin{equation}
    \phistar = \tan((\pi-\Delta\phi^{\ell\ell})/2)\sin(\theta^*_{\eta}),
\end{equation}
where $\Delta\phi^{\ell\ell}$ is the difference in azimuthal angle, $\phi$, between the two muons,
$\theta^*_{\eta}$ is the scattering angle of the muons with respect to the proton beam direction in the rest frame of the dimuon system. The variable $\theta^*_{\eta}$ is defined by $\cos(\theta^*_{\eta}) = \tanh [(\eta^{-} - \eta^{+}) /2]$, where $\eta^-$ and $\eta^+$ are the pseudorapidities of the negatively and positively charged muon, respectively.
Moreover, double differential cross-sections of \Z boson production in regions of \zy and \zpt, and of \zy and \phistar, are measured for the first time in the \lhcb forward acceptance.

\section{Detector and simulation}
\label{sec:Detector}
The \lhcb detector~\cite{LHCb-DP-2008-001,LHCb-DP-2014-002} is a single-arm forward
spectrometer covering the \mbox{pseudorapidity} range $2<\eta <5$,
designed for the study of particles containing \bquark or \cquark
quarks. The detector includes a high-precision tracking system
consisting of a silicon-strip vertex detector surrounding the $pp$
interaction region~\cite{LHCb-DP-2014-001}, a large-area silicon-strip detector (TT)~\cite{LHCb-TDR-009}, located
upstream of a dipole magnet with a bending power of about
$4{\mathrm{\,Tm}}$, and three stations of silicon-strip detectors and straw
drift tubes~\cite{LHCb-DP-2017-001} placed downstream of the magnet.
The tracking system provides a measurement of the momentum, \ptot, of charged particles with
a relative uncertainty that varies from 0.5\% at low momentum to 1.0\% at 200\gevc.
The minimum distance of a track to a primary $pp$ collision vertex (PV), the impact parameter (IP), 
is measured with a resolution of $(15+29/\pt)\mum$,
where \pt is the component of the momentum transverse to the beam, in\,\gevc.
Photons, electrons and hadrons are identified by a calorimeter system consisting of
scintillating-pad and preshower detectors, an electromagnetic
and a hadronic calorimeter. 
Muons are identified by a system composed of alternating layers of iron and multiwire
proportional chambers~\cite{LHCb-DP-2012-002}.
The online event selection is performed by a trigger~\cite{LHCb-DP-2012-004}, 
which consists of a hardware stage, based on information from the calorimeter and muon
systems, followed by a software stage, which applies a full event reconstruction.

Simulation is required to model the effects of the detector acceptance and the imposed selection requirements.
In the simulation, $pp$ collisions are generated using \pythia~\cite{Sjostrand:2006za} 
with a specific \lhcb configuration~\cite{LHCb-PROC-2010-056}.
The final state radiation is generated using \photos~\cite{davidson2015photos}.
The interaction of the generated particles with the detector, and its response,
are implemented using the \geant toolkit~\cite{Allison:2006ve} as described in Ref.~\cite{LHCb-PROC-2011-006}. 

\section{Reconstruction and selection}
\label{sec:sel}
The online event selection is performed by the muon triggers. 
At the hardware trigger stage, candidates are required to have a muon object with high \pt.
The muon candidate must satisfy $\pt>6\gevc$, $p>8\gevc$, with a good track fit quality
in the first software trigger stage.
While in the second software trigger stage, the muon candidate is further required to satisfy $\pt>12.5\gevc$.
For a \Zmm candidate, at least one of the muons is required to pass both hardware and software trigger
decision stages. 

\begin{figure}[tb]
\begin{center}
  \includegraphics[width=0.9\textwidth]{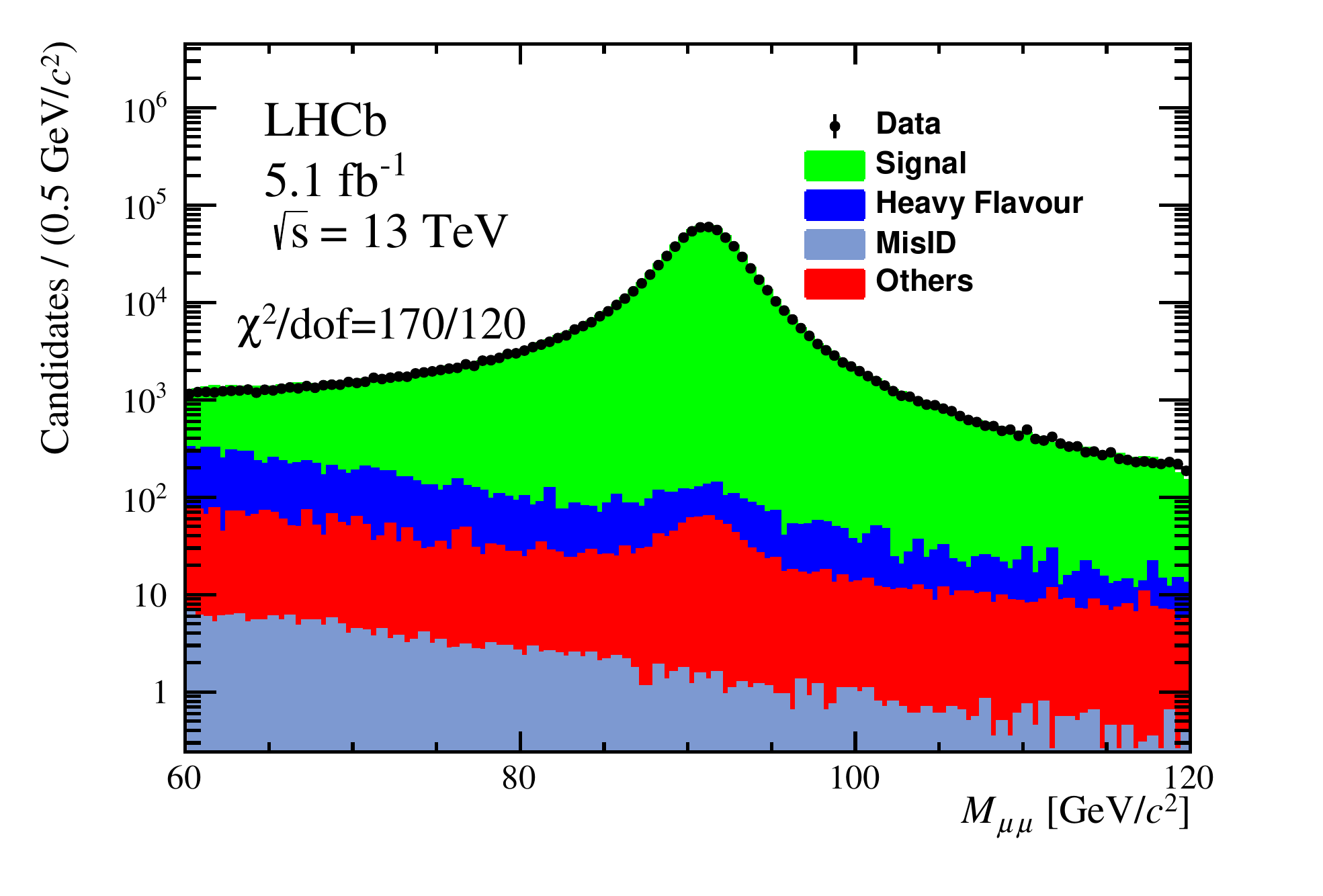}
\caption{Comparison of the invariant mass distribution between data and the sum of signal and background contributions for the selected \Zmm candidates.} 
\label{fig:mass}
\end{center}
\end{figure}

To select a \Zmm sample with high purity, candidates 
are required to have a pair of well-reconstructed tracks of opposite charge identified as muons.
The invariant mass of two muons must be in the range $60 <M_{\mu\mu}< 120 \gevcc$.
Muon tracks must have a transverse momentum $\pt > 20\gevc$ and pseudorapidity in the range $2.0<\eta<4.5$. 
The relative uncertainty in the momentum measurement for each muon is required to be less than 10\%. 
In total, 796 thousand \Zmm candidates are selected, 
and the dimuon invariant mass distribution of the selected candidates is shown in Fig.~\ref{fig:mass}.

\section{Background}

\subsection{Heavy flavour background}
\label{sec:heavyflavour}
Heavy flavour production (\bbbar and \ccbar quark pairs) has
a large branching fraction into semileptonic decays, and is
one of the largest background sources to the \Zmm process.
This contribution is estimated from data, 
using two control samples enriched in heavy flavour.

The event selection requirements described in Sec.~\ref{sec:sel} are used to select two control samples, in which 
the dimuon invariant mass requirement is changed to ${50<M_{\mu\mu}<80\gevcc}$. 
The first control sample is selected by requiring that the data candidate must have a primary vertex with a low fit quality. 
For signal events, the two muons originate at the primary vertex, while muons arising from decays of heavy hadrons do not, and thus have a low vertex fit quality.
The second sample is selected by requiring that the two muons are not spatially isolated ($I_{\mu}<0.7$) from other activity in the event. 
The muon isolation variable, $I_{\mu}$, is defined as the ratio of the muon \pt to the \pt of the vector sum of all charged particles \pt 
in a cone of size 0.5 in the $\eta-\phi$ coordinates around the muon, where $\phi$ is the azimuthal angle of the muon.

The event yields of these two control samples are determined by fitting the dimuon invariant mass distribution with an exponential function,
followed by an extrapolation of the fitted results to the signal region ($60 < M_{\mu\mu} < 120 \gevcc$).
These event yields are corrected 
with the corresponding efficiency of the vertex and muon isolation selections,
where the efficiency of the muon isolation (vertex) selection is calculated 
by applying the muon isolation requirement (vertex fit quality requirement) to 
the first (second) control sample. 

Studies on these two sub-samples are consistent, and the averaged value of the estimated background yields 
is taken as background contribution from the heavy flavour process,
which is determined to be $(1.0\pm0.1)$\% for the selected \Zmm sample.

\subsection{Background from misidentified hadrons}
Charged pions or kaons could be misidentified as muons and contribute to the selected \Zmm sample if they decay in flight before reaching the muon stations or if they have sufficient energy to traverse the calorimeters and be detected in the muon stations.
The contribution from the combinatorial background including misidentified hadrons and $B-\bar{B}$ mixing is determined using pairs of same-sign muons in the data. 
It is assumed that the charges of the selected muons are uncorrelated for these sources, which is validated by comparing the numbers of $\mu^+\mu^+$ and $\mu^-\mu^-$ candidates.
The difference between the number of $\mu^+\mu^+$ and $\mu^-\mu^-$ candidates is assigned as an additional uncertainty in the background contribution.

However, sizable contributions from heavy flavour processes (\ie, a muon from heavy flavour decay combined with a misidentified hadron) in the same-sign events are expected. 
To remove the double counting of the heavy flavour background in the background study,
a method similar to the one described in Sec.~\ref{sec:heavyflavour} is used,
by inverting the vertex fit quality and
the muon isolation requirements to obtain two background samples enhanced in same-sign events. 
The contribution from the heavy flavour processes in the same-sign events is determined to be $(95\pm4)\%$.

After removing the contribution from heavy flavour processes, the contribution from misidentified hadrons
is determined to be $0.04\%$ with negligibly small uncertainty.

\subsection{Background from other physics processes}
Background contributions from \ttbar, $\ww$, $\wz$, $\zz$, and \Ztt processes are estimated using simulation. 
The number of background events from the \Ztt process, which subsequently decay to dimuons that pass the event selection, 
is determined using simulation,
taking into account of the integrated luminosity and the predicted $Z$ boson production cross-section at next-to-next-to-leading-order (NNLO)~\cite{Gavin:2010az}. 
The contribution from \ttbar production, where both top quarks 
produce $W^{\pm}$ bosons and then decay to muons, is estimated using simulation and
the \ttbar production cross-section measured by the \lhcb collaboration~\cite{LHCb-PAPER-2017-050}.
The background contribution from diboson ($\ww$, $\wz$, and $\zz$) decays is estimated using a similar method. 

Summing contributions from the heavy flavour, misidentified hadrons and physics processes,
the total background contribution to the \Zmm sample in the mass range ${60-120\gevcc}$ is determined to be $(1.5\pm0.1)\%$. 

\section{Methods}
The differential cross-section is defined in interval regions of observable $a$ (\zy, \zpt or \phistar) as 
 \begin{equation}
\frac{d\sigma_{\Zmm}}{da}= \frac{N_{Z}\cdot f_{\rm{FSR}}^{Z}}{\lum \cdot \varepsilon^{Z} \cdot \Delta a},
\end{equation}
where $N_{Z}$ is the signal yield in a given region, 
$f_{\rm{FSR}}^{Z}$ is the final state radiation (FSR) correction factor (as discussed in Sec.~\ref{sec:fsr}), \lum is the integrated luminosity, $\Delta a$ is the interval width of the observable in a given region (as presented in Tables~\ref{tab:fsr_1D_y} through~\ref{tab:fsr_2D_yphi} in Appendix~\ref{app:fsr}), 
and $\varepsilon^{Z}$ is the total efficiency in this region.
The integrated cross-section is obtained by summing over all regions.

The differential production cross-section is measured in 18 
regions of \zy, from 2.0 to 4.5 with region width of 0.125, and of 0.25 above 4.0.
The differential production cross-section as a function of \zpt is measured in 14
regions [0.0, 2.2, 3.4, 4.6, 5.8, 7.2, 8.7, 10.5, 12.8, 15.4, 19.0, 24.5, 34.0, 63.0, 270.0]\gevc.
The differential production cross-section as a function of \phistar is measured in 15 regions [0.002, 0.01, 0.02, 0.03, 0.05, 0.07, 0.10, 0.15, 0.20, 0.30, 0.40, 0.60, 0.80, 1.20, 2.00, 4.00]. 
These region schemes are chosen based on the detector resolution and sample size of each region. 
The double-differential cross-section measurements are performed in five \zy regions of width 0.5. In each \zy region, the above 14 regions of \zpt and 15 regions of \phistar are used.

\subsection{Detector alignment and momentum scale calibration}
Starting with taking data in 2015, the \lhcb collaboration employs 
a novel online alignment procedure~\cite{LHCb-DP-2019-001}, 
which is used to obtain a stable performance of the detector. 
However, as the core physics programme is heavy flavour physics, 
the detector calibrations are not optimized for EW physics. 
In particular, the momentum scale calibration for the high-momentum muons that form the main signature of the $W^{\pm}$ and $Z$ boson decays,
can be improved significantly with an additional detector alignment. 

To correct the detector misalignment effects, 
the mass peak position of the selected \Zmm candidates is calibrated
in different kinematic and geometric regions to the world averaged value~\cite{PDG2020}.
The impact on the integrated cross-section measurement from this correction is found to be negligible. 
However, with finer region schemes, in all of the differential cross-section measurements,
this uncertainty has to be considered as one of the systematic uncertainties. 
\subsection{Efficiency}
Several corrections are developed and applied to the simulation, to achieve a better modelling of the \lhcb detector response.
The event selection efficiencies are determined for the muon trigger, as well as the tracking and identification requirements, using the \Zmm data candidates with the tag-and-probe method~\cite{LHCb-DP-2013-002}.

In the determination of the tracking efficiency, a particle reconstructed in all tracking subdetectors, having passed the muon trigger 
and muon identification requirements, is used as the tag. 
An object reconstructed by combining hits in the muon stations and the TT station, denoted as MuonTT track, acts as the probe.
The probe is then tested for the presence of an associated track, by searching for all reconstructed tracks linked to muon segments
which have more than 40\% of their hits in the muon stations and 60\% of their hits in the TT station in common with the MuonTT track. 
As described in Ref.~\cite{LHCb-DP-2013-002}, the tracking efficiency is calculated as
the fraction of probe candidates matched with a reconstructed track. 

This tag-and-probe tracking efficiency is further corrected to remove bias from the method itself. 
The correction is the ratio of the tracking efficiency estimated using truth level information to that of the tag-and-probe method, where the truth level tracking efficiency is defined as 
the fraction of simulated muon with sufficient hits in the muon and TT stations 
to satisfy the requirements of the track matching. 
There are two effects: a bias and a track matching correction. 
The bias correction takes into account the fact that the tracking efficiency is estimated using the MuonTT track, but not all of the muon tracks have an associated MuonTT track.
The track matching correction takes into account the inefficiency from the matching conditions. 
The determined muon tracking efficiency varies from 94\% to 97\% in different kinematic regions.

To determine the muon trigger and identification efficiency, the tag particle is selected from a particle reconstructed in all tracking subdetectors, by requiring it to be identified and triggered as muon, 
while the probe particle must be a track with good quality. 
The track must be identified as a muon when studying trigger efficiency. 
Both the tag and probe particles are further required to have \pt greater than 20\gevc, $\eta$ in a range from 2.0 to 4.5, and a relative momentum uncertainty
less than 10\%.
The invariant mass of the tag and probe candidates is required to be in the range from 60 to 120\gevcc. 
To suppress background further, 
the tag and probe are required to have
an azimuthal separation, $|\Delta\phi|$, greater than 2.7 radians. 
The efficiency is calculated as the ratio of the number of probes within the selected sample satisfying the muon trigger and identification requirements to the number of probes. 
The determined trigger and identification efficiency per-muon varies from 60\% to 85\%, and 65\% to 96\% in different kinematic regions, respectively. 

The total efficiency $\varepsilon^{Z}$ depends on the pseudorapidities of the two final-state muons and can be written as:
 \begin{equation}
\varepsilon^{Z} =  \left( \varepsilon_{\rm{track}}^{\mu^+} \cdot 
\varepsilon_{\rm{track}}^{\mu^-} \right) \cdot 
 \left( \varepsilon_{\rm{ID}}^{\mu^+} \cdot \varepsilon_{\rm{ID}}^{\mu^-} \right)\cdot 
 \left( \varepsilon_{\rm{trig}}^{\mu^+} + 
\varepsilon_{\rm{trig}}^{\mu^-} - \varepsilon_{\rm{trig}}^{\mu^+} \cdot \varepsilon_{\rm{trig}}^{\mu^-}\right),
\end{equation}
where $\varepsilon_{\rm{track}}^{\mu^{\pm}}$, $\varepsilon_{\rm{ID}}^{\mu^{\pm}}$, and $\varepsilon_{\rm{trig}}^{\mu^{\pm}}$ are the calculated efficiencies of muon track reconstruction, muon identification, and muon trigger, respectively.

\subsection{Unfolding}
The detector resolution effects could introduce a interval-to-interval migration between regions. 
This effect is corrected with the Bayesian unfolding method~\cite{DAgostini:1994fjx}.
Because of the good angular resolution of the \lhcb detector, negligible migration effects are observed in \zy and $\phistar$ measurements.
Therefore, unfolding correction is applied only to the differential cross-section measured as a function of $\zpt$. 

\subsection{Final state radiation correction}
\label{sec:fsr}
\begin{figure}[tb]
\begin{center}
\includegraphics[width=0.45\textwidth]{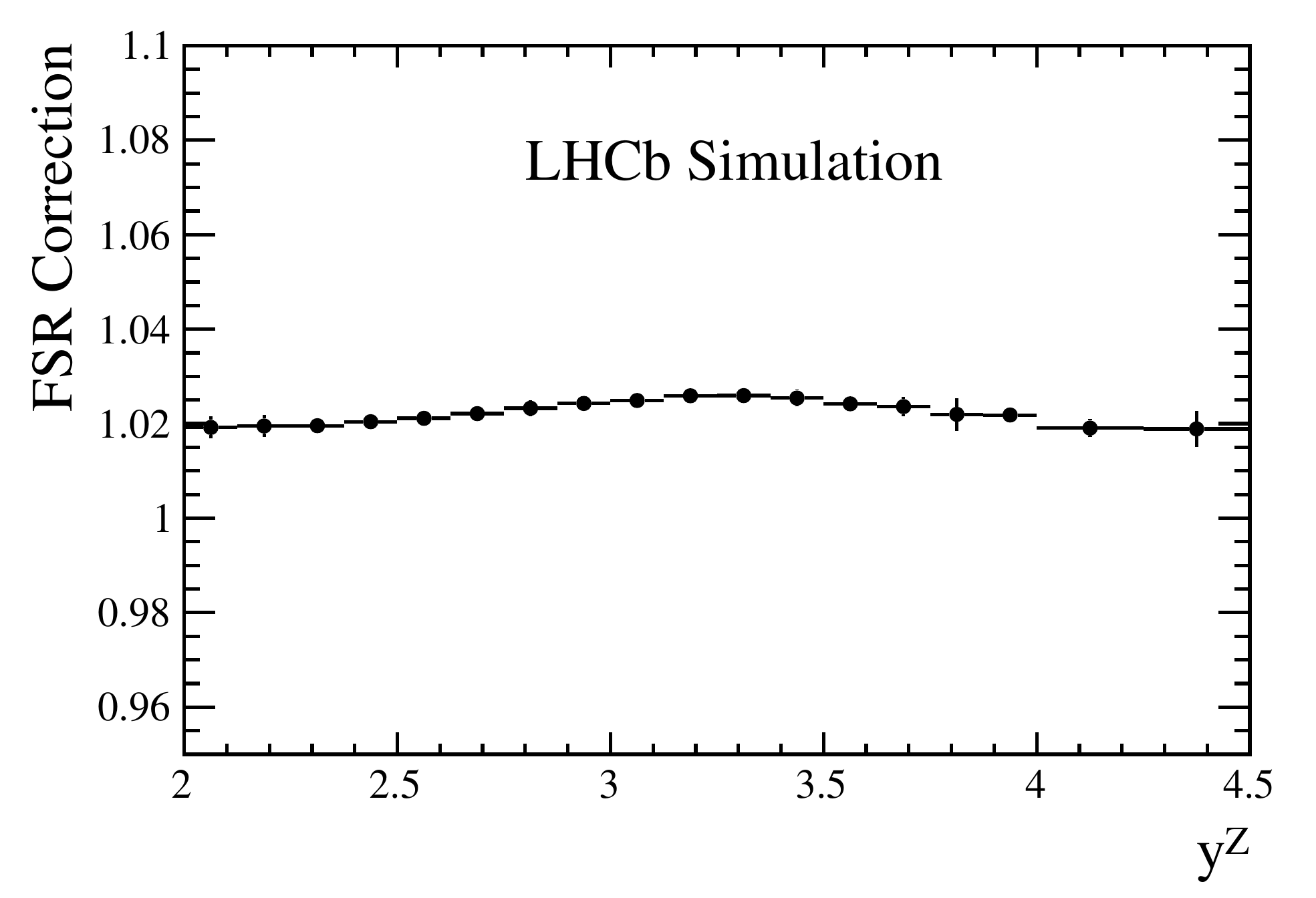}
\includegraphics[width=0.45\textwidth]{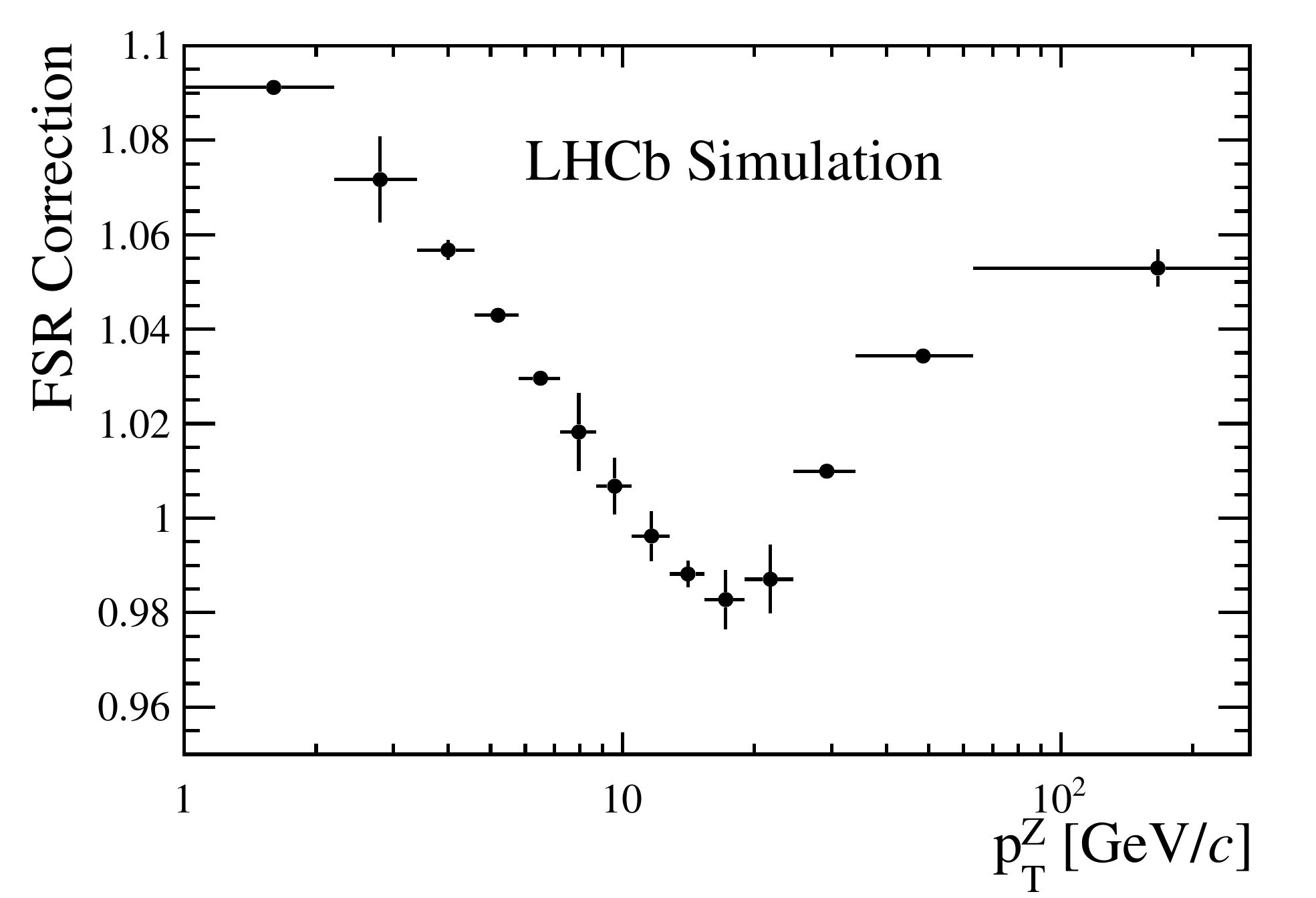}
\includegraphics[width=0.45\textwidth]{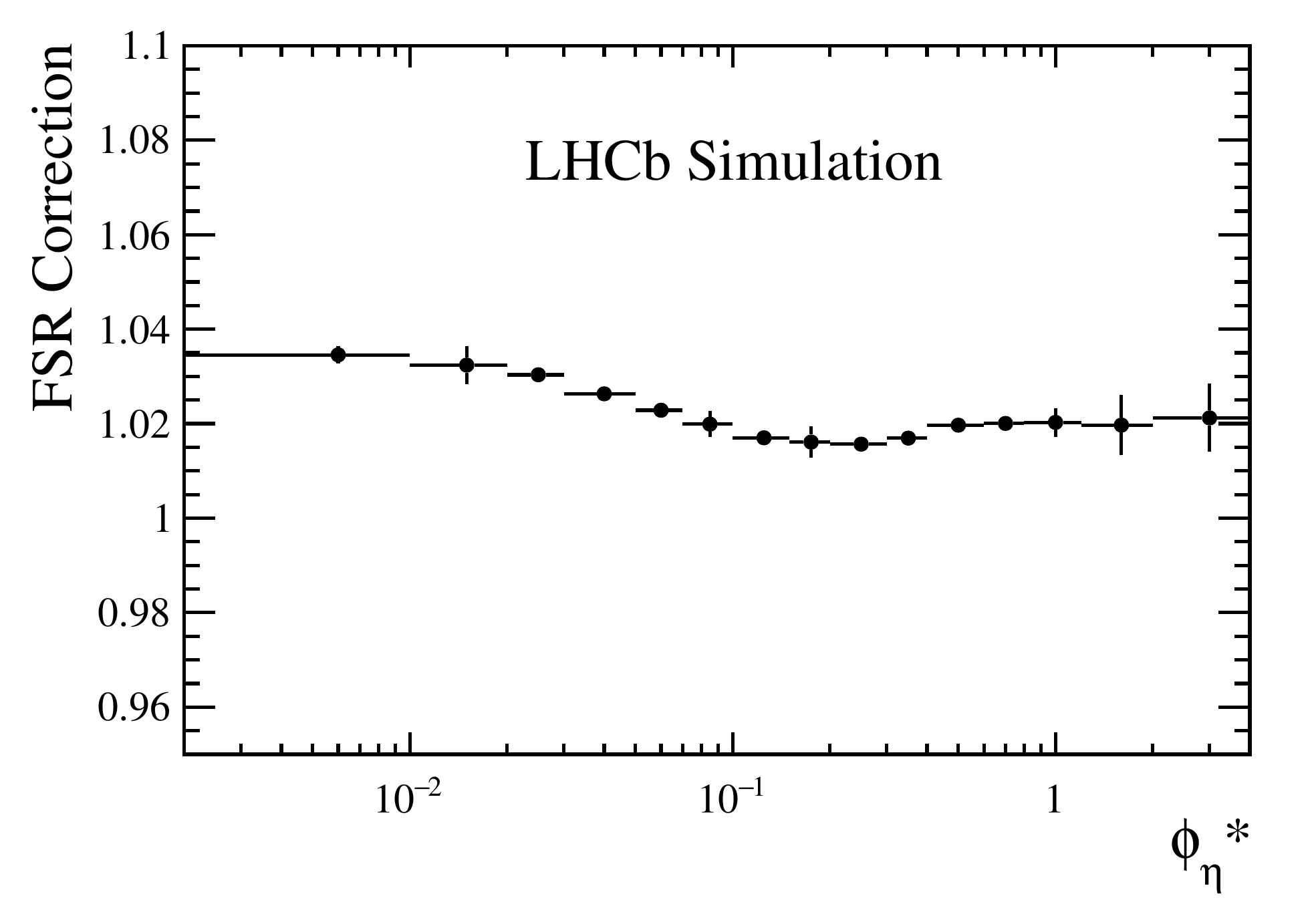}
\caption{Final state radiation correction estimated for the (top-left) \zy, (top-right) \zpt, and (bottom) \phistar differential cross-section measurements. The error bars represent the total (statistical and systematic) uncertainties.}
\label{fig:FSR_corr}
\end{center}
\end{figure}

The measured cross-section is corrected to the Born level in QED,
so that it can be
directly compared with different theoretical predictions. 
The final-state radiation correction is developed and applied to the measurements, 
by comparing the \resbos~\cite{Balazs:1997xd} predictions with and without the implementation of \photos~\cite{davidson2015photos}, which corrects the bare level muon to the Born level.
The FSR corrections in regions of \zy, \zpt, and \phistar 
are shown in Fig.~\ref{fig:FSR_corr}.
The corrections for single- and double-differential cross-section measurements are presented in Appendix~\ref{app:fsr}.

\section{Systematic uncertainty}

Various sources of systematic uncertainty in the cross-section measurement are estimated and combined in quadrature. These include uncertainty from background estimation, detector alignment and momentum scale calibration, 
efficiency, unfolding, closure test, the FSR correction and luminosity. 

\subsection{Background}
In the heavy flavour background determination, the averaged yield is used as the background contribution. 
Its uncertainty is taken as the difference of the background yields estimated using the two control samples.
Furthermore, the mass region and the selection requirements of the control samples are varied and the difference is taken as an additional systematic uncertainty. 

For the hadron misidentification and other background estimated from simulation, 
a systematic uncertainty is assigned to take into account the limited sample size 
of the same-sign data and simulation. 
Furthermore, the difference between the number of $\mu^+\mu^+$ and $\mu^-\mu^-$ events is taken
as an additional uncertainty.
For background estimated from the simulation,
uncertainty from the theoretical predictions are also taken into account.

\subsection{Detector alignment and momentum scale calibration}
To estimate a systematic uncertainty for the detector alignment, the data sample is divided into two independent sub-samples.
Then, a new alignment correction is developed using one of these sub-samples. The new determined correction is applied to the other sub-sample, and {\it vice versa}. 
The difference in the measurements using these two alignment corrections are taken as systematic uncertainty. 
 The uncertainty from alignment and calibration is found to be negligible in the integrated cross-section measurement, and is determined for the differential cross-section measurements.
 
\subsection{Efficiency correction}
The efficiencies of track reconstruction, identification and trigger of the high \pt muons are directly measured from data using the \Zmm events.
A systematic uncertainty is assigned for variations due to the limited size of the control samples, which
is determined to be 0.05\% for trigger efficiency, 0.11\% for identification efficiency, and 0.29\% for tracking efficiency.

The measured data efficiency is corrected for bias from the method itself. 
The correction developed from the simulation sample is broken down into the bias correction and the track matching correction. These two corrections are estimated using the simulation sample, and applied to the measured tag-and-probe
efficiency. Differences between the simulation truth level efficiency and the corrected tag-and-probe efficiency of the simulation sample
are assigned as systematic uncertainty. Furthermore, differences between the matching efficiency and the MuonTT track finding efficiency estimated from data and simulation samples are also considered as uncertainty.
Finally, the estimated systematic uncertainty in tracking efficiency for each muon is determined to be 0.47\%,
which is one of dominant source of systematic uncertainty.

In total, the uncertainty from the efficiency corrections is determined to be 0.77\% in the integrated cross-section measurement.

\subsection{Closure test}
As one-dimensional efficiency corrections in muon $\eta$ regions are used, 
possible additional dependence of efficiencies is not accounted for. 
To check effects from the multi-dimensional efficiency dependence, \ie muon $\eta$ and \pt regions, 
the number of reconstructed events in simulation is corrected using the efficiencies determined from the simulation, 
and compared to the yield at generator level. 
The differences, which show no evidence of a systematic trend across the regions, are assigned as uncertainty.

\subsection{Other sources of systematic uncertainty}
To estimate the uncertainty from unfolding, the \pt distribution is unfolded using the bin-by-bin correction approach~\cite{Adye:2011gm}. 
The difference of results with respect to the Bayesian approach is taken as a systematic uncertainty
on the differential cross-section measured in \zpt region.

The systematic uncertainty from the FSR correction is estimated by comparing the default correction with 
that calculated using the \powheg generator, with the \pythia showering. 
The differences of FSR corrections between \resbos with \photos and \powheg with \pythia are taken as systematic uncertainty. 

For the data sample used, the luminosity is determined with a precision of 2.0\%~\cite{LHCb-PAPER-2014-047}. 
The statistical and systematic uncertainties in the integrated cross-section measurement are presented in Table~\ref{tab:uncertainty}.

\begin{table}[tb]
\begin{center}
\caption{Relative uncertainty for the integrated $\Zmm$ cross-section measurement. The total uncertainty is the quadratic sum of uncertainties from statistical, systematic and luminosity contributions.}
\begin{tabular}{lc}
\hline
 Source               & $\Delta\sigma/\sigma$ [\%]   \\
\hline
Statistical    & 0.11 \\ \hline
Background     & 0.06 \\
Alignment \& calibration & - \\
Efficiency       & 0.77 \\
Closure & 0.23 \\
FSR            & 0.15 \\ \hline
Total Systematic (excl. lumi.) & 0.82  \\ \hline
Luminosity     & 2.00 \\ \hline
Total          & 2.16 \\ 
\hline
\end{tabular}
\label{tab:uncertainty}
\end{center}
\end{table}

\section{Results}

The datasets that were collected in 2016, 2017 and 2018 are considered as independent datasets, which are used to perform cross-section measurements and combined to get results of the full dataset.
In the combination of integrated cross-section measurements obtained from the different datasets, the systematic uncertainties from FSR, background modelling, luminosity, and closure test, 
are treated as 100\% correlated between different datasets, and 
other systematic uncertainties are assumed to be uncorrelated.
The combination is performed using the Best Linear Unbiased Estimator (BLUE) method~\cite{Lyons:1988rp,Valassi:2003mu}.

\subsection{Differential cross-section results}

\begin{figure}[tb]
\begin{center}
\includegraphics[width=0.9\textwidth]{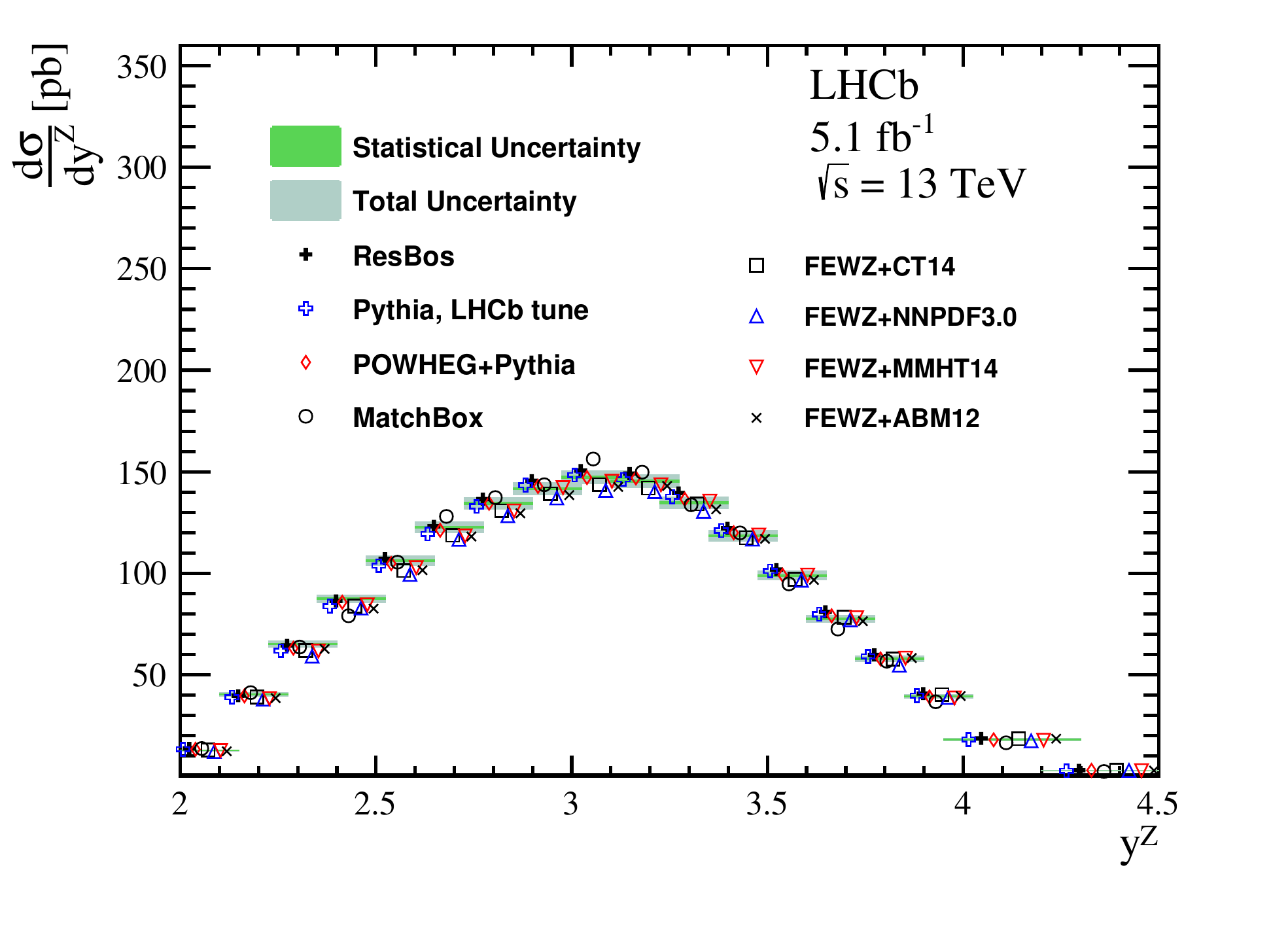}
\includegraphics[width=0.9\textwidth]{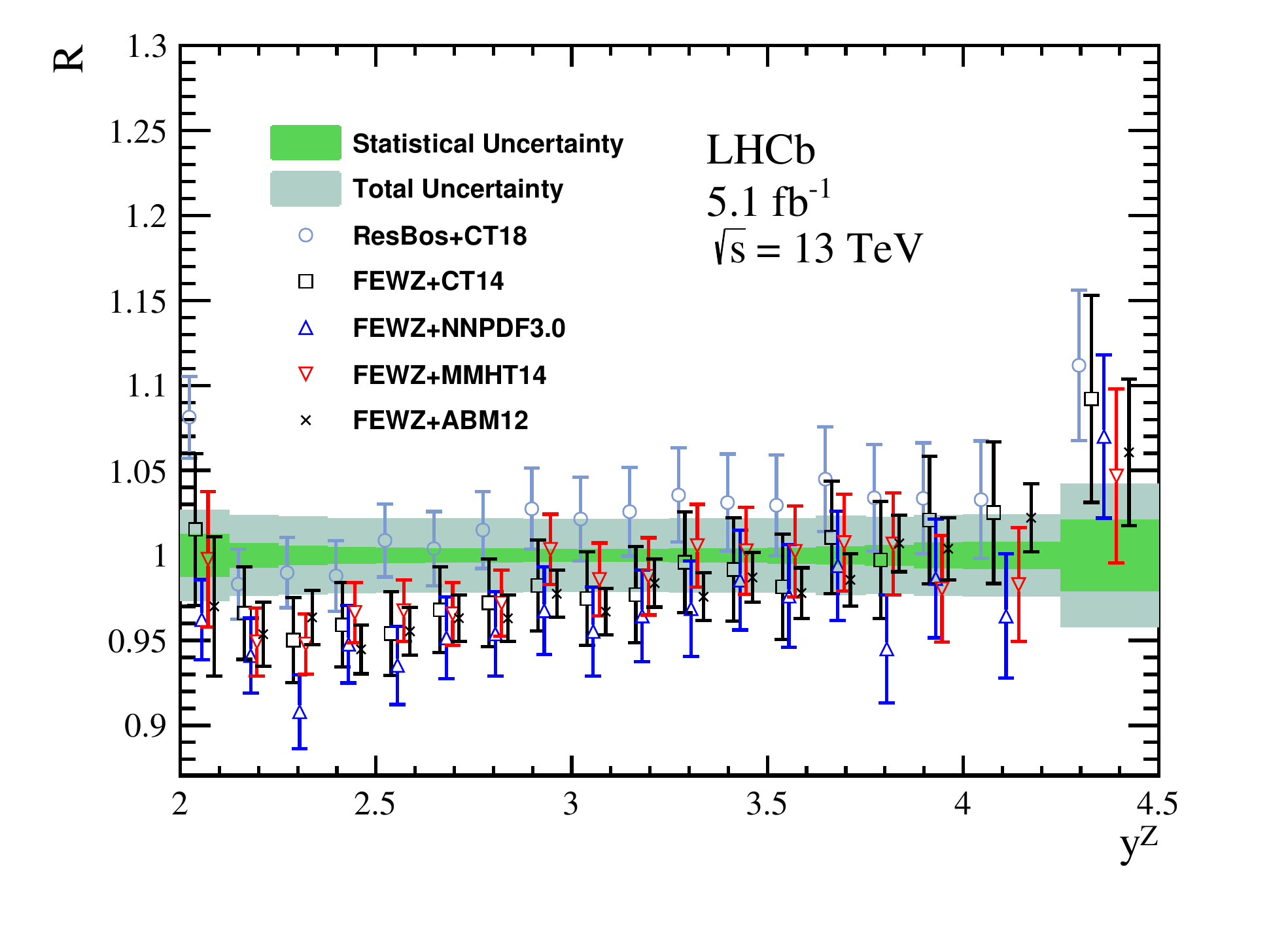}
\caption{(Top) Measured single differential cross-section in regions of \zy, compared with different theoretical predictions. In order to present the measurements more clearly, data bands are drawn wider than the width of the interval. (Bottom) Ratio of theoretical predictions to measured values, with the horizontal bars showing the uncertainty from the PDFs.}
\label{fig:differential_xsec_y}
\end{center}
\end{figure}

The measured differential cross-section in regions of \zy is shown in Fig.~\ref{fig:differential_xsec_y}.
Different theoretical predictions are compared with the measurements, and ratios ($R$) between predictions and data are also shown.
The \resbos~\cite{Balazs:1997xd} prediction combines a next-to-leading order (NLO) fixed-order calculation at high \Z boson \pt with the
Collins-Soper-Sterman resummation formalism~\cite{Collins:1984kg,Collins:1981uk,Collins:1981va} 
at low boson \pt, which is an all-order summation of large terms from gluon emission.
\resbos is used to get predictions for all measurements, by generating a \Zmm sample using the CT18NNLO PDFs~\cite{Hou:2019efy}.
\powheg-BOX~\cite{Nason:2004rx,Frixione:2007vw,Alioli:2008gx,Alioli:2010xd} 
can be interfaced with \pythia for QCD showering. 
\fewz~\cite{Gavin:2010az} is a fixed-order generator for hadron collider production of lepton pairs through the Drell-Yan process at NNLO in the strong coupling constant.
\herwig~\cite{Bahr:2008pv,Bellm:2015jjp} with \matchbox mode is also used to 
get predictions, where \matchbox is a generator interfaced with higher-order corrections 
provided by \herwig. 
As in the \zy measurement the resummation effects are integrated, the measurements are compared with
the predictions from a resummation calculation (\resbos) and other higher-order calculations (\fewz with CT14NNLO~\cite{Dulat:2015mca}, NNPDF3.0NNLO~\cite{NNPDF:2014otw}, 
MMHT14NNLO~\cite{Harland-Lang:2014zoa} and ABM12NNLO~\cite{Alekhin:2013nda}, 
\powheg, \matchbox with NNPDF3.1NLO~\cite{NNPDF:2017mvq}, \pythia with CT09MCS~\cite{Lai:2009ne}, and $\as=0.118$). 
Measurements are in good agreement with theoretical predictions. 
However, the \fewz predictions for the ratios are systematically smaller than the measured results in the lower \zy region, from 2.0 to 3.0. The ratio predicted by \resbos using CT18NNLO is consistent with the meausred results. 
A consistency check has been performed, by fitting the measured differential cross-section
of \Z boson with the framework of the published \W boson mass result~\cite{LHCb-PAPER-2021-024}, where the fitted $\alpha_S^Z$ from two analyses are consistent with each other within uncertainty. 

\begin{figure}[tb]
\begin{center}
\includegraphics[width=0.9\textwidth]{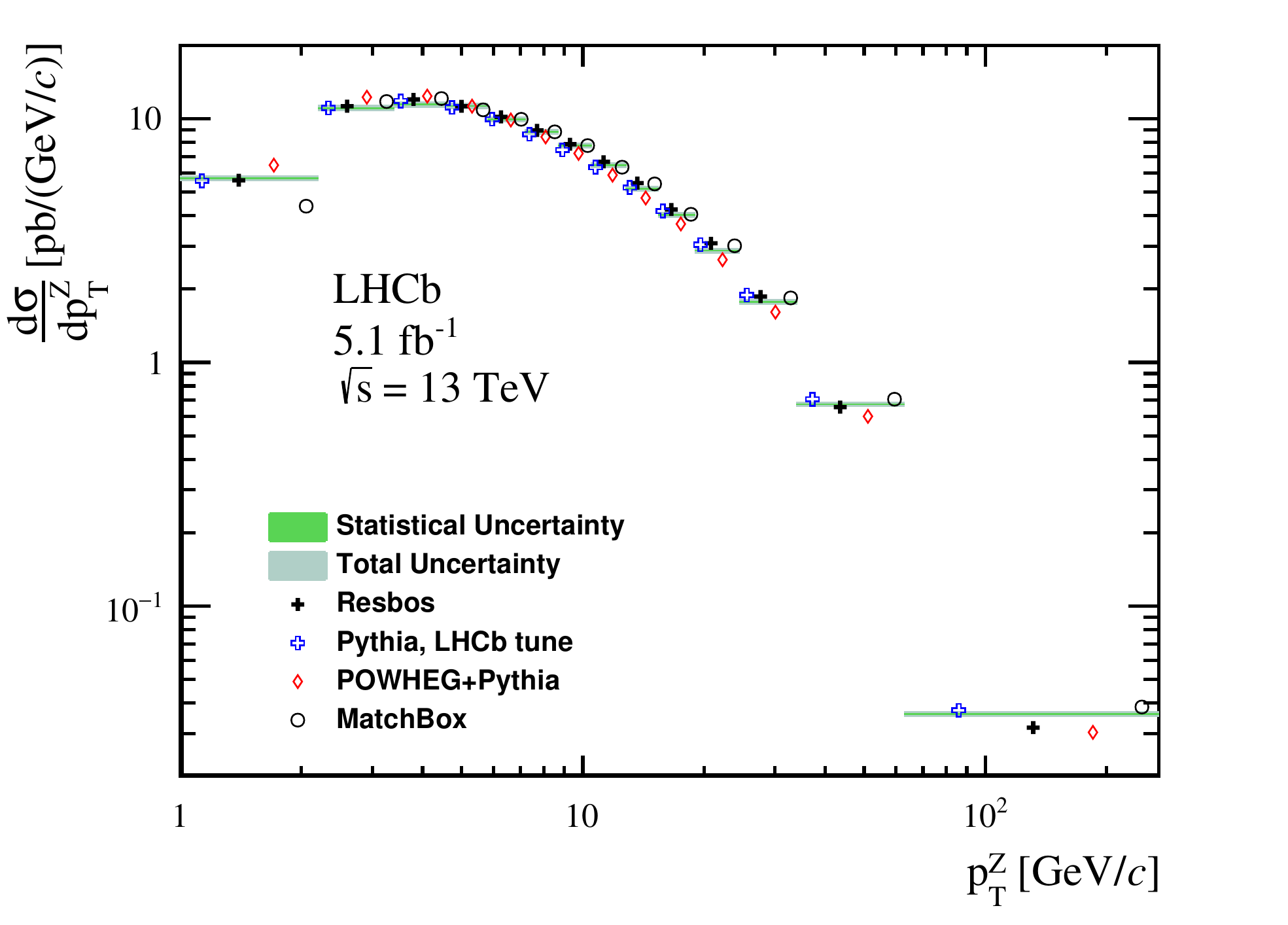}
\includegraphics[width=0.9\textwidth]{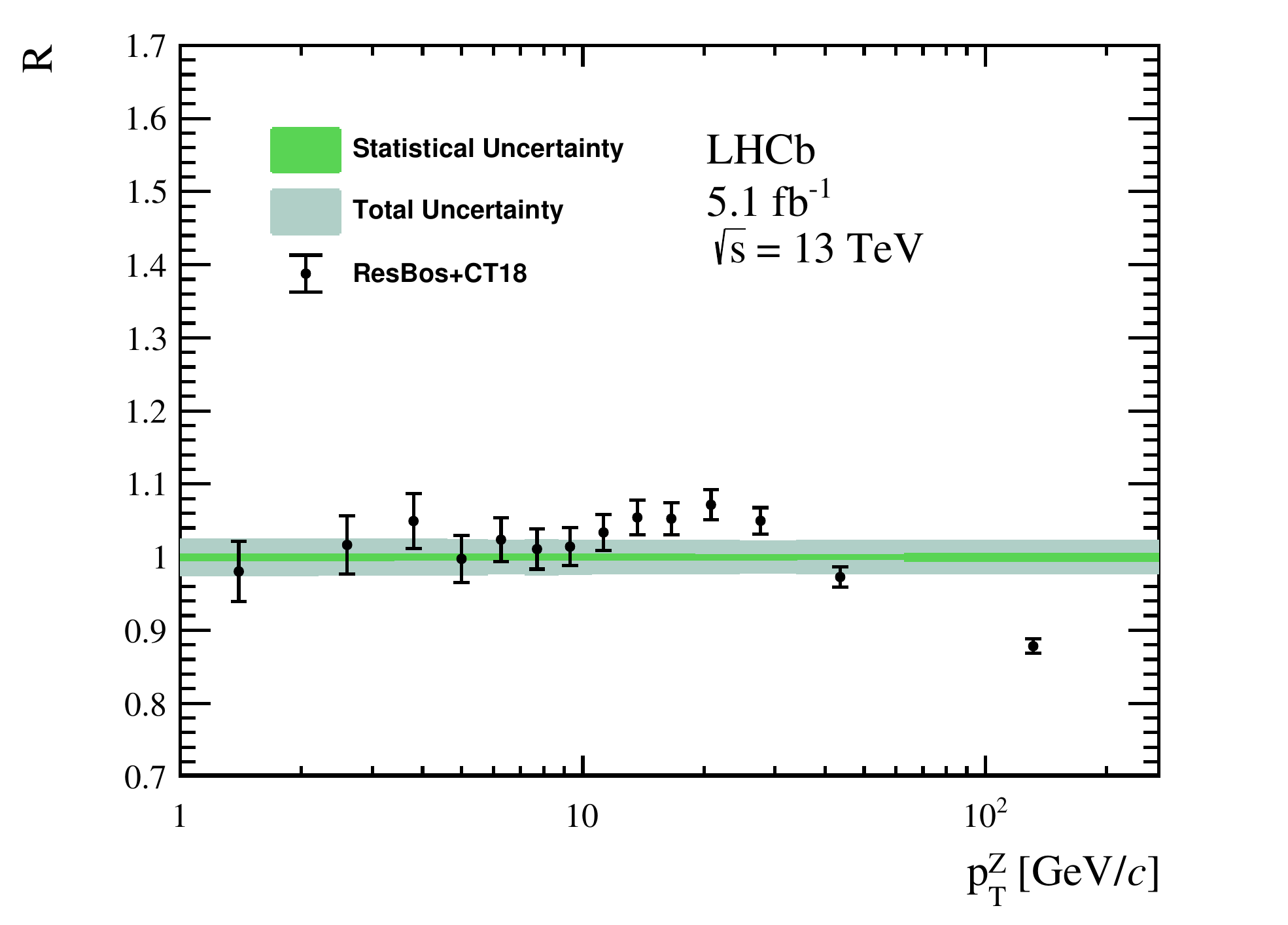}
\caption{(Top) Measured single differential cross-section in regions of \zpt, compared with different theoretical predictions. (Bottom) Ratio of \resbos predictions to measurement, 
with the horizontal bars showing the uncertainty from the PDFs. }
\label{fig:differential_xsec}
\end{center}
\end{figure}

\begin{figure}[tb]
\begin{center}
\includegraphics[width=0.9\textwidth]{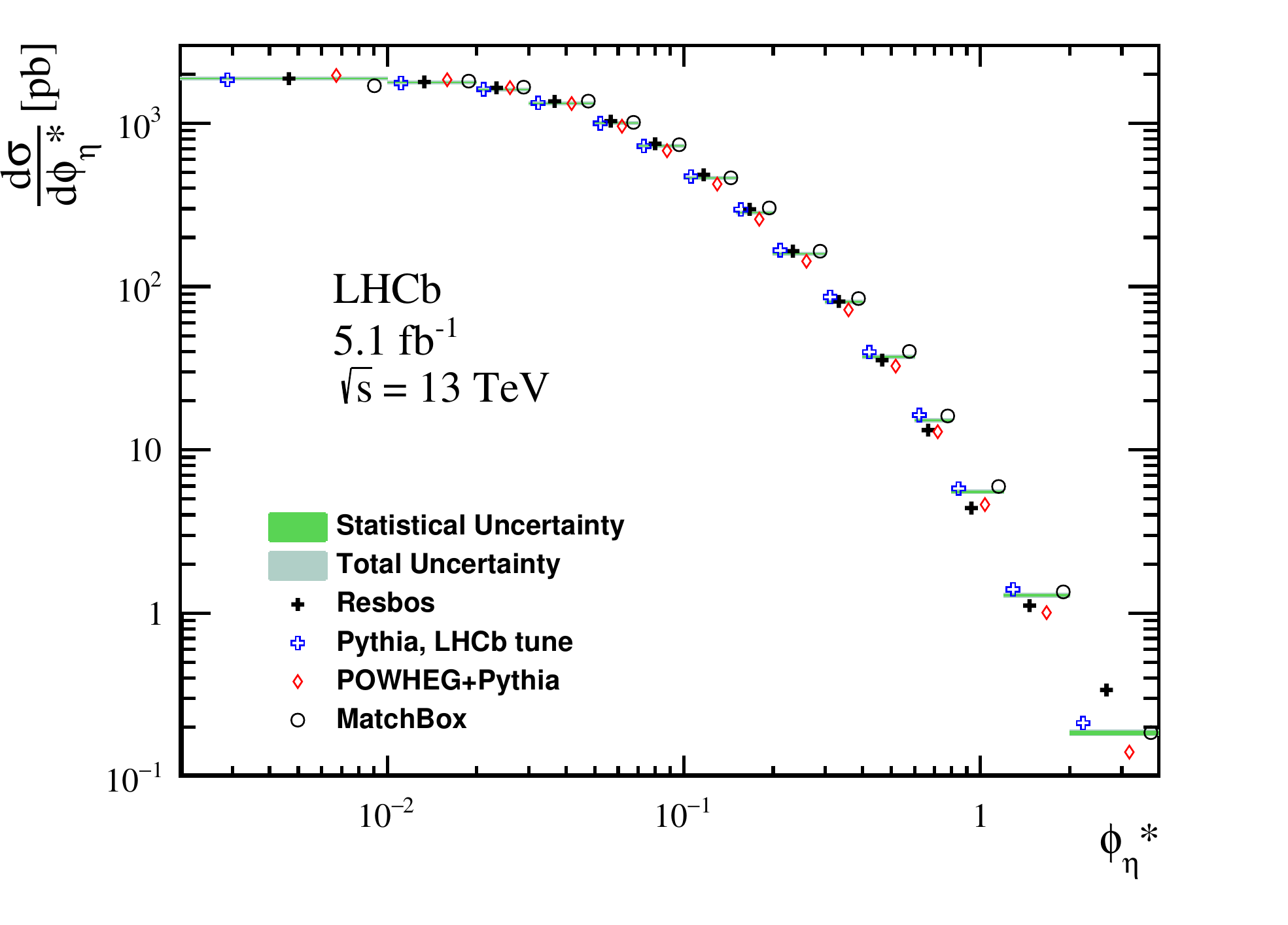}
\includegraphics[width=0.9\textwidth]{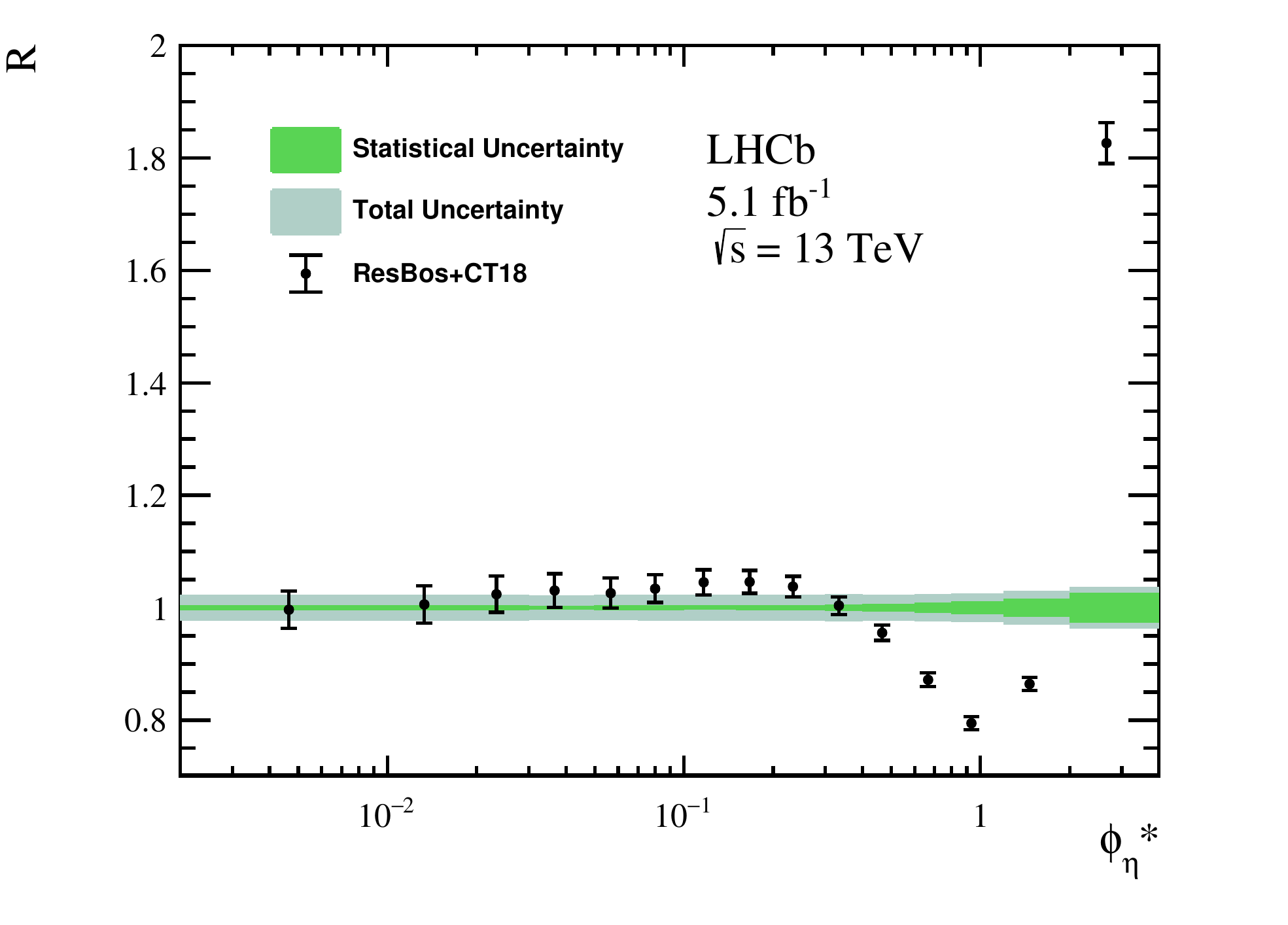}
\caption{(Top) Measured single differential cross-section in regions of \phistar, compared with different theoretical predictions. (Bottom) Ratio of \resbos predictions to measurement, with the horizontal bars showing the uncertainty from the PDFs. }
\label{fig:differential_xsec_phi}
\end{center}
\end{figure}

The single differential cross-sections in regions of \zpt and \phistar are shown in Fig.~\ref{fig:differential_xsec} and Fig.~\ref{fig:differential_xsec_phi}, with ratios ($R$) of predictions to data are shown.
Measurements are in reasonable agreement with the different theoretical predictions. 
In the lower \pt region, the measurements agree with predictions from \resbos and \pythia with the \lhcb tune~\cite{Skands:2014pea}, but disagree with other predictions.
In the large \zpt and \phistar regions, the \resbos predictions are in disagreement with the measured data.
The \powheg with \pythia prediction is larger than the measurements in the small \zpt region, and smaller in the middle and large \zpt region, indicating that the \powheg
prediction cannot describe the data. The predictions from \matchbox are smaller than the data in the first \zpt region, and larger than data in other \zpt regions. Similar conclusions are obtained for the predictions as a function of \phistar.

\begin{figure}[tb]
\begin{center}
\includegraphics[width=0.95\textwidth]{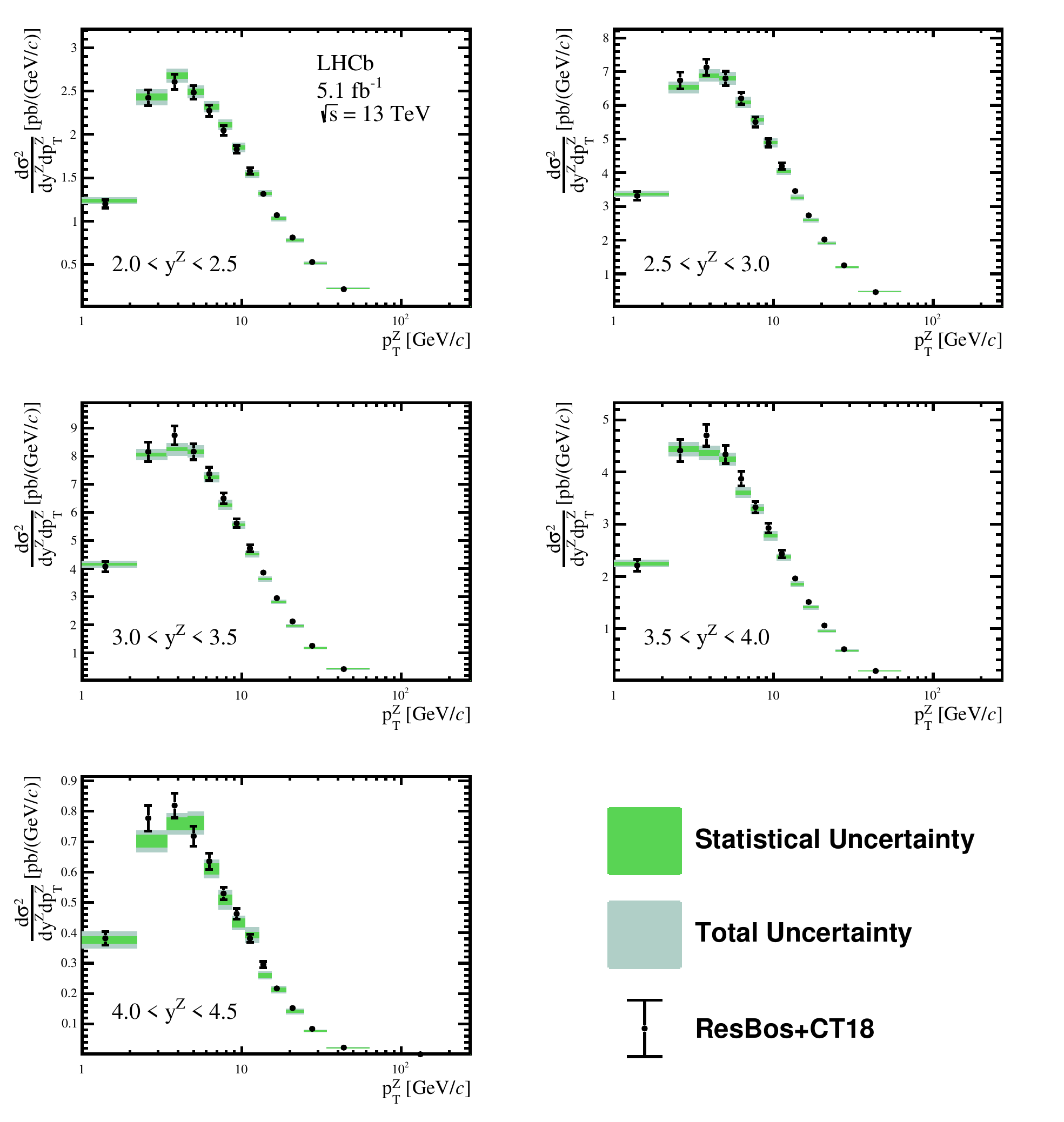}
\caption{Measured double differential cross-section as a function of \zpt in regions of \zy, compared with \resbos predictions, with the horizontal bars showing the uncertainty from the PDFs. }
\label{fig:double_differential_xsec}
\end{center}
\end{figure}

\begin{figure}[tb]
\begin{center}
\includegraphics[width=0.95\textwidth]{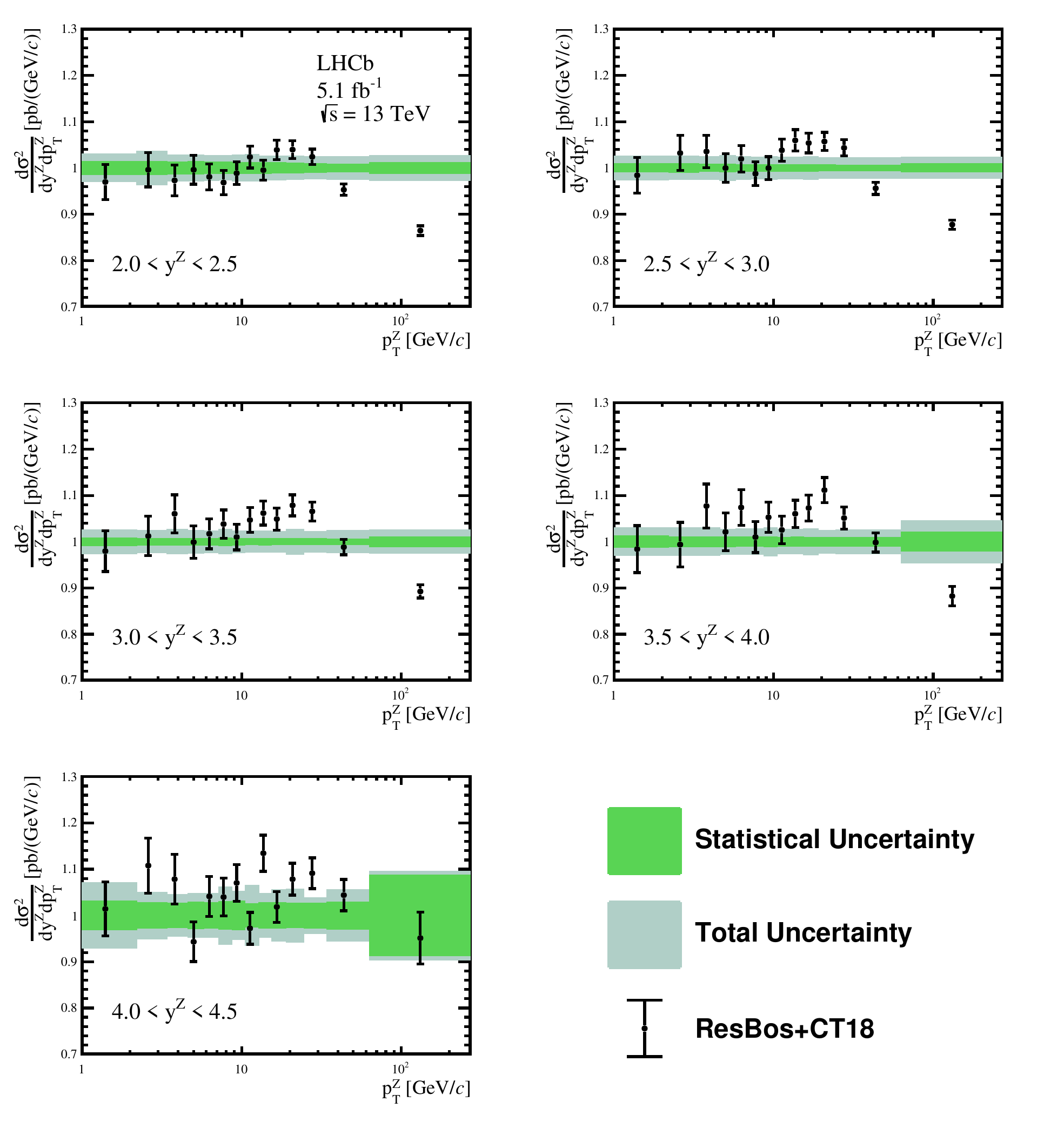}
\caption{Ratios of \resbos predictions to measurements as a function of \zpt in regions of \zy, with the horizontal bars showing the uncertainty from the PDFs.}
\label{fig:double_differential_xsec_ratio}
\end{center}
\end{figure}

\begin{figure}[tb]
\begin{center}
\includegraphics[width=0.95\textwidth]{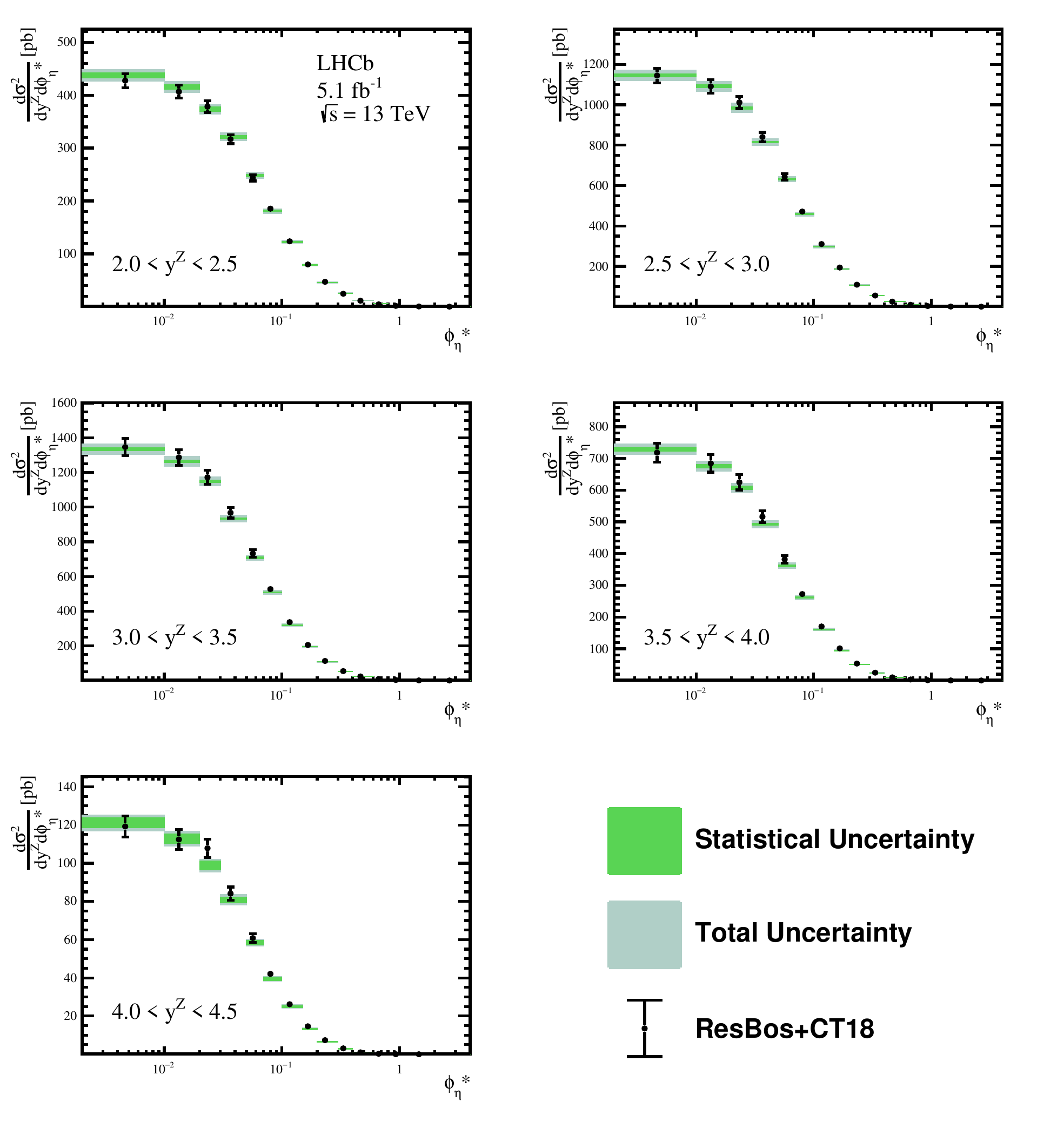}
\caption{Measured double differential cross-section as a function of \phistar in regions of \zy, compared with \resbos predictions, with the horizontal bars showing the uncertainty from the PDFs.}
\label{fig:double_differential_xsec_phi}
\end{center}
\end{figure}

\begin{figure}[tb]
\begin{center}
\includegraphics[width=0.95\textwidth]{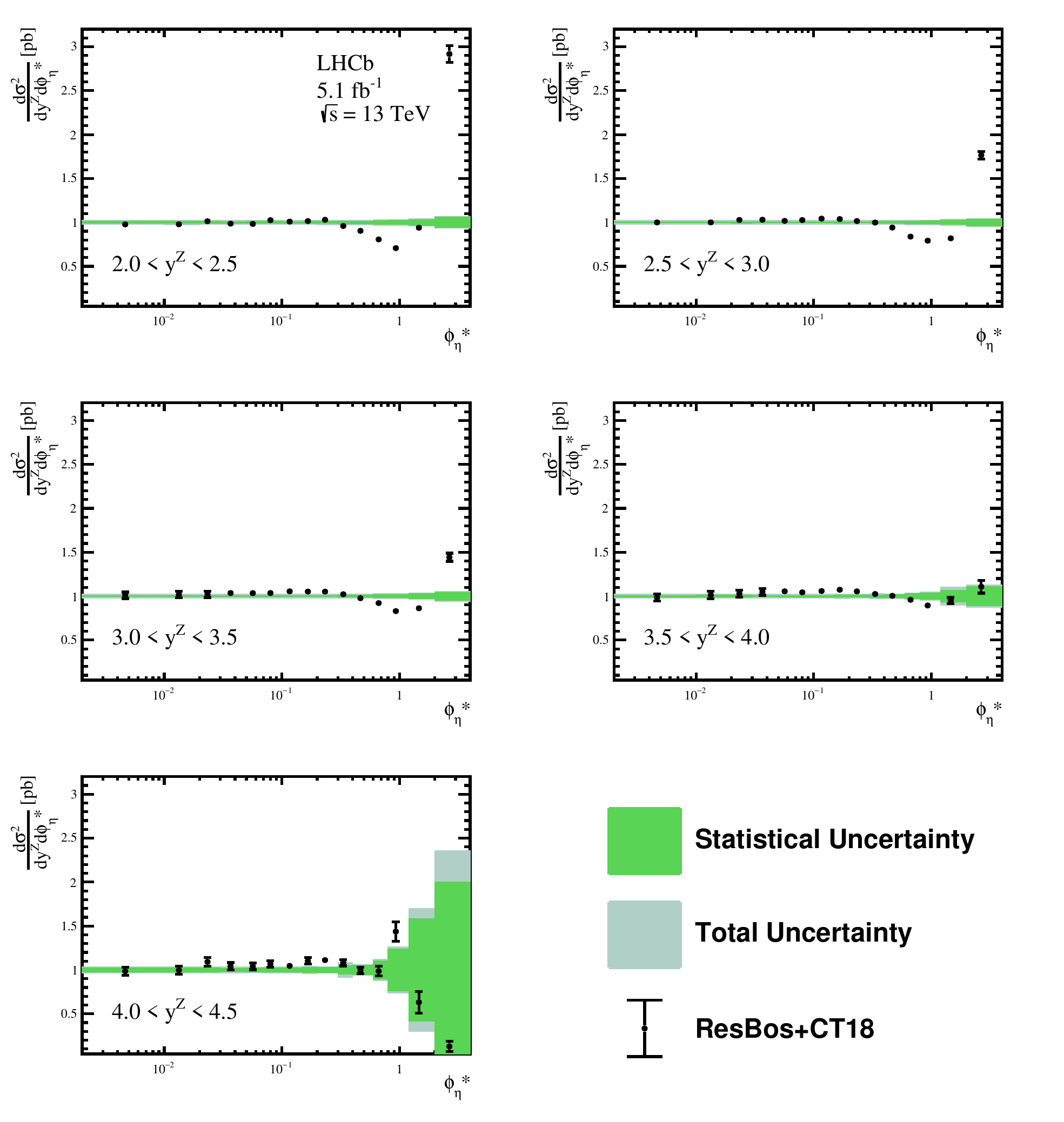}
\caption{Ratios of \resbos predictions to measurements as a function of \phistar in regions of \zy, with the horizontal bars showing the uncertainty from the PDFs.}
\label{fig:double_differential_xsec_phi_ratio}
\end{center}
\end{figure}

Thanks to the large size of the data sample, double differential cross-section measurements are performed in regions of \zy and \zpt, and \zy and \phistar. 
The measurements are compared with the \resbos predictions in Fig.~\ref{fig:double_differential_xsec}, with ratios ($R$) of predictions to data shown in Fig.~\ref{fig:double_differential_xsec_ratio}, for $\zy-\zpt$. The corresponding results for $\zy-\phistar$ are shown in Figs.~\ref{fig:double_differential_xsec_phi} and~\ref{fig:double_differential_xsec_phi_ratio}. 
The \resbos predictions are consistent with the measured results within the uncertainty.
In the large \zpt and \phistar regions, there are sizable disagreements between data and predictions. 
Numerical results and systematic uncertainties are shown in Appendix~\ref{app:results}.

\subsection{Correlation matrices}
The statistical correlation due to the event migration between regions is determined using simulation, 
where the numbers of signal events in different generator-level and reconstruction-level regions are used. 
The calculated correlation matrices are shown in Appendix~\ref{app:correlation}.
There are large correlations in the low 
\zpt region, and small correlations in the high \zpt region. 
On the other hand, the statistical correlation between regions in \zy or in \phistar is found to be negligible. 

In the differential cross-section measurements, the systematic uncertainties from background, alignment, efficiency closure test, and FSR
are considered to be 50\% correlated between different regions. The luminosity uncertainty is considered to be 100\% correlated between different regions.
Regarding uncertainties from the selection efficiencies, the correlation between regions is determined by varying the 
efficiencies within their uncertainty, as
\begin{equation}
    {\rm{cov}}(f_k, f_l)= \sum_i\sum_j\left(\frac{\partial f_k}{\partial x_i}\right)\left(\frac{\partial f_l}{\partial x_j}\right)\cdot {\rm{cov}}(x_i,x_j),
\end{equation}
where $x_{i,j}$ is the determined efficiencies in $i,j$-th muon $\eta$ region, $f_{k,l}$ is the measured \Z boson cross-section with given efficiencies, and ${\rm{cov}}(x_i,x_j)$ the correlation coefficient between the $i$- and $j$-th region. 
The calculated correlation matrices for efficiencies are presented in Appendix~\ref{app:correlation}. 
Because of the presence of two muons in the final state, there are large correlations between different regions in \zpt measurement. 
However, for the \zy and \phistar measurements, small correlations are presented in most off-diagonal regions.

\subsection{Integrated cross-section results}
The measurements of integrated cross-section from different datasets are shown in Appendix~\ref{app:results}.
The $\chisq$ per degree-of-freedom of the combination is determined to be 0.9/2, using the BLUE method. 

\begin{figure}[tb]
\begin{center}
\includegraphics[width=0.8\textwidth]{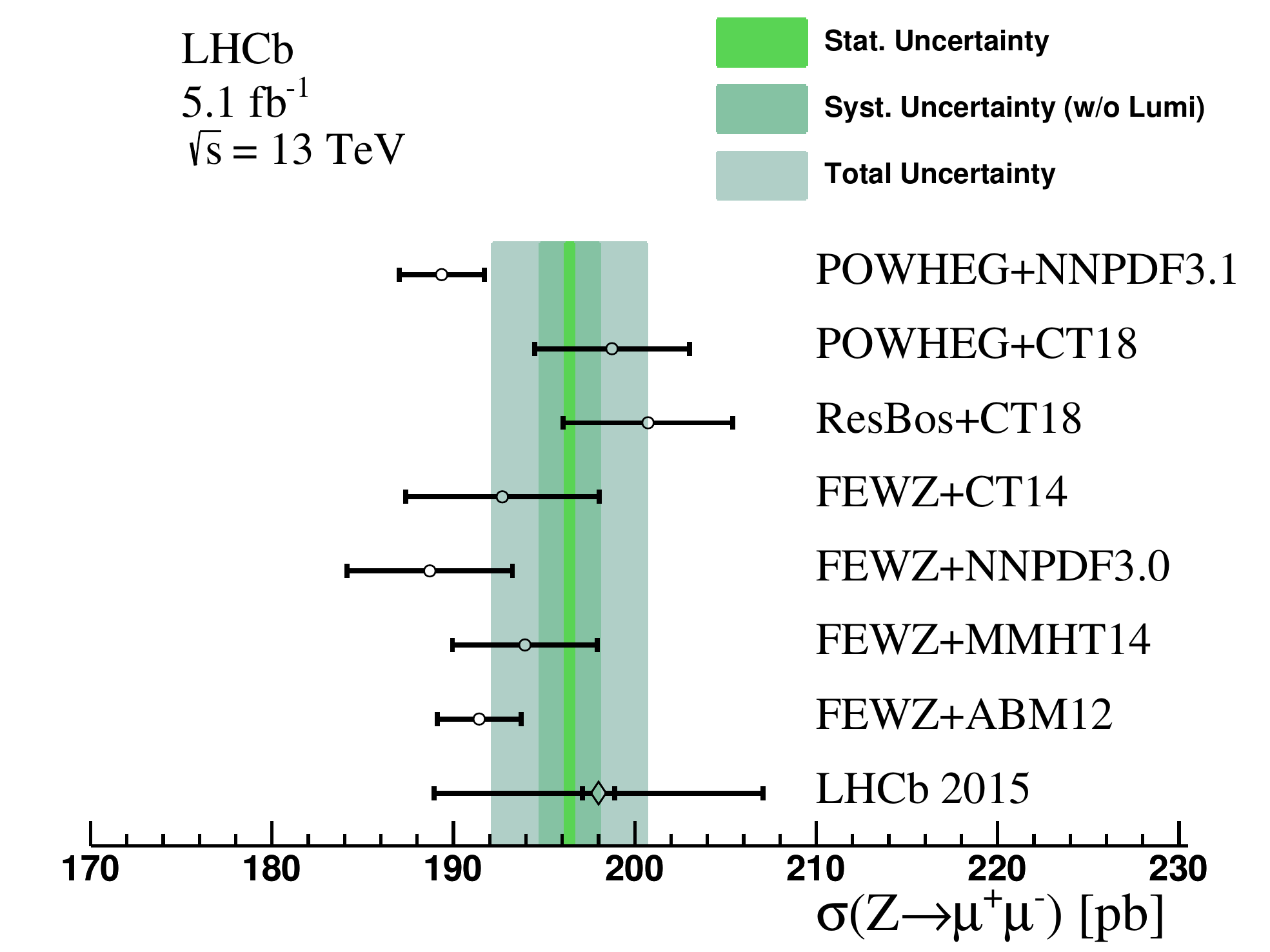}
\caption{Comparison of the integrated cross-section, $\sigma_{\Zmm}$, between data and theoretical predictions. 
The bands correspond to the data, with the inner band corresponding to the statistical uncertainty and the outer bands corresponding to the systematic uncertainty and total uncertainty. 
The open circles correspond to the different theoretical predictions. 
The diamond point corresponds to the previous \lhcb measurement~\cite{LHCb-PAPER-2016-021}.}
\label{fig:total_xsec}
\end{center}
\end{figure}

In the \lhcb detector fiducial region, the \Z boson integrated production cross-section is measured to be
\begin{equation}
 \sigma(\Zmm) = \totxsec\pb, \nonumber
\end{equation}
where uncertainties are statistical, systematic, and due to the luminosity measurement, respectively. 
In Fig.~\ref{fig:total_xsec} the determined integrated cross-section is compared with different theoretical predictions and the previous \lhcb measurement~\cite{LHCb-PAPER-2016-021}.
In the comparison, the predictions are calculated using \powheg with NNPDF3.1NLO~\cite{NNPDF:2017mvq}, 
\powheg with CT18NLO~\cite{Hou:2019efy}, \resbos with CT18NNLO, \fewz with CT14NNLO~\cite{Dulat:2015mca}, \fewz with NNPDF3.0NNLO~\cite{NNPDF:2014otw}, 
\fewz with MMHT14NNLO~\cite{Harland-Lang:2014zoa}, and \fewz with ABM12NNLO~\cite{Alekhin:2013nda}, with both their statistical and PDF uncertainties.
Measurements are in reasonable agreement with all theoretical predictions.

\section{Conclusion}
In summary, the most precise measurement to date of the \Z boson production cross-section in the forward region at $\sqs=13\tev$ is presented,
using $pp$ collision data collected with the \lhcb detector. 
The dataset corresponds to an integrated luminosity of ${\runtwolumi\invfb}$.
The integrated cross-section in fiducial region is measured to be
\begin{equation*}
    \sigma_{\Zmm} = \totxsec\pb,
\end{equation*}
where the first uncertainty is statistical, the second is systematic, and the third is due to the luminosity determination.
The single differential and the double differential cross-sections are measured. This is the first measurement of the double differential cross-section in the forward region. 
Overall, reasonable agreement between measured results and the theoretical predictions are seen. However,
there are sizable disagreements in the large \zpt and \phistar regions, which need more investigations in future. 
These measurements provide important and unique information to the PDF determination, especially in the large and small $x$ regions.
\clearpage

\section*{Acknowledgements}
%
%
\noindent We thank C.-P. Yuan for frequent and interesting discussions on the PDFs.
We express our gratitude to our colleagues in the CERN
accelerator departments for the excellent performance of the LHC. We
thank the technical and administrative staff at the LHCb
institutes.
We acknowledge support from CERN and from the national agencies:
CAPES, CNPq, FAPERJ and FINEP (Brazil); 
MOST and NSFC (China); 
CNRS/IN2P3 (France); 
BMBF, DFG and MPG (Germany); 
INFN (Italy); 
NWO (Netherlands); 
MNiSW and NCN (Poland); 
MEN/IFA (Romania); 
MSHE (Russia); 
MICINN (Spain); 
SNSF and SER (Switzerland); 
NASU (Ukraine); 
STFC (United Kingdom); 
DOE NP and NSF (USA).
We acknowledge the computing resources that are provided by CERN, IN2P3
(France), KIT and DESY (Germany), INFN (Italy), SURF (Netherlands),
PIC (Spain), GridPP (United Kingdom), RRCKI and Yandex
LLC (Russia), CSCS (Switzerland), IFIN-HH (Romania), CBPF (Brazil),
PL-GRID (Poland) and NERSC (USA).
We are indebted to the communities behind the multiple open-source
software packages on which we depend.
Individual groups or members have received support from
ARC and ARDC (Australia);
AvH Foundation (Germany);
EPLANET, Marie Sk\l{}odowska-Curie Actions and ERC (European Union);
A*MIDEX, ANR, IPhU and Labex P2IO, and R\'{e}gion Auvergne-Rh\^{o}ne-Alpes (France);
Key Research Program of Frontier Sciences of CAS, CAS PIFI, CAS CCEPP, 
Fundamental Research Funds for the Central Universities, 
and Sci. \& Tech. Program of Guangzhou (China);
RFBR, RSF and Yandex LLC (Russia);
GVA, XuntaGal and GENCAT (Spain);
the Leverhulme Trust, the Royal Society
 and UKRI (United Kingdom).



\section*{Appendices}

\appendix

\section{Final state radiation corrections}
\label{app:fsr}
\begin{figure}[tb]
\begin{center}
\includegraphics[width=0.45\textwidth]{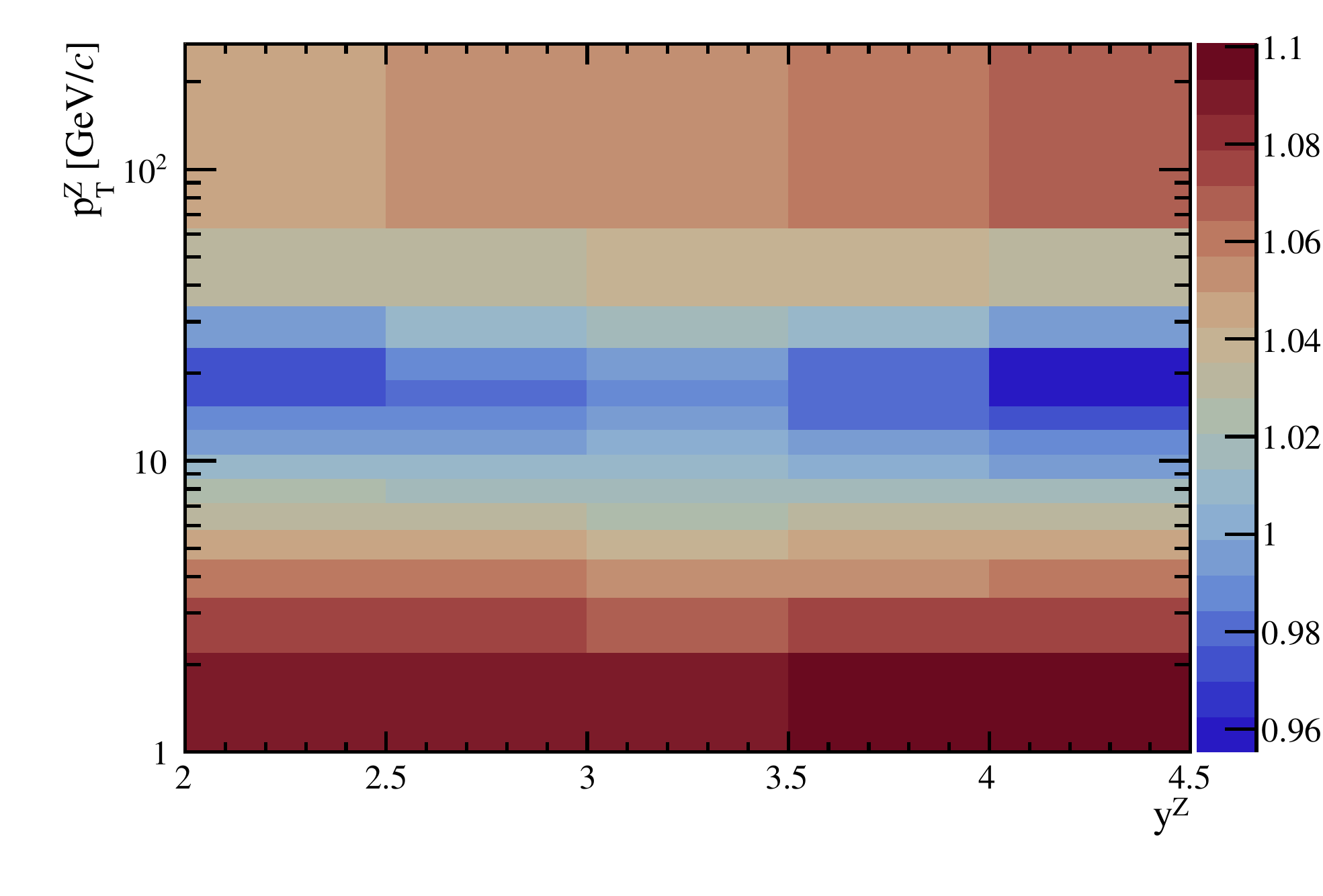}
\includegraphics[width=0.45\textwidth]{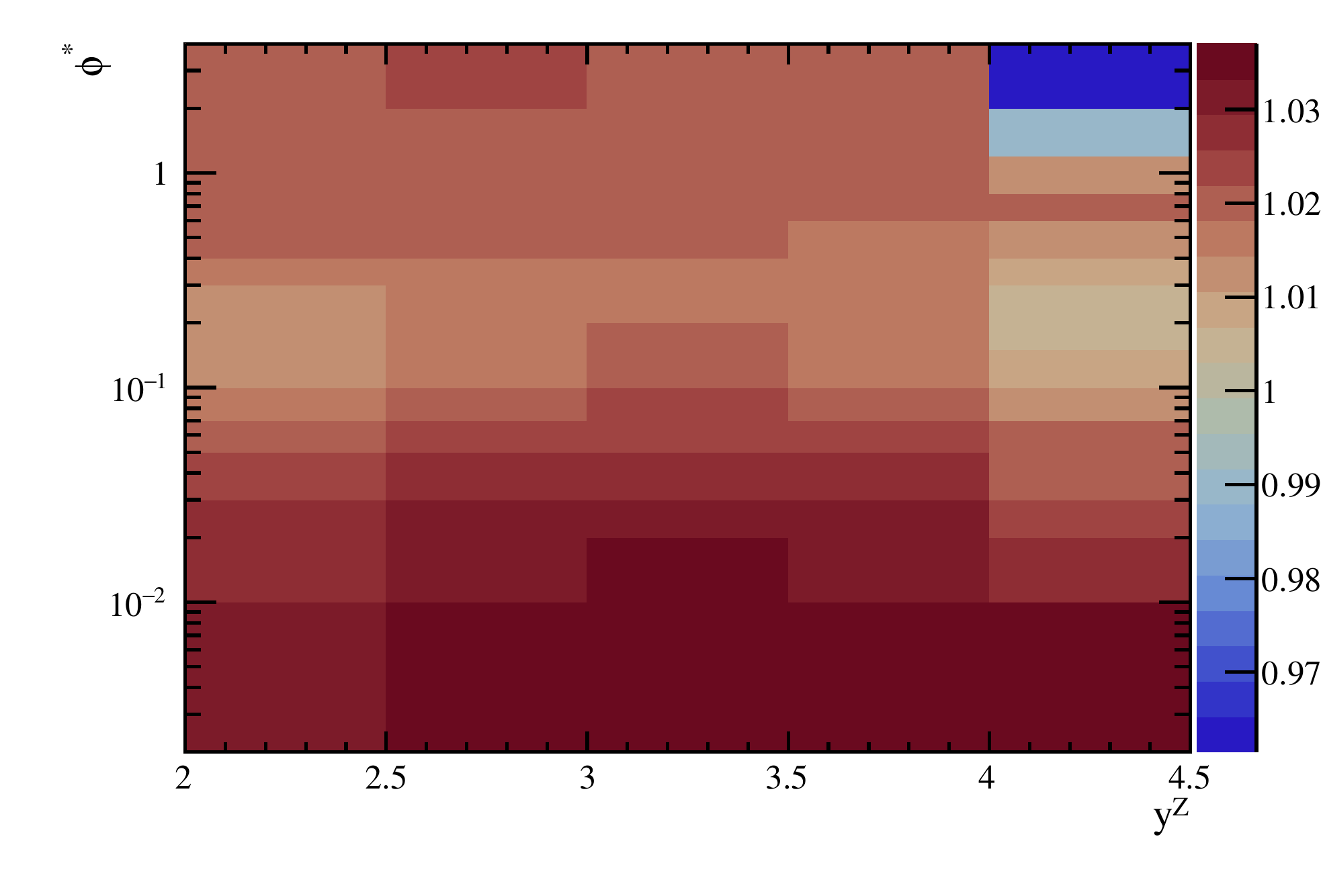}
\caption{Final state radiation correction estimated for double differential cross section measurement for (left) $\zy-\zpt$ measurement, and for (right) $\zy-\phistar$ measurement.}
\label{fig:FSR_corr_2D}
\end{center}
\end{figure}

Final state radiation corrections for double differential cross-section measurement are shown in 
Fig.~\ref{fig:FSR_corr_2D}. 
Tabled results of final state radiation corrections used in the single differential cross-section measurements
are presented from Table~\ref{tab:fsr_1D_y} to Table~\ref{tab:fsr_1D_phi}.
Final state radiation corrections used in the double differential cross-section measurements
are presented in Tables~\ref{tab:fsr_2D_ypt} and~\ref{tab:fsr_2D_yphi}.

\begin{table}[tb]
\begin{center}
\caption{Final state radiation correction used in the \zy cross-section measurement. The first uncertainty is statistical and the second is systematic.}
\begin{tabular}{rcl|ccccc}
\hline
 \multicolumn{3}{c|}{\zy}               & \multicolumn{5}{c}{Correction}  \\ \hline
2.000 & $-$ &  2.125 &     1.019 & $\pm$ &    0.002 & $\pm$ &    0.001  \\ 
2.125 & $-$ &  2.250 &     1.020 & $\pm$ &    0.001 & $\pm$ &    0.002  \\ 
2.250 & $-$ &  2.375 &     1.020 & $\pm$ &    0.001 & $\pm$ &    0.000  \\ 
2.375 & $-$ &  2.500 &     1.020 & $\pm$ &    0.001 & $\pm$ &    0.001  \\ 
2.500 & $-$ &  2.625 &     1.021 & $\pm$ &    0.001 & $\pm$ &    0.001  \\ 
2.625 & $-$ &  2.750 &     1.022 & $\pm$ &    0.001 & $\pm$ &    0.000  \\ 
2.750 & $-$ &  2.875 &     1.023 & $\pm$ &    0.001 & $\pm$ &    0.002  \\ 
2.875 & $-$ &  3.000 &     1.024 & $\pm$ &    0.001 & $\pm$ &    0.000  \\ 
3.000 & $-$ &  3.125 &     1.025 & $\pm$ &    0.001 & $\pm$ &    0.000  \\ 
3.125 & $-$ &  3.250 &     1.026 & $\pm$ &    0.001 & $\pm$ &    0.000  \\ 
3.250 & $-$ &  3.375 &     1.026 & $\pm$ &    0.001 & $\pm$ &    0.001  \\ 
3.375 & $-$ &  3.500 &     1.025 & $\pm$ &    0.001 & $\pm$ &    0.002  \\ 
3.500 & $-$ &  3.625 &     1.024 & $\pm$ &    0.001 & $\pm$ &    0.001  \\ 
3.625 & $-$ &  3.750 &     1.024 & $\pm$ &    0.001 & $\pm$ &    0.002  \\ 
3.750 & $-$ &  3.875 &     1.022 & $\pm$ &    0.001 & $\pm$ &    0.003  \\ 
3.875 & $-$ &  4.000 &     1.022 & $\pm$ &    0.001 & $\pm$ &    0.000  \\ 
4.000 & $-$ &  4.250 &     1.019 & $\pm$ &    0.001 & $\pm$ &    0.001  \\ 
4.250 & $-$ &  4.500 &     1.019 & $\pm$ &    0.003 & $\pm$ &    0.003  \\ 
\hline
\end{tabular}
\label{tab:fsr_1D_y}
\end{center}
\end{table}

\begin{table}[tb]
\begin{center}
\caption{Final state radiation correction used in the \zpt cross-section measurement. The first uncertainty is statistical and the second is systematic.}
\begin{tabular}{rcl|ccccc}
\hline
 \multicolumn{3}{c|}{\zpt~[\gevc]}              & \multicolumn{5}{c}{Correction}  \\ \hline
 0.0 & $-$ &   2.2 &     1.091 & $\pm$ &    0.001 & $\pm$ &    0.000  \\ 
 2.2 & $-$ &   3.4 &     1.072 & $\pm$ &    0.001 & $\pm$ &    0.009  \\ 
 3.4 & $-$ &   4.6 &     1.057 & $\pm$ &    0.001 & $\pm$ &    0.002  \\ 
 4.6 & $-$ &   5.8 &     1.043 & $\pm$ &    0.001 & $\pm$ &    0.001  \\ 
 5.8 & $-$ &   7.2 &     1.030 & $\pm$ &    0.001 & $\pm$ &    0.001  \\ 
 7.2 & $-$ &   8.7 &     1.018 & $\pm$ &    0.001 & $\pm$ &    0.008  \\ 
 8.7 & $-$ &  10.5 &     1.007 & $\pm$ &    0.001 & $\pm$ &    0.006  \\ 
10.5 & $-$ &  12.8 &     0.996 & $\pm$ &    0.001 & $\pm$ &    0.005  \\ 
12.8 & $-$ &  15.4 &     0.988 & $\pm$ &    0.001 & $\pm$ &    0.003  \\ 
15.4 & $-$ &  19.0 &     0.983 & $\pm$ &    0.001 & $\pm$ &    0.006  \\ 
19.0 & $-$ &  24.5 &     0.987 & $\pm$ &    0.001 & $\pm$ &    0.007  \\ 
24.5 & $-$ &  34.0 &     1.010 & $\pm$ &    0.001 & $\pm$ &    0.001  \\ 
34.0 & $-$ &  63.0 &     1.034 & $\pm$ &    0.001 & $\pm$ &    0.000  \\ 
63.0 & $-$ &  270.0 &     1.053 & $\pm$ &    0.001 & $\pm$ &    0.004  \\ 
\hline
\end{tabular}
\label{tab:fsr_1D_pt}
\end{center}
\end{table}

\begin{table}[tb]
\begin{center}
\caption{Final state radiation correction used in the \phistar cross-section measurement. The first uncertainty is statistical and the second is systematic.}
\begin{tabular}{rcl|ccccc}
\hline
 \multicolumn{3}{c|}{\phistar}              & \multicolumn{5}{c}{Correction}  \\ \hline
0.00 & $-$ &  0.01 &     1.035 & $\pm$ &    0.001 & $\pm$ &    0.002  \\ 
0.01 & $-$ &  0.02 &     1.032 & $\pm$ &    0.001 & $\pm$ &    0.004  \\ 
0.02 & $-$ &  0.03 &     1.030 & $\pm$ &    0.001 & $\pm$ &    0.001  \\ 
0.03 & $-$ &  0.05 &     1.026 & $\pm$ &    0.001 & $\pm$ &    0.001  \\ 
0.05 & $-$ &  0.07 &     1.023 & $\pm$ &    0.001 & $\pm$ &    0.000  \\ 
0.07 & $-$ &  0.10 &     1.020 & $\pm$ &    0.001 & $\pm$ &    0.003  \\ 
0.10 & $-$ &  0.15 &     1.017 & $\pm$ &    0.001 & $\pm$ &    0.001  \\ 
0.15 & $-$ &  0.20 &     1.016 & $\pm$ &    0.001 & $\pm$ &    0.003  \\ 
0.20 & $-$ &  0.30 &     1.016 & $\pm$ &    0.001 & $\pm$ &    0.001  \\ 
0.30 & $-$ &  0.40 &     1.017 & $\pm$ &    0.001 & $\pm$ &    0.001  \\ 
0.40 & $-$ &  0.60 &     1.020 & $\pm$ &    0.001 & $\pm$ &    0.000  \\ 
0.60 & $-$ &  0.80 &     1.020 & $\pm$ &    0.001 & $\pm$ &    0.001  \\ 
0.80 & $-$ &  1.20 &     1.020 & $\pm$ &    0.002 & $\pm$ &    0.002  \\ 
1.20 & $-$ &  2.00 &     1.020 & $\pm$ &    0.003 & $\pm$ &    0.006  \\ 
2.00 & $-$ &  4.00 &     1.021 & $\pm$ &    0.003 & $\pm$ &    0.006  \\ 
\hline
\end{tabular}
\label{tab:fsr_1D_phi}
\end{center}
\end{table}

\clearpage

\begin{center}
\begin{longtable}{rclrcl|ccccc}
\caption{Final state radiation correction used in the $\zy-\zpt$ double cross-section measurement. The first uncertainty is statistical and the second is systematic.}
\label{tab:fsr_2D_ypt}\\
\hline
 \multicolumn{3}{c}{\zy} & \multicolumn{3}{c|}{\zpt~[\gevc]}         & \multicolumn{5}{c}{Correction}  \\ \hline
2.0 & $-$ &   2.5 &  0.0 & $-$ &   2.2 &     1.092 & $\pm$ &    0.002 & $\pm$ 
&    0.007  \\ 
 2.0 & $-$ &   2.5 &  2.2 & $-$ &   3.4 &     1.074 & $\pm$ &    0.002 & $\pm$ 
&    0.023  \\ 
 2.0 & $-$ &   2.5 &  3.4 & $-$ &   4.6 &     1.060 & $\pm$ &    0.002 & $\pm$ 
&    0.008  \\ 
 2.0 & $-$ &   2.5 &  4.6 & $-$ &   5.8 &     1.045 & $\pm$ &    0.002 & $\pm$ 
&    0.000  \\ 
 2.0 & $-$ &   2.5 &  5.8 & $-$ &   7.2 &     1.033 & $\pm$ &    0.002 & $\pm$ 
&    0.004  \\ 
 2.0 & $-$ &   2.5 &  7.2 & $-$ &   8.7 &     1.021 & $\pm$ &    0.002 & $\pm$ 
&    0.003  \\ 
 2.0 & $-$ &   2.5 &  8.7 & $-$ &  10.5 &     1.010 & $\pm$ &    0.002 & $\pm$ 
&    0.004  \\ 
 2.0 & $-$ &   2.5 & 10.5 & $-$ &  12.8 &     0.995 & $\pm$ &    0.002 & $\pm$ 
&    0.014  \\ 
 2.0 & $-$ &   2.5 & 12.8 & $-$ &  15.4 &     0.986 & $\pm$ &    0.002 & $\pm$ 
&    0.007  \\ 
 2.0 & $-$ &   2.5 & 15.4 & $-$ &  19.0 &     0.976 & $\pm$ &    0.002 & $\pm$ 
&    0.005  \\ 
 2.0 & $-$ &   2.5 & 19.0 & $-$ &  24.5 &     0.977 & $\pm$ &    0.002 & $\pm$ 
&    0.003  \\ 
 2.0 & $-$ &   2.5 & 24.5 & $-$ &  34.0 &     0.998 & $\pm$ &    0.002 & $\pm$ 
&    0.001  \\ 
 2.0 & $-$ &   2.5 & 34.0 & $-$ &  63.0 &     1.028 & $\pm$ &    0.001 & $\pm$ 
&    0.003  \\ 
 2.0 & $-$ &   2.5 & 63.0 & $-$ &  270.0 &     1.049 & $\pm$ &    0.003 & $\pm$
 &    0.007  \\ 
 2.5 & $-$ &   3.0 &  0.0 & $-$ &   2.2 &     1.090 & $\pm$ &    0.001 & $\pm$ 
&    0.001  \\ 
 2.5 & $-$ &   3.0 &  2.2 & $-$ &   3.4 &     1.072 & $\pm$ &    0.001 & $\pm$ 
&    0.006  \\ 
 2.5 & $-$ &   3.0 &  3.4 & $-$ &   4.6 &     1.057 & $\pm$ &    0.001 & $\pm$ 
&    0.002  \\ 
 2.5 & $-$ &   3.0 &  4.6 & $-$ &   5.8 &     1.044 & $\pm$ &    0.001 & $\pm$ 
&    0.004  \\ 
 2.5 & $-$ &   3.0 &  5.8 & $-$ &   7.2 &     1.031 & $\pm$ &    0.001 & $\pm$ 
&    0.000  \\ 
 2.5 & $-$ &   3.0 &  7.2 & $-$ &   8.7 &     1.018 & $\pm$ &    0.001 & $\pm$ 
&    0.003  \\ 
 2.5 & $-$ &   3.0 &  8.7 & $-$ &  10.5 &     1.007 & $\pm$ &    0.001 & $\pm$ 
&    0.009  \\ 
 2.5 & $-$ &   3.0 & 10.5 & $-$ &  12.8 &     0.997 & $\pm$ &    0.001 & $\pm$ 
&    0.001  \\ 
 2.5 & $-$ &   3.0 & 12.8 & $-$ &  15.4 &     0.987 & $\pm$ &    0.001 & $\pm$ 
&    0.003  \\ 
 2.5 & $-$ &   3.0 & 15.4 & $-$ &  19.0 &     0.984 & $\pm$ &    0.001 & $\pm$ 
&    0.009  \\ 
 2.5 & $-$ &   3.0 & 19.0 & $-$ &  24.5 &     0.988 & $\pm$ &    0.001 & $\pm$ 
&    0.005  \\ 
 2.5 & $-$ &   3.0 & 24.5 & $-$ &  34.0 &     1.011 & $\pm$ &    0.001 & $\pm$ 
&    0.001  \\ 
 2.5 & $-$ &   3.0 & 34.0 & $-$ &  63.0 &     1.034 & $\pm$ &    0.001 & $\pm$ 
&    0.002  \\ 
 2.5 & $-$ &   3.0 & 63.0 & $-$ &  270.0 &     1.052 & $\pm$ &    0.002 & $\pm$
 &    0.001  \\ 
 3.0 & $-$ &   3.5 &  0.0 & $-$ &   2.2 &     1.089 & $\pm$ &    0.001 & $\pm$ 
&    0.002  \\ 
 3.0 & $-$ &   3.5 &  2.2 & $-$ &   3.4 &     1.069 & $\pm$ &    0.001 & $\pm$ 
&    0.006  \\ 
 3.0 & $-$ &   3.5 &  3.4 & $-$ &   4.6 &     1.055 & $\pm$ &    0.001 & $\pm$ 
&    0.010  \\ 
 3.0 & $-$ &   3.5 &  4.6 & $-$ &   5.8 &     1.041 & $\pm$ &    0.001 & $\pm$ 
&    0.002  \\ 
 3.0 & $-$ &   3.5 &  5.8 & $-$ &   7.2 &     1.028 & $\pm$ &    0.001 & $\pm$ 
&    0.002  \\ 
 3.0 & $-$ &   3.5 &  7.2 & $-$ &   8.7 &     1.019 & $\pm$ &    0.001 & $\pm$ 
&    0.012  \\ 
 3.0 & $-$ &   3.5 &  8.7 & $-$ &  10.5 &     1.008 & $\pm$ &    0.001 & $\pm$ 
&    0.003  \\ 
 3.0 & $-$ &   3.5 & 10.5 & $-$ &  12.8 &     0.999 & $\pm$ &    0.001 & $\pm$ 
&    0.005  \\ 
 3.0 & $-$ &   3.5 & 12.8 & $-$ &  15.4 &     0.993 & $\pm$ &    0.001 & $\pm$ 
&    0.003  \\ 
 3.0 & $-$ &   3.5 & 15.4 & $-$ &  19.0 &     0.989 & $\pm$ &    0.001 & $\pm$ 
&    0.002  \\ 
 3.0 & $-$ &   3.5 & 19.0 & $-$ &  24.5 &     0.995 & $\pm$ &    0.001 & $\pm$ 
&    0.012  \\ 
 3.0 & $-$ &   3.5 & 24.5 & $-$ &  34.0 &     1.015 & $\pm$ &    0.001 & $\pm$ 
&    0.002  \\ 
 3.0 & $-$ &   3.5 & 34.0 & $-$ &  63.0 &     1.037 & $\pm$ &    0.001 & $\pm$ 
&    0.001  \\ 
 3.0 & $-$ &   3.5 & 63.0 & $-$ &  270.0 &     1.055 & $\pm$ &    0.002 & $\pm$
 &    0.008  \\ 
 3.5 & $-$ &   4.0 &  0.0 & $-$ &   2.2 &     1.094 & $\pm$ &    0.002 & $\pm$ 
&    0.004  \\ 
 3.5 & $-$ &   4.0 &  2.2 & $-$ &   3.4 &     1.074 & $\pm$ &    0.002 & $\pm$ 
&    0.012  \\ 
 3.5 & $-$ &   4.0 &  3.4 & $-$ &   4.6 &     1.057 & $\pm$ &    0.002 & $\pm$ 
&    0.005  \\ 
 3.5 & $-$ &   4.0 &  4.6 & $-$ &   5.8 &     1.044 & $\pm$ &    0.002 & $\pm$ 
&    0.001  \\ 
 3.5 & $-$ &   4.0 &  5.8 & $-$ &   7.2 &     1.029 & $\pm$ &    0.002 & $\pm$ 
&    0.008  \\ 
 3.5 & $-$ &   4.0 &  7.2 & $-$ &   8.7 &     1.016 & $\pm$ &    0.002 & $\pm$ 
&    0.009  \\ 
 3.5 & $-$ &   4.0 &  8.7 & $-$ &  10.5 &     1.004 & $\pm$ &    0.002 & $\pm$ 
&    0.009  \\ 
 3.5 & $-$ &   4.0 & 10.5 & $-$ &  12.8 &     0.992 & $\pm$ &    0.002 & $\pm$ 
&    0.008  \\ 
 3.5 & $-$ &   4.0 & 12.8 & $-$ &  15.4 &     0.984 & $\pm$ &    0.002 & $\pm$ 
&    0.009  \\ 
 3.5 & $-$ &   4.0 & 15.4 & $-$ &  19.0 &     0.979 & $\pm$ &    0.002 & $\pm$ 
&    0.007  \\ 
 3.5 & $-$ &   4.0 & 19.0 & $-$ &  24.5 &     0.984 & $\pm$ &    0.002 & $\pm$ 
&    0.003  \\ 
 3.5 & $-$ &   4.0 & 24.5 & $-$ &  34.0 &     1.011 & $\pm$ &    0.002 & $\pm$ 
&    0.003  \\ 
 3.5 & $-$ &   4.0 & 34.0 & $-$ &  63.0 &     1.038 & $\pm$ &    0.001 & $\pm$ 
&    0.002  \\ 
 3.5 & $-$ &   4.0 & 63.0 & $-$ &  270.0 &     1.060 & $\pm$ &    0.003 & $\pm$
 &    0.008  \\ 
 4.0 & $-$ &   4.5 &  0.0 & $-$ &   2.2 &     1.101 & $\pm$ &    0.005 & $\pm$ 
&    0.016  \\ 
 4.0 & $-$ &   4.5 &  2.2 & $-$ &   3.4 &     1.078 & $\pm$ &    0.005 & $\pm$ 
&    0.010  \\ 
 4.0 & $-$ &   4.5 &  3.4 & $-$ &   4.6 &     1.061 & $\pm$ &    0.005 & $\pm$ 
&    0.013  \\ 
 4.0 & $-$ &   4.5 &  4.6 & $-$ &   5.8 &     1.046 & $\pm$ &    0.004 & $\pm$ 
&    0.007  \\ 
 4.0 & $-$ &   4.5 &  5.8 & $-$ &   7.2 &     1.030 & $\pm$ &    0.004 & $\pm$ 
&    0.007  \\ 
 4.0 & $-$ &   4.5 &  7.2 & $-$ &   8.7 &     1.015 & $\pm$ &    0.004 & $\pm$ 
&    0.038  \\ 
 4.0 & $-$ &   4.5 &  8.7 & $-$ &  10.5 &     0.996 & $\pm$ &    0.004 & $\pm$ 
&    0.004  \\ 
 4.0 & $-$ &   4.5 & 10.5 & $-$ &  12.8 &     0.988 & $\pm$ &    0.004 & $\pm$ 
&    0.004  \\ 
 4.0 & $-$ &   4.5 & 12.8 & $-$ &  15.4 &     0.971 & $\pm$ &    0.004 & $\pm$ 
&    0.009  \\ 
 4.0 & $-$ &   4.5 & 15.4 & $-$ &  19.0 &     0.959 & $\pm$ &    0.004 & $\pm$ 
&    0.036  \\ 
 4.0 & $-$ &   4.5 & 19.0 & $-$ &  24.5 &     0.955 & $\pm$ &    0.003 & $\pm$ 
&    0.025  \\ 
 4.0 & $-$ &   4.5 & 24.5 & $-$ &  34.0 &     0.992 & $\pm$ &    0.004 & $\pm$ 
&    0.009  \\ 
 4.0 & $-$ &   4.5 & 34.0 & $-$ &  63.0 &     1.033 & $\pm$ &    0.004 & $\pm$ 
&    0.006  \\ 
 4.0 & $-$ &   4.5 & 63.0 & $-$ &  270.0 &     1.066 & $\pm$ &    0.008 & $\pm$
 &    0.015  \\ 
\hline
\end{longtable}
\end{center}

\begin{center}
\begin{longtable}{rclrcl|ccccc}
\caption{Final state radiation correction used in the $\zy-\phistar$ double cross-section measurement. The first uncertainty is statistical and the second is systematic.}
\label{tab:fsr_2D_yphi}\\ \hline
 \multicolumn{3}{c}{\zy} & \multicolumn{3}{c|}{\phistar}       & \multicolumn{5}{c}{Correction}  \\ \hline
 2.0 & $-$ &   2.5 & 0.00 & $-$ &  0.01 &     1.030 & $\pm$ &    0.002 & $\pm$ 
&    0.000  \\ 
 2.0 & $-$ &   2.5 & 0.01 & $-$ &  0.02 &     1.028 & $\pm$ &    0.002 & $\pm$ 
&    0.005  \\ 
 2.0 & $-$ &   2.5 & 0.02 & $-$ &  0.03 &     1.026 & $\pm$ &    0.002 & $\pm$ 
&    0.004  \\ 
 2.0 & $-$ &   2.5 & 0.03 & $-$ &  0.05 &     1.024 & $\pm$ &    0.001 & $\pm$ 
&    0.002  \\ 
 2.0 & $-$ &   2.5 & 0.05 & $-$ &  0.07 &     1.019 & $\pm$ &    0.002 & $\pm$ 
&    0.003  \\ 
 2.0 & $-$ &   2.5 & 0.07 & $-$ &  0.10 &     1.017 & $\pm$ &    0.001 & $\pm$ 
&    0.003  \\ 
 2.0 & $-$ &   2.5 & 0.10 & $-$ &  0.15 &     1.014 & $\pm$ &    0.001 & $\pm$ 
&    0.000  \\ 
 2.0 & $-$ &   2.5 & 0.15 & $-$ &  0.20 &     1.013 & $\pm$ &    0.002 & $\pm$ 
&    0.005  \\ 
 2.0 & $-$ &   2.5 & 0.20 & $-$ &  0.30 &     1.014 & $\pm$ &    0.002 & $\pm$ 
&    0.003  \\ 
 2.0 & $-$ &   2.5 & 0.30 & $-$ &  0.40 &     1.017 & $\pm$ &    0.002 & $\pm$ 
&    0.004  \\ 
 2.0 & $-$ &   2.5 & 0.40 & $-$ &  0.60 &     1.019 & $\pm$ &    0.002 & $\pm$ 
&    0.002  \\ 
 2.0 & $-$ &   2.5 & 0.60 & $-$ &  0.80 &     1.018 & $\pm$ &    0.004 & $\pm$ 
&    0.009  \\ 
 2.0 & $-$ &   2.5 & 0.80 & $-$ &  1.20 &     1.019 & $\pm$ &    0.005 & $\pm$ 
&    0.002  \\ 
 2.0 & $-$ &   2.5 & 1.20 & $-$ &  2.00 &     1.018 & $\pm$ &    0.007 & $\pm$ 
&    0.015  \\ 
 2.0 & $-$ &   2.5 & 2.00 & $-$ &  4.00 &     1.019 & $\pm$ &    0.006 & $\pm$ 
&    0.005  \\ 
 2.5 & $-$ &   3.0 & 0.00 & $-$ &  0.01 &     1.034 & $\pm$ &    0.001 & $\pm$ 
&    0.004  \\ 
 2.5 & $-$ &   3.0 & 0.01 & $-$ &  0.02 &     1.031 & $\pm$ &    0.001 & $\pm$ 
&    0.005  \\ 
 2.5 & $-$ &   3.0 & 0.02 & $-$ &  0.03 &     1.030 & $\pm$ &    0.001 & $\pm$ 
&    0.003  \\ 
 2.5 & $-$ &   3.0 & 0.03 & $-$ &  0.05 &     1.026 & $\pm$ &    0.001 & $\pm$ 
&    0.002  \\ 
 2.5 & $-$ &   3.0 & 0.05 & $-$ &  0.07 &     1.022 & $\pm$ &    0.001 & $\pm$ 
&    0.002  \\ 
 2.5 & $-$ &   3.0 & 0.07 & $-$ &  0.10 &     1.019 & $\pm$ &    0.001 & $\pm$ 
&    0.001  \\ 
 2.5 & $-$ &   3.0 & 0.10 & $-$ &  0.15 &     1.017 & $\pm$ &    0.001 & $\pm$ 
&    0.001  \\ 
 2.5 & $-$ &   3.0 & 0.15 & $-$ &  0.20 &     1.017 & $\pm$ &    0.001 & $\pm$ 
&    0.004  \\ 
 2.5 & $-$ &   3.0 & 0.20 & $-$ &  0.30 &     1.016 & $\pm$ &    0.001 & $\pm$ 
&    0.001  \\ 
 2.5 & $-$ &   3.0 & 0.30 & $-$ &  0.40 &     1.018 & $\pm$ &    0.001 & $\pm$ 
&    0.002  \\ 
 2.5 & $-$ &   3.0 & 0.40 & $-$ &  0.60 &     1.020 & $\pm$ &    0.002 & $\pm$ 
&    0.003  \\ 
 2.5 & $-$ &   3.0 & 0.60 & $-$ &  0.80 &     1.020 & $\pm$ &    0.002 & $\pm$ 
&    0.000  \\ 
 2.5 & $-$ &   3.0 & 0.80 & $-$ &  1.20 &     1.020 & $\pm$ &    0.003 & $\pm$ 
&    0.005  \\ 
 2.5 & $-$ &   3.0 & 1.20 & $-$ &  2.00 &     1.021 & $\pm$ &    0.004 & $\pm$ 
&    0.020  \\ 
 2.5 & $-$ &   3.0 & 2.00 & $-$ &  4.00 &     1.022 & $\pm$ &    0.005 & $\pm$ 
&    0.004  \\ 
 3.0 & $-$ &   3.5 & 0.00 & $-$ &  0.01 &     1.037 & $\pm$ &    0.001 & $\pm$ 
&    0.001  \\ 
 3.0 & $-$ &   3.5 & 0.01 & $-$ &  0.02 &     1.035 & $\pm$ &    0.001 & $\pm$ 
&    0.001  \\ 
 3.0 & $-$ &   3.5 & 0.02 & $-$ &  0.03 &     1.033 & $\pm$ &    0.001 & $\pm$ 
&    0.001  \\ 
 3.0 & $-$ &   3.5 & 0.03 & $-$ &  0.05 &     1.028 & $\pm$ &    0.001 & $\pm$ 
&    0.002  \\ 
 3.0 & $-$ &   3.5 & 0.05 & $-$ &  0.07 &     1.025 & $\pm$ &    0.001 & $\pm$ 
&    0.003  \\ 
 3.0 & $-$ &   3.5 & 0.07 & $-$ &  0.10 &     1.022 & $\pm$ &    0.001 & $\pm$ 
&    0.004  \\ 
 3.0 & $-$ &   3.5 & 0.10 & $-$ &  0.15 &     1.019 & $\pm$ &    0.001 & $\pm$ 
&    0.000  \\ 
 3.0 & $-$ &   3.5 & 0.15 & $-$ &  0.20 &     1.018 & $\pm$ &    0.001 & $\pm$ 
&    0.002  \\ 
 3.0 & $-$ &   3.5 & 0.20 & $-$ &  0.30 &     1.017 & $\pm$ &    0.001 & $\pm$ 
&    0.005  \\ 
 3.0 & $-$ &   3.5 & 0.30 & $-$ &  0.40 &     1.017 & $\pm$ &    0.001 & $\pm$ 
&    0.001  \\ 
 3.0 & $-$ &   3.5 & 0.40 & $-$ &  0.60 &     1.021 & $\pm$ &    0.002 & $\pm$ 
&    0.004  \\ 
 3.0 & $-$ &   3.5 & 0.60 & $-$ &  0.80 &     1.021 & $\pm$ &    0.002 & $\pm$ 
&    0.005  \\ 
 3.0 & $-$ &   3.5 & 0.80 & $-$ &  1.20 &     1.021 & $\pm$ &    0.003 & $\pm$ 
&    0.001  \\ 
 3.0 & $-$ &   3.5 & 1.20 & $-$ &  2.00 &     1.019 & $\pm$ &    0.005 & $\pm$ 
&    0.004  \\ 
 3.0 & $-$ &   3.5 & 2.00 & $-$ &  4.00 &     1.022 & $\pm$ &    0.006 & $\pm$ 
&    0.026  \\ 
 3.5 & $-$ &   4.0 & 0.00 & $-$ &  0.01 &     1.034 & $\pm$ &    0.002 & $\pm$ 
&    0.004  \\ 
 3.5 & $-$ &   4.0 & 0.01 & $-$ &  0.02 &     1.033 & $\pm$ &    0.002 & $\pm$ 
&    0.004  \\ 
 3.5 & $-$ &   4.0 & 0.02 & $-$ &  0.03 &     1.030 & $\pm$ &    0.002 & $\pm$ 
&    0.001  \\ 
 3.5 & $-$ &   4.0 & 0.03 & $-$ &  0.05 &     1.026 & $\pm$ &    0.001 & $\pm$ 
&    0.001  \\ 
 3.5 & $-$ &   4.0 & 0.05 & $-$ &  0.07 &     1.022 & $\pm$ &    0.001 & $\pm$ 
&    0.002  \\ 
 3.5 & $-$ &   4.0 & 0.07 & $-$ &  0.10 &     1.019 & $\pm$ &    0.001 & $\pm$ 
&    0.004  \\ 
 3.5 & $-$ &   4.0 & 0.10 & $-$ &  0.15 &     1.016 & $\pm$ &    0.001 & $\pm$ 
&    0.003  \\ 
 3.5 & $-$ &   4.0 & 0.15 & $-$ &  0.20 &     1.015 & $\pm$ &    0.002 & $\pm$ 
&    0.002  \\ 
 3.5 & $-$ &   4.0 & 0.20 & $-$ &  0.30 &     1.015 & $\pm$ &    0.002 & $\pm$ 
&    0.005  \\ 
 3.5 & $-$ &   4.0 & 0.30 & $-$ &  0.40 &     1.016 & $\pm$ &    0.002 & $\pm$ 
&    0.001  \\ 
 3.5 & $-$ &   4.0 & 0.40 & $-$ &  0.60 &     1.018 & $\pm$ &    0.002 & $\pm$ 
&    0.002  \\ 
 3.5 & $-$ &   4.0 & 0.60 & $-$ &  0.80 &     1.020 & $\pm$ &    0.004 & $\pm$ 
&    0.007  \\ 
 3.5 & $-$ &   4.0 & 0.80 & $-$ &  1.20 &     1.021 & $\pm$ &    0.005 & $\pm$ 
&    0.003  \\ 
 3.5 & $-$ &   4.0 & 1.20 & $-$ &  2.00 &     1.020 & $\pm$ &    0.009 & $\pm$ 
&    0.020  \\ 
 3.5 & $-$ &   4.0 & 2.00 & $-$ &  4.00 &     1.018 & $\pm$ &    0.016 & $\pm$ 
&    0.018  \\ 
 4.0 & $-$ &   4.5 & 0.00 & $-$ &  0.01 &     1.035 & $\pm$ &    0.004 & $\pm$ 
&    0.002  \\ 
 4.0 & $-$ &   4.5 & 0.01 & $-$ &  0.02 &     1.029 & $\pm$ &    0.004 & $\pm$ 
&    0.016  \\ 
 4.0 & $-$ &   4.5 & 0.02 & $-$ &  0.03 &     1.024 & $\pm$ &    0.004 & $\pm$ 
&    0.002  \\ 
 4.0 & $-$ &   4.5 & 0.03 & $-$ &  0.05 &     1.021 & $\pm$ &    0.003 & $\pm$ 
&    0.004  \\ 
 4.0 & $-$ &   4.5 & 0.05 & $-$ &  0.07 &     1.020 & $\pm$ &    0.003 & $\pm$ 
&    0.006  \\ 
 4.0 & $-$ &   4.5 & 0.07 & $-$ &  0.10 &     1.014 & $\pm$ &    0.003 & $\pm$ 
&    0.007  \\ 
 4.0 & $-$ &   4.5 & 0.10 & $-$ &  0.15 &     1.009 & $\pm$ &    0.003 & $\pm$ 
&    0.004  \\ 
 4.0 & $-$ &   4.5 & 0.15 & $-$ &  0.20 &     1.005 & $\pm$ &    0.004 & $\pm$ 
&    0.016  \\ 
 4.0 & $-$ &   4.5 & 0.20 & $-$ &  0.30 &     1.007 & $\pm$ &    0.004 & $\pm$ 
&    0.009  \\ 
 4.0 & $-$ &   4.5 & 0.30 & $-$ &  0.40 &     1.007 & $\pm$ &    0.005 & $\pm$ 
&    0.011  \\ 
 4.0 & $-$ &   4.5 & 0.40 & $-$ &  0.60 &     1.013 & $\pm$ &    0.006 & $\pm$ 
&    0.008  \\ 
 4.0 & $-$ &   4.5 & 0.60 & $-$ &  0.80 &     1.019 & $\pm$ &    0.010 & $\pm$ 
&    0.001  \\ 
 4.0 & $-$ &   4.5 & 0.80 & $-$ &  1.20 &     1.011 & $\pm$ &    0.017 & $\pm$ 
&    0.075  \\ 
 4.0 & $-$ &   4.5 & 1.20 & $-$ &  2.00 &     0.988 & $\pm$ &    0.037 & $\pm$ 
&    0.238  \\ 
 4.0 & $-$ &   4.5 & 2.00 & $-$ &  4.00 &     0.961 & $\pm$ &    0.134 & $\pm$ 
&    0.039  \\
\hline
\end{longtable}
\end{center}

\clearpage

\section{Correlation matrices}
\label{app:correlation}
\begin{figure}[h]
\begin{center}
\includegraphics[width=0.45\textwidth]{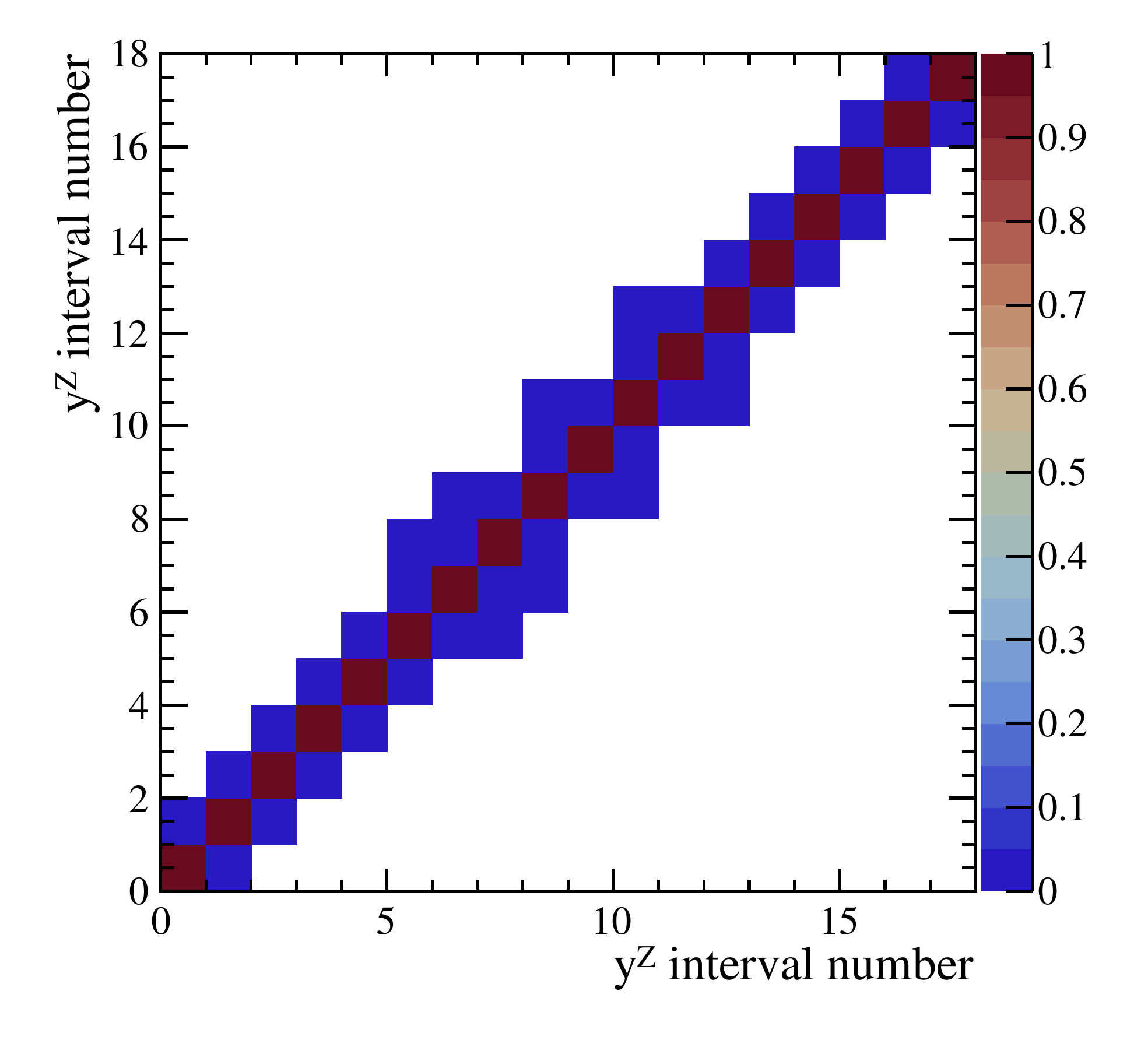}
\includegraphics[width=0.45\textwidth]{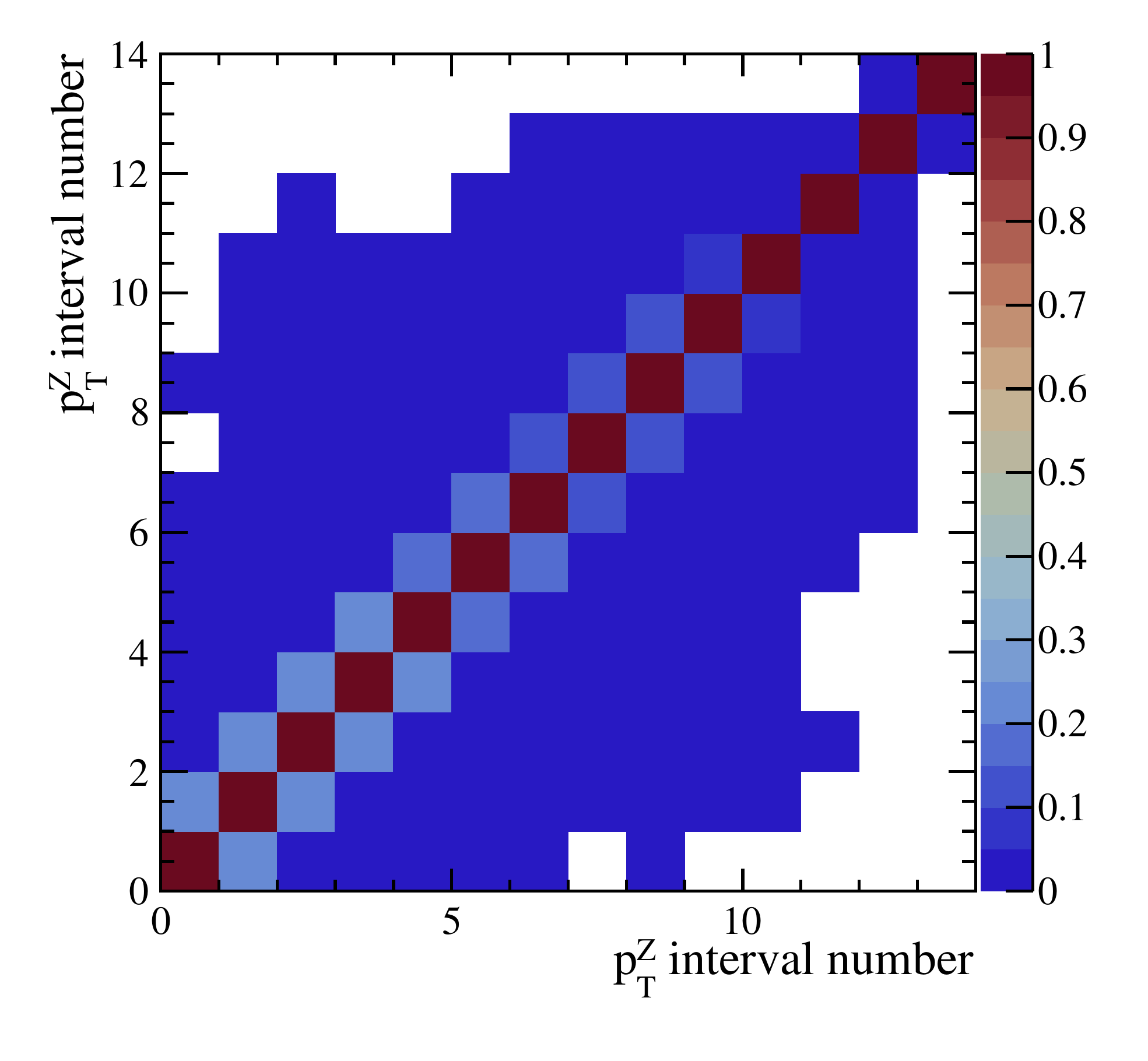}
\includegraphics[width=0.45\textwidth]{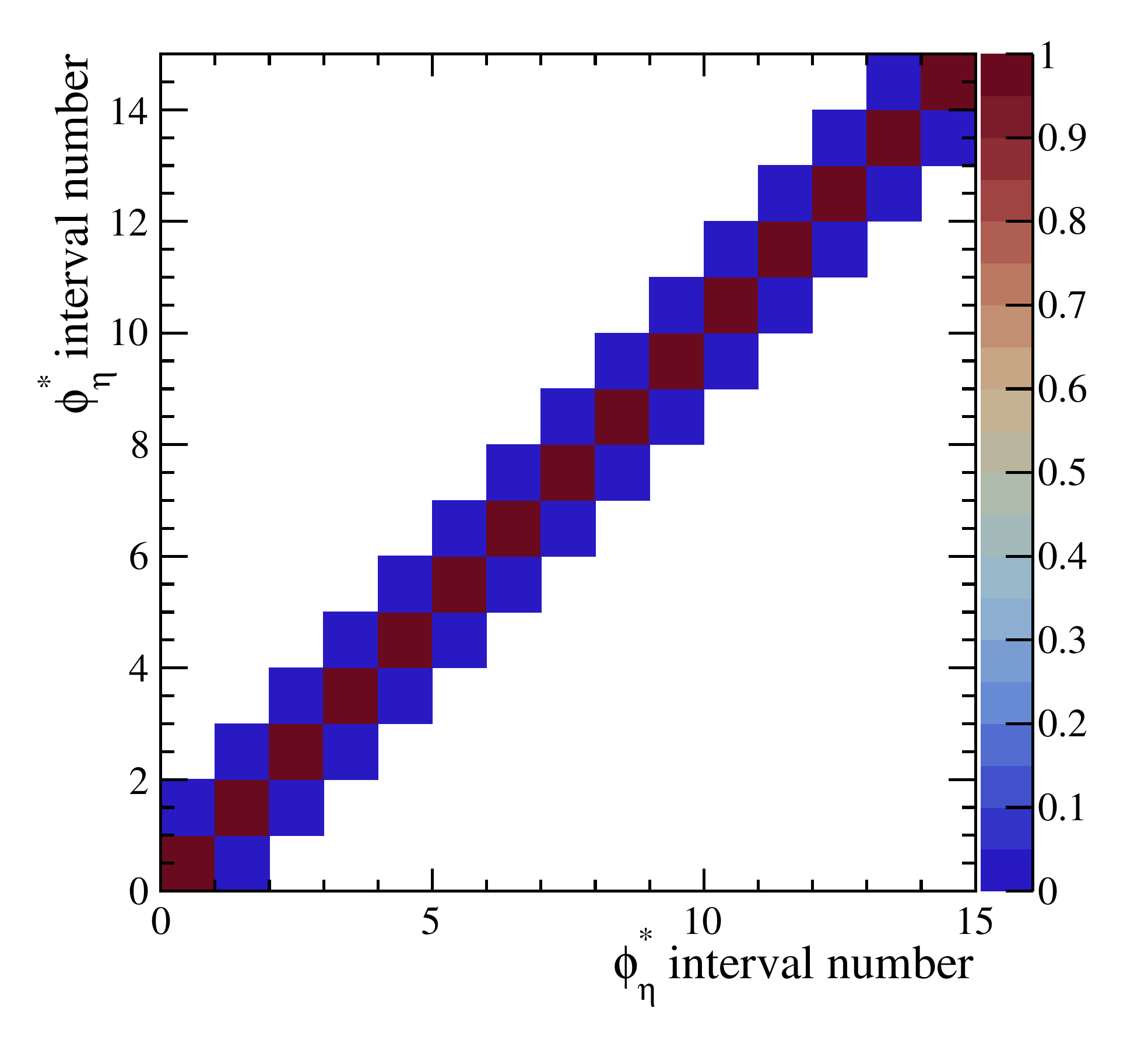}
\caption{Statistical correlation matrix of the cross-section measurements in one-dimensional interval regions of (top-left) \zy, (top-right) \zpt and (bottom) \phistar. 
More details about the `interval number' can be found from Table~\ref{tab:cen_y} and Table~\ref{tab:cen_phi}.}
\label{fig:corr_stat_1D}
\end{center}
\end{figure}

\begin{figure}[h]
\begin{center}
\includegraphics[width=0.45\textwidth]{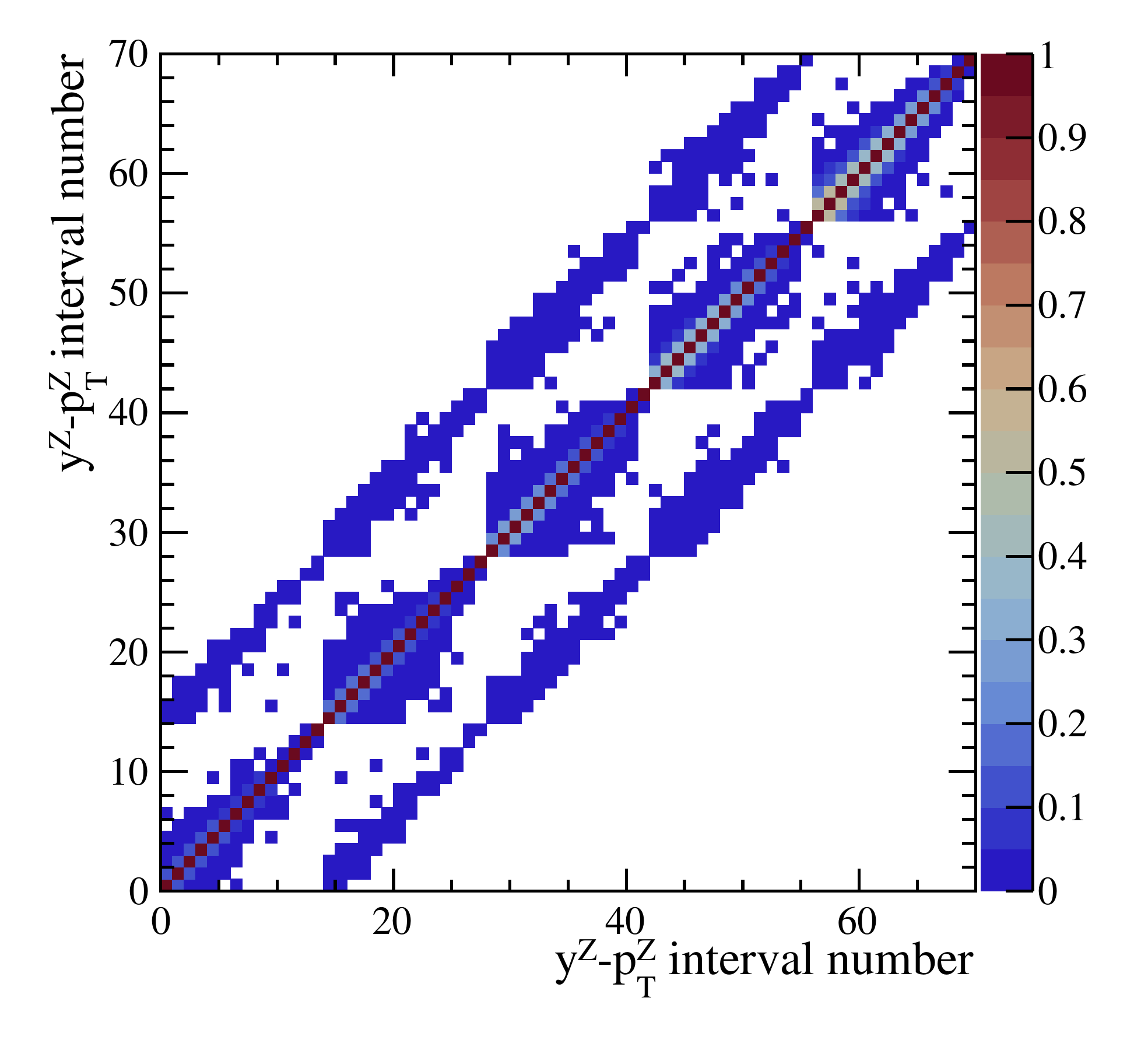}
\includegraphics[width=0.45\textwidth]{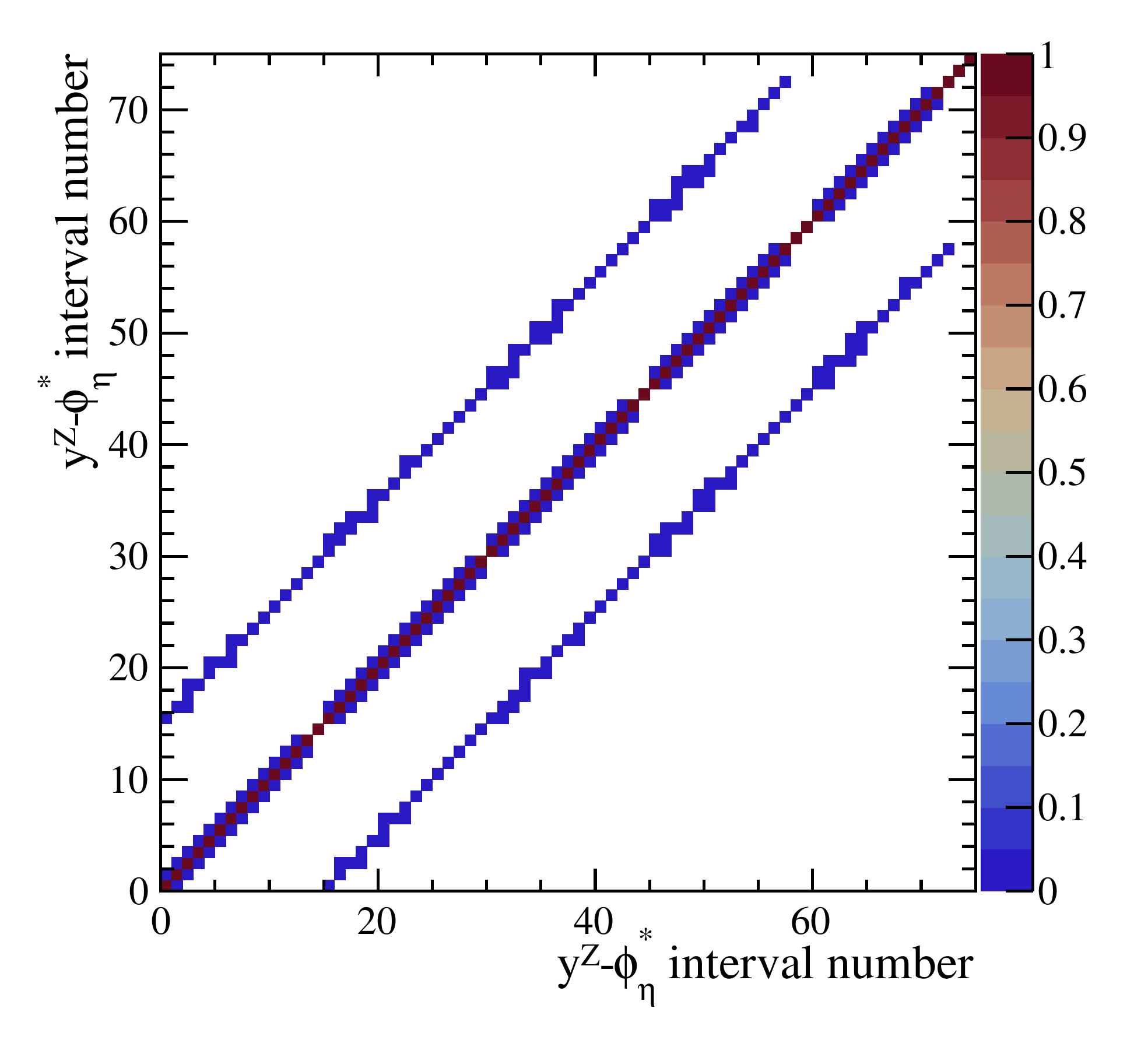}
\caption{Statistical correlation matrix of the cross-section measurements in two-dimensional interval regions of (left) $\zy-\zpt$ and (right) $\zy-\phistar$. More details about the `interval number' can be found in Table~\ref{tab:cen_Zpty} and Table~\ref{tab:cen_ZY_ZPHI}. }
\label{fig:corr_stat_2D}
\end{center}
\end{figure}

\begin{figure}[h]
\begin{center}
\includegraphics[width=0.45\textwidth]{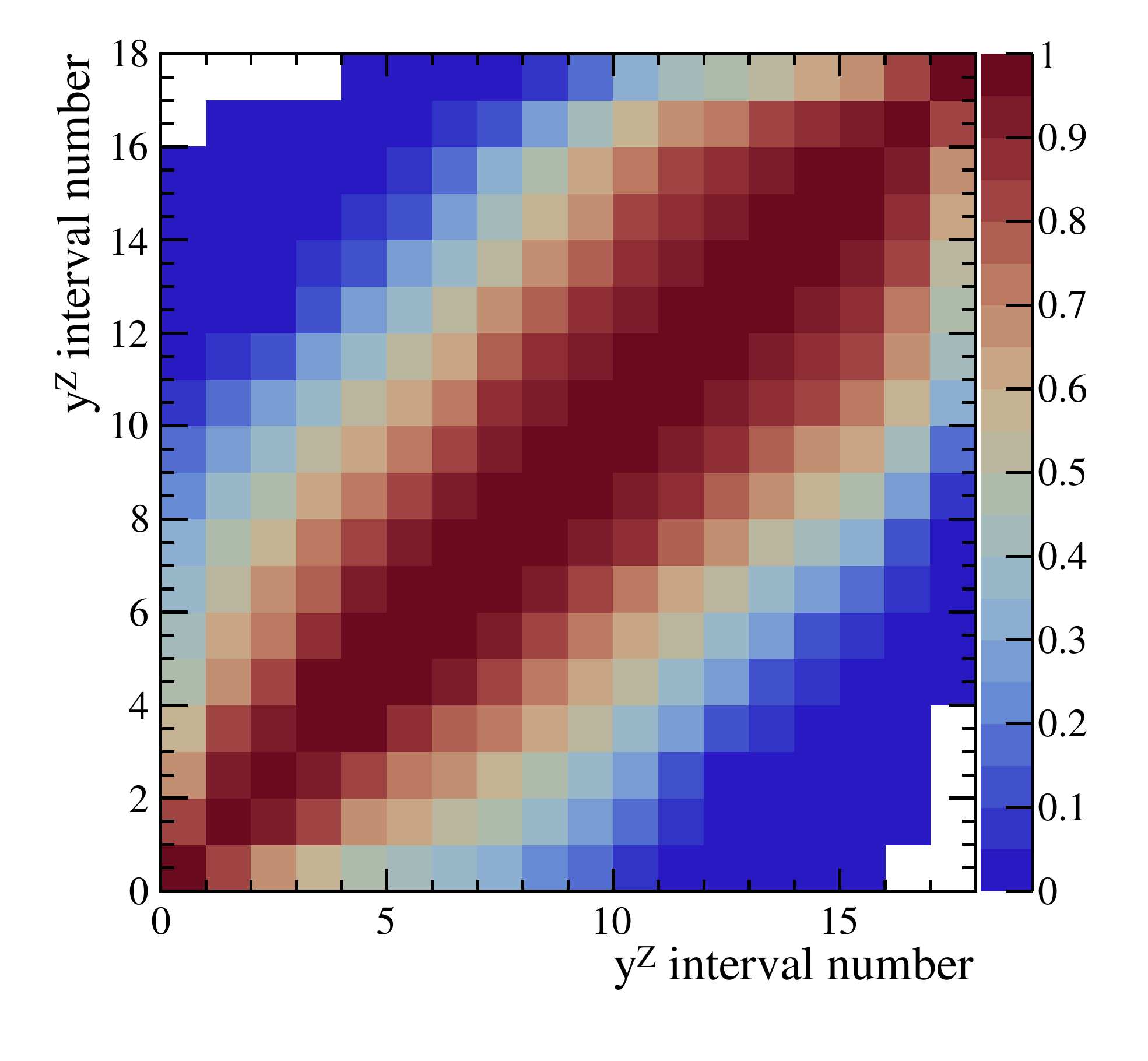}
\includegraphics[width=0.45\textwidth]{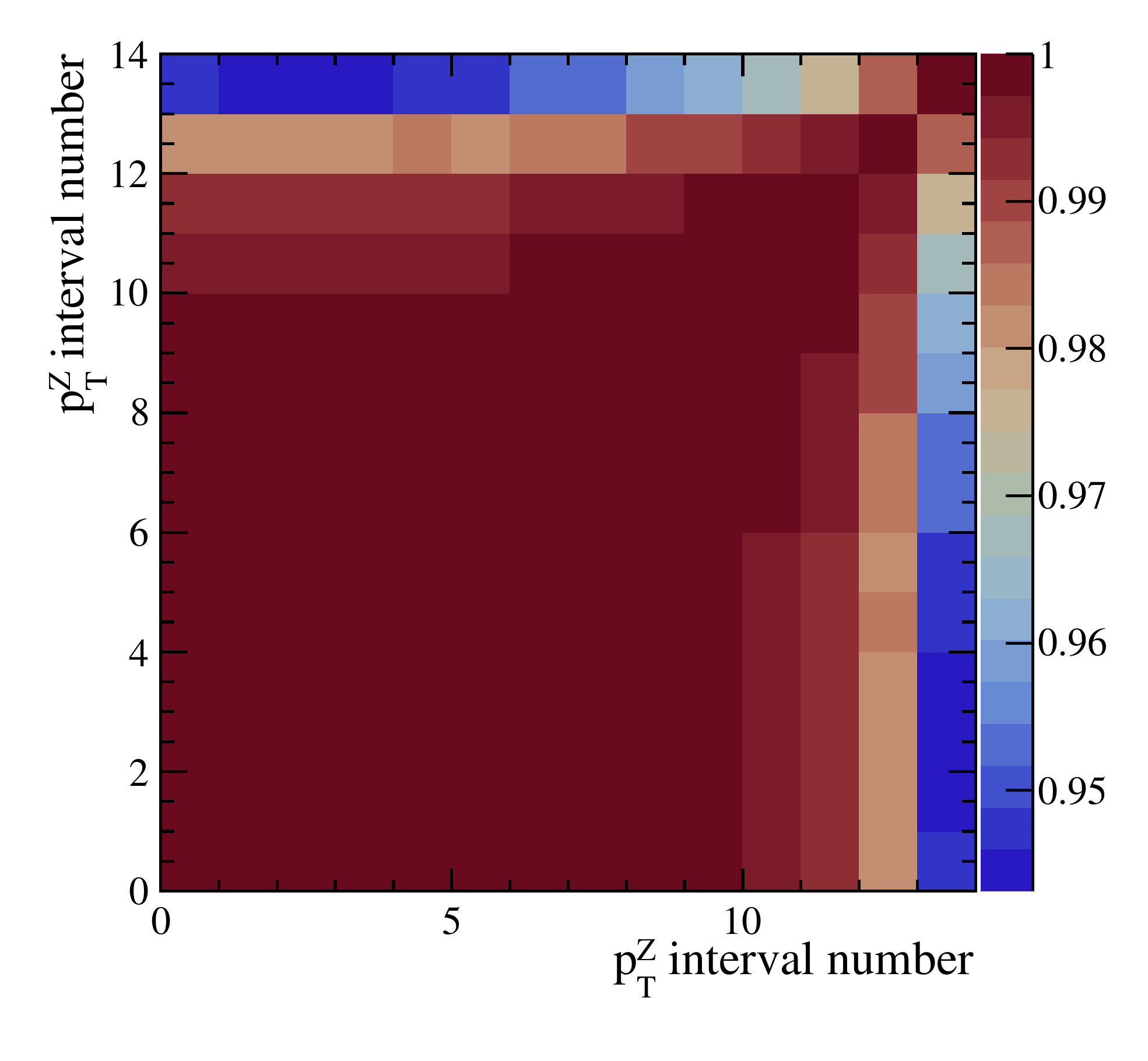}
\includegraphics[width=0.45\textwidth]{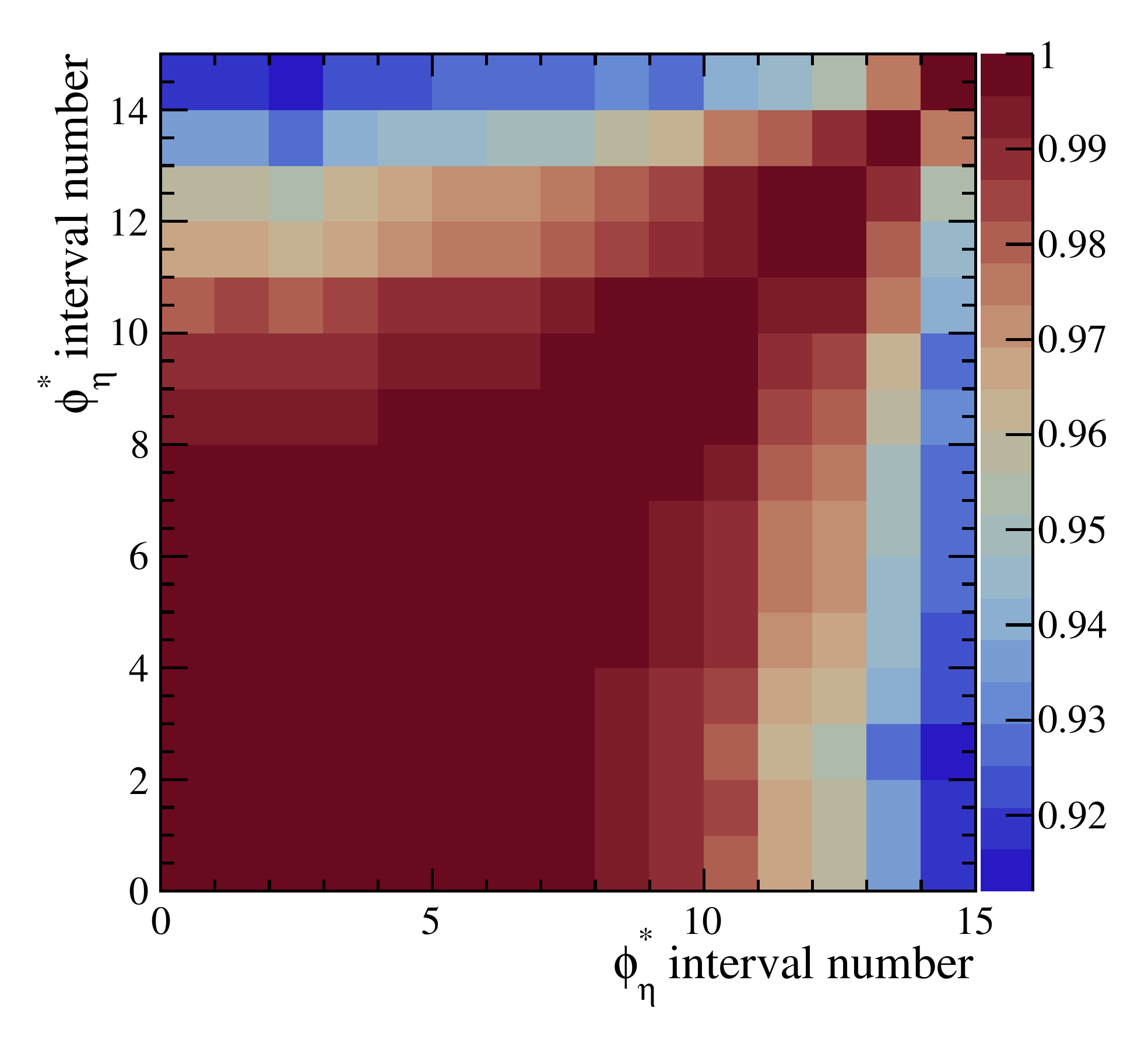}
\caption{Correlation matrix of efficiencies uncertainty for one-dimensional (top-left) \zy, (top-right) \zpt and (bottom) \phistar measurements. More details about the `interval number' can be found in Table~\ref{tab:cen_y} and Table~\ref{tab:cen_phi}.}
\label{fig:corr_syst_1D}
\end{center}
\end{figure}

\begin{figure}[h]
\begin{center}
\includegraphics[width=0.45\textwidth]{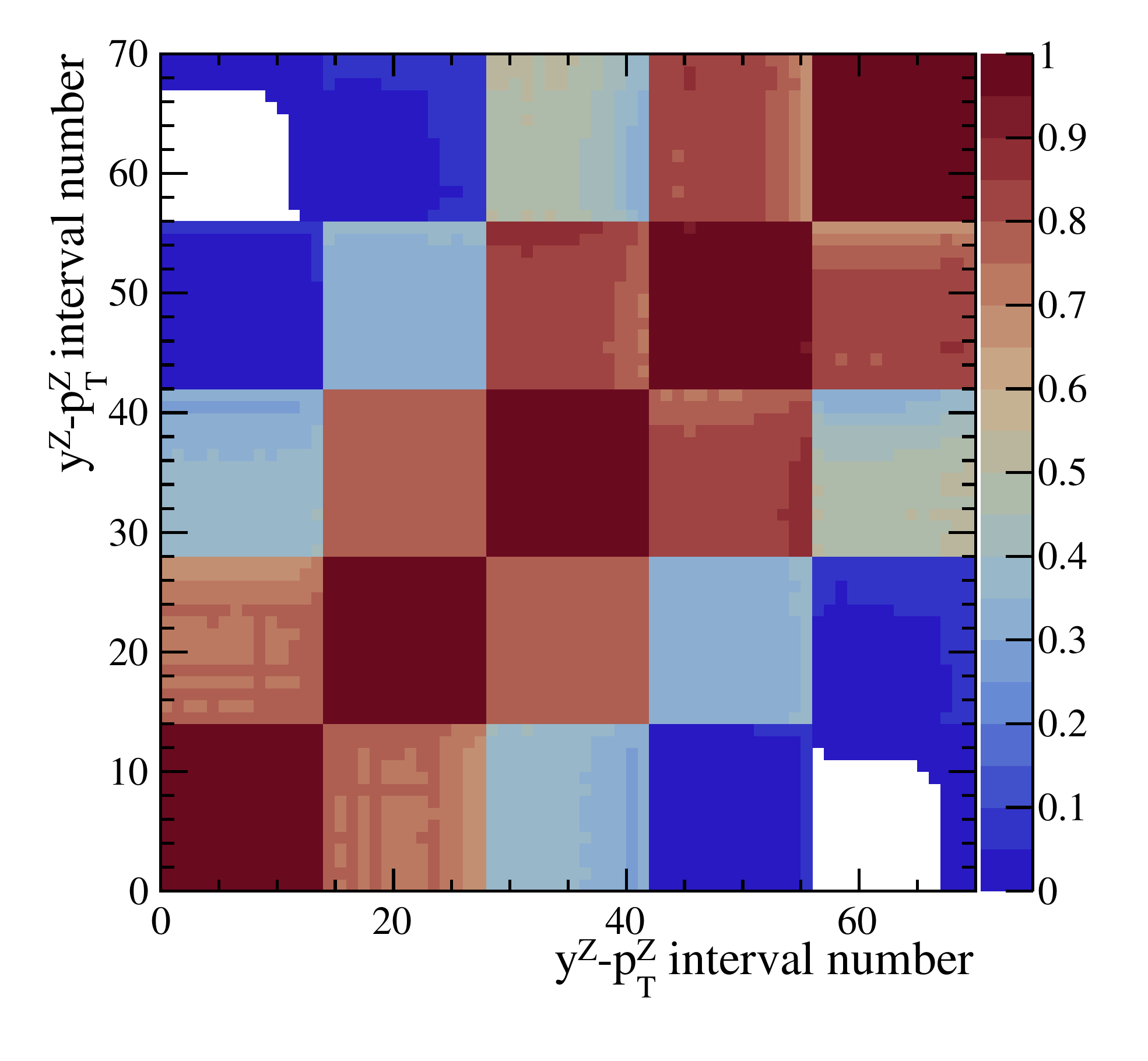}
\includegraphics[width=0.45\textwidth]{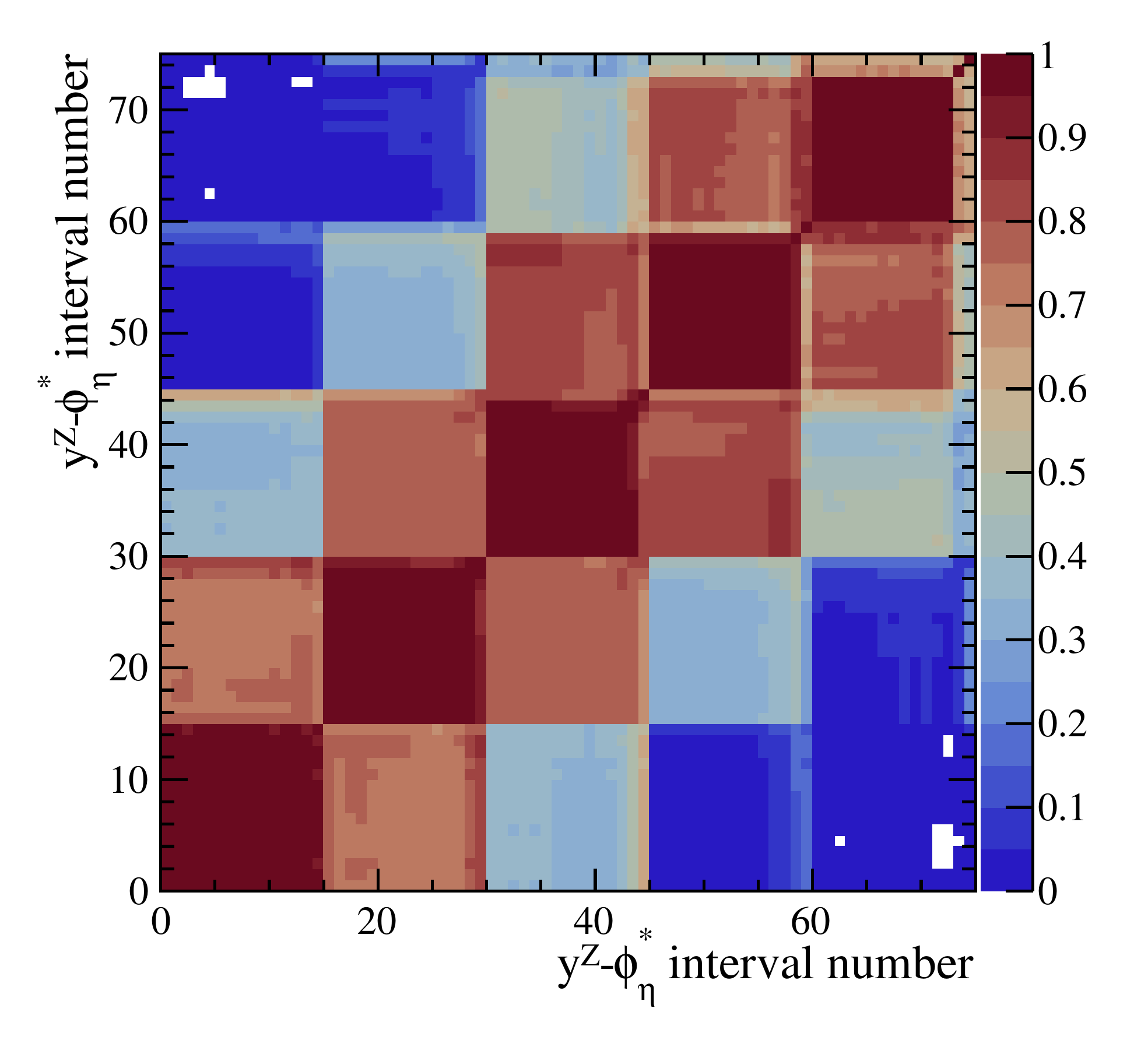}
\caption{Correlation matrix of efficiencies uncertainty for two-dimensional (left) $\zy-\zpt$ and (right) $\zy-\phistar$ measurements. More details about the `interval number' can be found in Table~\ref{tab:cen_Zpty} and Table~\ref{tab:cen_ZY_ZPHI}.}
\label{fig:corr_syst_2D}
\end{center}
\end{figure}

The calculated correlation matrices for statistical uncertainty are shown in Figs.~\ref{fig:corr_stat_1D} and~\ref{fig:corr_stat_2D},
and presented from Table~\ref{tab:LPer_ystat} to~\ref{tab:LPer_phistat}, 
and the correlation matrices for efficiency uncertainty
are shown in Figs.~\ref{fig:corr_syst_1D} and~\ref{fig:corr_syst_2D} for single and double 
differential cross-section measurements, and
presented from Table~\ref{tab:LPer_ysyst} to~\ref{tab:LPer_phisyst}. 

\begin{sidewaystable}[h]
\caption{Correlation matrix of statistical uncertainty for one-dimensional \zy measurement.} 
\centering 
\begin{tabular}{c|cccccccccccccccccc} 
\hline
Interval & 1 & 2 & 3 & 4 & 5 & 6 & 7 & 8 & 9 & 10 & 11 & 12 & 13 & 14 & 15 & 16 & 17 & 18  \\ \hline
1  & 1.00 &      &      &      &      &      &      &      &      &      &      &      &      &      &      &      &      &      \\ 
2  & 0.00 & 1.00 &      &      &      &      &      &      &      &      &      &      &      &      &      &      &      &      \\ 
3  & 0.00 & 0.00 & 1.00 &      &      &      &      &      &      &      &      &      &      &      &      &      &      &      \\ 
4  & 0.00 & 0.00 & 0.00 & 1.00 &      &      &      &      &      &      &      &      &      &      &      &      &      &      \\ 
5  & 0.00 & 0.00 & 0.00 & 0.01 & 1.00 &      &      &      &      &      &      &      &      &      &      &      &      &      \\ 
6  & 0.00 & 0.00 & 0.00 & 0.00 & 0.01 & 1.00 &      &      &      &      &      &      &      &      &      &      &      &      \\ 
7  & 0.00 & 0.00 & 0.00 & 0.00 & 0.00 & 0.01 & 1.00 &      &      &      &      &      &      &      &      &      &      &      \\ 
8  & 0.00 & 0.00 & 0.00 & 0.00 & 0.00 & 0.00 & 0.01 & 1.00 &      &      &      &      &      &      &      &      &      &      \\ 
9  & 0.00 & 0.00 & 0.00 & 0.00 & 0.00 & 0.00 & 0.00 & 0.01 & 1.00 &      &      &      &      &      &      &      &      &      \\ 
10  & 0.00 & 0.00 & 0.00 & 0.00 & 0.00 & 0.00 & 0.00 & 0.00 & 0.02 & 1.00 &      &      &      &      &      &      &      &      \\ 
11  & 0.00 & 0.00 & 0.00 & 0.00 & 0.00 & 0.00 & 0.00 & 0.00 & 0.00 & 0.02 & 1.00 &      &      &      &      &      &      &      \\ 
12  & 0.00 & 0.00 & 0.00 & 0.00 & 0.00 & 0.00 & 0.00 & 0.00 & 0.00 & 0.00 & 0.02 & 1.00 &      &      &      &      &      &      \\ 
13  & 0.00 & 0.00 & 0.00 & 0.00 & 0.00 & 0.00 & 0.00 & 0.00 & 0.00 & 0.00 & 0.00 & 0.02 & 1.00 &      &      &      &      &      \\ 
14  & 0.00 & 0.00 & 0.00 & 0.00 & 0.00 & 0.00 & 0.00 & 0.00 & 0.00 & 0.00 & 0.00 & 0.00 & 0.02 & 1.00 &      &      &      &      \\ 
15  & 0.00 & 0.00 & 0.00 & 0.00 & 0.00 & 0.00 & 0.00 & 0.00 & 0.00 & 0.00 & 0.00 & 0.00 & 0.00 & 0.02 & 1.00 &      &      &      \\ 
16  & 0.00 & 0.00 & 0.00 & 0.00 & 0.00 & 0.00 & 0.00 & 0.00 & 0.00 & 0.00 & 0.00 & 0.00 & 0.00 & 0.00 & 0.02 & 1.00 &      &      \\ 
17  & 0.00 & 0.00 & 0.00 & 0.00 & 0.00 & 0.00 & 0.00 & 0.00 & 0.00 & 0.00 & 0.00 & 0.00 & 0.00 & 0.00 & 0.00 & 0.01 & 1.00 &      \\ 
18  & 0.00 & 0.00 & 0.00 & 0.00 & 0.00 & 0.00 & 0.00 & 0.00 & 0.00 & 0.00 & 0.00 & 0.00 & 0.00 & 0.00 & 0.00 & 0.00 & 0.00 & 1.00   \\
\hline 
\end{tabular}
\label{tab:LPer_ystat}
\end{sidewaystable}

\begin{sidewaystable}[h]
\caption{Correlation matrix of statistical uncertainty for one-dimensional \zpt measurement.} 
\centering 
\begin{tabular}{c|cccccccccccccccccc} 
\hline
Interval & 1 & 2 & 3 & 4 & 5 & 6 & 7 & 8 & 9 & 10 & 11 & 12 & 13 & 14    \\ \hline
1  & 1.00 &      &      &      &      &      &      &      &      &      &      &      &      &      \\ 
2  & 0.22 & 1.00 &      &      &      &      &      &      &      &      &      &      &      &      \\ 
3  & 0.03 & 0.24 & 1.00 &      &      &      &      &      &      &      &      &      &      &      \\ 
4  & 0.00 & 0.02 & 0.22 & 1.00 &      &      &      &      &      &      &      &      &      &      \\ 
5  & 0.00 & 0.00 & 0.02 & 0.22 & 1.00 &      &      &      &      &      &      &      &      &      \\ 
6  & 0.00 & 0.00 & 0.00 & 0.02 & 0.19 & 1.00 &      &      &      &      &      &      &      &      \\ 
7  & 0.00 & 0.00 & 0.00 & 0.00 & 0.01 & 0.18 & 1.00 &      &      &      &      &      &      &      \\ 
8  & 0.00 & 0.00 & 0.00 & 0.00 & 0.00 & 0.01 & 0.15 & 1.00 &      &      &      &      &      &      \\ 
9  & 0.00 & 0.00 & 0.00 & 0.00 & 0.00 & 0.00 & 0.00 & 0.11 & 1.00 &      &      &      &      &      \\ 
10  & 0.00 & 0.00 & 0.00 & 0.00 & 0.00 & 0.00 & 0.00 & 0.00 & 0.11 & 1.00 &      &      &      &      \\ 
11  & 0.00 & 0.00 & 0.00 & 0.00 & 0.00 & 0.00 & 0.00 & 0.00 & 0.00 & 0.08 & 1.00 &      &      &      \\ 
12  & 0.00 & 0.00 & 0.00 & 0.00 & 0.00 & 0.00 & 0.00 & 0.00 & 0.00 & 0.00 & 0.05 & 1.00 &      &      \\ 
13  & 0.00 & 0.00 & 0.00 & 0.00 & 0.00 & 0.00 & 0.00 & 0.00 & 0.00 & 0.00 & 0.00 & 0.03 & 1.00 &      \\ 
14  & 0.00 & 0.00 & 0.00 & 0.00 & 0.00 & 0.00 & 0.00 & 0.00 & 0.00 & 0.00 & 0.00 & 0.00 & 0.01 & 1.00   \\ 
\hline 
\end{tabular}
\label{tab:LPer_ptstat}
\end{sidewaystable}

\begin{sidewaystable}[h]
\caption{Correlation matrix of statistical uncertainty for one-dimensional \phistar measurement.} 
\centering 
\begin{tabular}{c|cccccccccccccccccc} 
\hline
Interval & 1 & 2 & 3 & 4 & 5 & 6 & 7 & 8 & 9 & 10 & 11 & 12 & 13 & 14 & 15   \\ \hline
1  & 1.00 &      &      &      &      &      &      &      &      &      &      &      &      &      &      \\ 
2  & 0.01 & 1.00 &      &      &      &      &      &      &      &      &      &      &      &      &      \\ 
3  & 0.00 & 0.01 & 1.00 &      &      &      &      &      &      &      &      &      &      &      &      \\ 
4  & 0.00 & 0.00 & 0.01 & 1.00 &      &      &      &      &      &      &      &      &      &      &      \\ 
5  & 0.00 & 0.00 & 0.00 & 0.00 & 1.00 &      &      &      &      &      &      &      &      &      &      \\ 
6  & 0.00 & 0.00 & 0.00 & 0.00 & 0.00 & 1.00 &      &      &      &      &      &      &      &      &      \\ 
7  & 0.00 & 0.00 & 0.00 & 0.00 & 0.00 & 0.00 & 1.00 &      &      &      &      &      &      &      &      \\ 
8  & 0.00 & 0.00 & 0.00 & 0.00 & 0.00 & 0.00 & 0.00 & 1.00 &      &      &      &      &      &      &      \\ 
9  & 0.00 & 0.00 & 0.00 & 0.00 & 0.00 & 0.00 & 0.00 & 0.00 & 1.00 &      &      &      &      &      &      \\ 
10  & 0.00 & 0.00 & 0.00 & 0.00 & 0.00 & 0.00 & 0.00 & 0.00 & 0.00 & 1.00 &      &      &      &      &      \\ 
11  & 0.00 & 0.00 & 0.00 & 0.00 & 0.00 & 0.00 & 0.00 & 0.00 & 0.00 & 0.00 & 1.00 &      &      &      &      \\ 
12  & 0.00 & 0.00 & 0.00 & 0.00 & 0.00 & 0.00 & 0.00 & 0.00 & 0.00 & 0.00 & 0.00 & 1.00 &      &      &      \\ 
13  & 0.00 & 0.00 & 0.00 & 0.00 & 0.00 & 0.00 & 0.00 & 0.00 & 0.00 & 0.00 & 0.00 & 0.00 & 1.00 &      &      \\ 
14  & 0.00 & 0.00 & 0.00 & 0.00 & 0.00 & 0.00 & 0.00 & 0.00 & 0.00 & 0.00 & 0.00 & 0.00 & 0.00 & 1.00 &      \\ 
15  & 0.00 & 0.00 & 0.00 & 0.00 & 0.00 & 0.00 & 0.00 & 0.00 & 0.00 & 0.00 & 0.00 & 0.00 & 0.00 & 0.00 & 1.00   \\ 
\hline
\end{tabular}
\label{tab:LPer_phistat}
\end{sidewaystable}

\begin{sidewaystable}[h]
\caption{Correlation matrix of efficiency uncertainty for one-dimensional \zy measurement.} 
\centering 
\begin{tabular}{c|cccccccccccccccccc} 
\hline
Interval & 1 & 2 & 3 & 4 & 5 & 6 & 7 & 8 & 9 & 10 & 11 & 12 & 13 & 14 & 15 & 16 & 17 & 18  \\ \hline
1  & 1.00 &      &      &      &      &      &      &      &      &      &      &      &      &      &      &      &      &      \\
2  & 0.84 & 1.00 &      &      &      &      &      &      &      &      &      &      &      &      &      &      &      &      \\
3  & 0.67 & 0.92 & 1.00 &      &      &      &      &      &      &      &      &      &      &      &      &      &      &      \\
4  & 0.57 & 0.81 & 0.95 & 1.00 &      &      &      &      &      &      &      &      &      &      &      &      &      &      \\
5  & 0.49 & 0.70 & 0.84 & 0.95 & 1.00 &      &      &      &      &      &      &      &      &      &      &      &      &      \\
6  & 0.43 & 0.62 & 0.75 & 0.87 & 0.97 & 1.00 &      &      &      &      &      &      &      &      &      &      &      &      \\
7  & 0.37 & 0.55 & 0.67 & 0.79 & 0.91 & 0.98 & 1.00 &      &      &      &      &      &      &      &      &      &      &      \\
8  & 0.31 & 0.47 & 0.59 & 0.71 & 0.83 & 0.91 & 0.97 & 1.00 &      &      &      &      &      &      &      &      &      &      \\
9  & 0.25 & 0.38 & 0.49 & 0.62 & 0.74 & 0.83 & 0.92 & 0.98 & 1.00 &      &      &      &      &      &      &      &      &      \\
10  & 0.17 & 0.29 & 0.40 & 0.51 & 0.64 & 0.74 & 0.84 & 0.93 & 0.98 & 1.00 &      &      &      &      &      &      &      &      \\
11  & 0.09 & 0.18 & 0.28 & 0.39 & 0.53 & 0.63 & 0.75 & 0.85 & 0.93 & 0.98 & 1.00 &      &      &      &      &      &      &      \\
12  & 0.02 & 0.06 & 0.14 & 0.25 & 0.39 & 0.50 & 0.63 & 0.75 & 0.85 & 0.93 & 0.98 & 1.00 &      &      &      &      &      &      \\
13  & 0.00 & 0.01 & 0.05 & 0.13 & 0.26 & 0.38 & 0.52 & 0.65 & 0.77 & 0.87 & 0.95 & 0.99 & 1.00 &      &      &      &      &      \\
14  & 0.00 & 0.00 & 0.01 & 0.06 & 0.15 & 0.26 & 0.40 & 0.55 & 0.68 & 0.80 & 0.89 & 0.95 & 0.98 & 1.00 &      &      &      &      \\
15  & 0.00 & 0.00 & 0.00 & 0.02 & 0.06 & 0.14 & 0.27 & 0.42 & 0.57 & 0.70 & 0.81 & 0.88 & 0.93 & 0.98 & 1.00 &      &      &      \\
16  & 0.00 & 0.00 & 0.00 & 0.00 & 0.02 & 0.06 & 0.16 & 0.31 & 0.46 & 0.60 & 0.73 & 0.81 & 0.87 & 0.93 & 0.98 & 1.00 &      &      \\
17  & 0.00 & 0.00 & 0.00 & 0.00 & 0.00 & 0.01 & 0.05 & 0.14 & 0.28 & 0.43 & 0.57 & 0.66 & 0.73 & 0.80 & 0.87 & 0.93 & 1.00 &      \\
18  & 0.00 & 0.00 & 0.00 & 0.00 & 0.00 & 0.00 & 0.00 & 0.01 & 0.07 & 0.19 & 0.33 & 0.42 & 0.48 & 0.55 & 0.62 & 0.69 & 0.85 & 1.00   \\
\hline 
\end{tabular}
\label{tab:LPer_ysyst}
\end{sidewaystable}

\begin{sidewaystable}[h]
\caption{Correlation matrix of efficiency uncertainty for one-dimensional \zpt measurement.} 
\centering 
\begin{tabular}{c|cccccccccccccccccc} 
\hline
Interval & 1 & 2 & 3 & 4 & 5 & 6 & 7 & 8 & 9 & 10 & 11 & 12 & 13 & 14    \\ \hline
1  & 1.00 &      &      &      &      &      &      &      &      &      &      &      &      &      \\
2  & 1.00 & 1.00 &      &      &      &      &      &      &      &      &      &      &      &      \\
3  & 1.00 & 1.00 & 1.00 &      &      &      &      &      &      &      &      &      &      &      \\
4  & 1.00 & 1.00 & 1.00 & 1.00 &      &      &      &      &      &      &      &      &      &      \\
5  & 1.00 & 1.00 & 1.00 & 1.00 & 1.00 &      &      &      &      &      &      &      &      &      \\
6  & 1.00 & 1.00 & 1.00 & 1.00 & 1.00 & 1.00 &      &      &      &      &      &      &      &      \\
7  & 1.00 & 1.00 & 1.00 & 1.00 & 1.00 & 1.00 & 1.00 &      &      &      &      &      &      &      \\
8  & 1.00 & 1.00 & 1.00 & 1.00 & 1.00 & 1.00 & 1.00 & 1.00 &      &      &      &      &      &      \\
9  & 1.00 & 1.00 & 1.00 & 1.00 & 1.00 & 1.00 & 1.00 & 1.00 & 1.00 &      &      &      &      &      \\
10  & 1.00 & 1.00 & 1.00 & 1.00 & 1.00 & 1.00 & 1.00 & 1.00 & 1.00 & 1.00 &      &      &      &      \\
11  & 1.00 & 1.00 & 1.00 & 1.00 & 1.00 & 1.00 & 1.00 & 1.00 & 1.00 & 1.00 & 1.00 &      &      &      \\
12  & 0.99 & 0.99 & 0.99 & 0.99 & 0.99 & 0.99 & 0.99 & 0.99 & 1.00 & 1.00 & 1.00 & 1.00 &      &      \\
13  & 0.98 & 0.98 & 0.98 & 0.98 & 0.98 & 0.98 & 0.99 & 0.99 & 0.99 & 0.99 & 0.99 & 1.00 & 1.00 &      \\
14  & 0.95 & 0.94 & 0.94 & 0.95 & 0.95 & 0.95 & 0.95 & 0.95 & 0.96 & 0.96 & 0.97 & 0.98 & 0.99 & 1.00   \\ 
\hline 
\end{tabular}
\label{tab:LPer_ptsyst}
\end{sidewaystable}

\begin{sidewaystable}[h]
\caption{Correlation matrix of efficiency uncertainty for one-dimensional \phistar measurement.} 
\centering 
\begin{tabular}{c|cccccccccccccccccc} 
\hline
Interval & 1 & 2 & 3 & 4 & 5 & 6 & 7 & 8 & 9 & 10 & 11 & 12 & 13 & 14 & 15   \\ \hline
1  & 1.00 &      &      &      &      &      &      &      &      &      &      &      &      &      &      \\
2  & 1.00 & 1.00 &      &      &      &      &      &      &      &      &      &      &      &      &      \\
3  & 1.00 & 1.00 & 1.00 &      &      &      &      &      &      &      &      &      &      &      &      \\
4  & 1.00 & 1.00 & 1.00 & 1.00 &      &      &      &      &      &      &      &      &      &      &      \\
5  & 1.00 & 1.00 & 1.00 & 1.00 & 1.00 &      &      &      &      &      &      &      &      &      &      \\
6  & 1.00 & 1.00 & 1.00 & 1.00 & 1.00 & 1.00 &      &      &      &      &      &      &      &      &      \\
7  & 1.00 & 1.00 & 1.00 & 1.00 & 1.00 & 1.00 & 1.00 &      &      &      &      &      &      &      &      \\
8  & 1.00 & 1.00 & 1.00 & 1.00 & 1.00 & 1.00 & 1.00 & 1.00 &      &      &      &      &      &      &      \\
9  & 0.99 & 0.99 & 0.99 & 1.00 & 1.00 & 1.00 & 1.00 & 1.00 & 1.00 &      &      &      &      &      &      \\
10  & 0.99 & 0.99 & 0.99 & 0.99 & 0.99 & 0.99 & 1.00 & 1.00 & 1.00 & 1.00 &      &      &      &      &      \\
11  & 0.98 & 0.98 & 0.98 & 0.98 & 0.99 & 0.99 & 0.99 & 0.99 & 1.00 & 1.00 & 1.00 &      &      &      &      \\
12  & 0.97 & 0.97 & 0.96 & 0.97 & 0.97 & 0.97 & 0.98 & 0.98 & 0.99 & 0.99 & 1.00 & 1.00 &      &      &      \\
13  & 0.96 & 0.96 & 0.96 & 0.96 & 0.97 & 0.97 & 0.97 & 0.98 & 0.98 & 0.99 & 0.99 & 1.00 & 1.00 &      &      \\
14  & 0.93 & 0.93 & 0.93 & 0.94 & 0.94 & 0.95 & 0.95 & 0.95 & 0.96 & 0.96 & 0.98 & 0.98 & 0.99 & 1.00 &      \\
15  & 0.92 & 0.92 & 0.91 & 0.92 & 0.92 & 0.93 & 0.93 & 0.93 & 0.93 & 0.93 & 0.94 & 0.94 & 0.95 & 0.98 & 1.00   \\ 
\hline 
\end{tabular}
\label{tab:LPer_phisyst}
\end{sidewaystable}

\clearpage

\section{Numerical results}
\label{app:results}
The measured total cross-sections using different data sets are presented in Table~\ref{tab:xsec_tot}.
The measured single differential cross-sections in interval regions of \zy, \zpt and \phistar are presented from
Table~\ref{tab:cen_y} to~\ref{tab:cen_phi}.
The measured double differential cross-section as a function of \zpt and \phistar in interval regions of \zy are presented in
Tables~\ref{tab:cen_Zpty} and~\ref{tab:cen_ZY_ZPHI}.

The summarized systematic uncertainties for single differential cross-sections are shown in 
Table~\ref{tab:err_y} to~\ref{tab:err_phi}, and in Tables~\ref{tab:err_Zpty} and~\ref{tab:err_ZY_ZPHI}
for double differential cross-section measurements.

\begin{table}[h]
\begin{center}
\caption{Measured total \Z-boson cross-section for different datasets.
The first uncertainty is statistical, the second systematic, and the third is due to the luminosity. }
\begin{tabular}{c|c}
\hline
  Year  & $\sigma(\Zmm)$ [$\pb$] \\
\hline
2016    &  $195.0 \pm 0.4 \pm 2.0 \pm 3.9$ \\
2017    &  $197.0 \pm 0.4 \pm 1.9 \pm 3.9$ \\
2018    &  $197.3 \pm 0.4 \pm 1.9 \pm 3.9$ \\
Run II    &  $\totxsec$ \\
\hline
 \end{tabular}
\label{tab:xsec_tot}
\end{center}
\end{table}

\begin{table}[h]
\caption{Measured single differential cross-sections in interval regions of \zy. 
The first uncertainty is statistical, the second systematic, and the third is due to the luminosity. }
\begin{center}
\begin{tabular}{rcl|ccccccc}
\hline
  \multicolumn{3}{c|}{\zy}  & \multicolumn{7}{c}{$d\sigma(\Zmm)/d{\zy}$ [$\pb$]} \\
\hline
2.000 & $-$ & 2.125 & 12.8 &$\pm$& 0.2 &$\pm$& 0.2 &$\pm$& 0.3 \\
2.125 & $-$ & 2.250 & 40.4 &$\pm$& 0.3 &$\pm$& 0.4 &$\pm$& 0.8 \\
2.250 & $-$ & 2.375 & 65.2 &$\pm$& 0.4 &$\pm$& 0.6 &$\pm$& 1.3 \\
2.375 & $-$ & 2.500 & 87.5 &$\pm$& 0.4 &$\pm$& 0.6 &$\pm$& 1.7 \\
2.500 & $-$ & 2.625 & 106.3 &$\pm$& 0.5 &$\pm$& 0.8 &$\pm$& 2.1 \\
2.625 & $-$ & 2.750 & 122.7 &$\pm$& 0.5 &$\pm$& 0.9 &$\pm$& 2.5 \\
2.750 & $-$ & 2.875 & 134.5 &$\pm$& 0.5 &$\pm$& 0.9 &$\pm$& 2.7 \\
2.875 & $-$ & 3.000 & 141.7 &$\pm$& 0.5 &$\pm$& 0.9 &$\pm$& 2.8 \\
3.000 & $-$ & 3.125 & 147.5 &$\pm$& 0.5 &$\pm$& 1.0 &$\pm$& 2.9 \\
3.125 & $-$ & 3.250 & 145.4 &$\pm$& 0.5 &$\pm$& 1.0 &$\pm$& 2.9 \\
3.250 & $-$ & 3.375 & 134.8 &$\pm$& 0.5 &$\pm$& 1.0 &$\pm$& 2.7 \\
3.375 & $-$ & 3.500 & 118.5 &$\pm$& 0.5 &$\pm$& 0.9 &$\pm$& 2.4 \\
3.500 & $-$ & 3.625 & 99.0 &$\pm$& 0.4 &$\pm$& 0.7 &$\pm$& 2.0 \\
3.625 & $-$ & 3.750 & 77.6 &$\pm$& 0.4 &$\pm$& 0.8 &$\pm$& 1.6 \\
3.750 & $-$ & 3.875 & 57.9 &$\pm$& 0.3 &$\pm$& 0.5 &$\pm$& 1.2 \\
3.875 & $-$ & 4.000 & 39.5 &$\pm$& 0.3 &$\pm$& 0.4 &$\pm$& 0.8 \\
4.000 & $-$ & 4.250 & 18.2 &$\pm$& 0.1 &$\pm$& 0.2 &$\pm$& 0.4 \\
4.250 & $-$ & 4.500 & 2.7 &$\pm$& 0.1 &$\pm$& 0.1 &$\pm$& 0.1 \\
\hline
\end{tabular}
\label{tab:cen_y}
\end{center}
\end{table}

\begin{table}[h]
\caption{Measured single differential cross-sections in interval regions of \zpt. 
The first uncertainty is statistical, the second systematic, and the third is due to the luminosity. }
\begin{center}
\begin{tabular}{rcl|ccccccc}
\hline
  \multicolumn{3}{c|}{\zpt~[\gevc]}  & \multicolumn{7}{c}{$d\sigma(\Zmm)/d{\zpt}$ [$\pb$]} \\
\hline
0.0 & $-$ & 2.2 & 5.70 &$\pm$& 0.03 &$\pm$& 0.08 &$\pm$& 0.11 \\
2.2 & $-$ & 3.4 & 11.07 &$\pm$& 0.05 &$\pm$& 0.16 &$\pm$& 0.22 \\
3.4 & $-$ & 4.6 & 11.44 &$\pm$& 0.05 &$\pm$& 0.16 &$\pm$& 0.23 \\
4.6 & $-$ & 5.8 & 11.28 &$\pm$& 0.05 &$\pm$& 0.15 &$\pm$& 0.23 \\
5.8 & $-$ & 7.2 & 9.94 &$\pm$& 0.04 &$\pm$& 0.11 &$\pm$& 0.20 \\
7.2 & $-$ & 8.7 & 8.86 &$\pm$& 0.04 &$\pm$& 0.12 &$\pm$& 0.18 \\
8.7 & $-$ & 10.5 & 7.75 &$\pm$& 0.03 &$\pm$& 0.09 &$\pm$& 0.15 \\
10.5 & $-$ & 12.8 & 6.44 &$\pm$& 0.03 &$\pm$& 0.07 &$\pm$& 0.13 \\
12.8 & $-$ & 15.4 & 5.16 &$\pm$& 0.02 &$\pm$& 0.05 &$\pm$& 0.10 \\
15.4 & $-$ & 19.0 & 4.03 &$\pm$& 0.02 &$\pm$& 0.04 &$\pm$& 0.08 \\
19.0 & $-$ & 24.5 & 2.88 &$\pm$& 0.01 &$\pm$& 0.03 &$\pm$& 0.06 \\
24.5 & $-$ & 34.0 & 1.774 &$\pm$& 0.007 &$\pm$& 0.016 &$\pm$& 0.035 \\
34.0 & $-$ & 63.0 & 0.674 &$\pm$& 0.002 &$\pm$& 0.007 &$\pm$& 0.013 \\
63.0 & $-$ & 270.0 & 0.0361 &$\pm$& 0.0002 &$\pm$& 0.0004 &$\pm$& 0.0007 \\
\hline
\end{tabular}
\label{tab:cen_Zpt}
\end{center}
\end{table}

\begin{table}[h]
\caption{Measured single differential cross-sections in interval regions of \phistar. 
The first uncertainty is statistical, the second systematic, and the third is due to the luminosity.}
\begin{center}
\begin{tabular}{rcl|ccccccc}
\hline
  \multicolumn{3}{c|}{\phistar}  & \multicolumn{7}{c}{$d\sigma(\Zmm)/d{\phistar}$ [$\pb$]} \\
\hline
0.00 & $-$ & 0.01 & 1885 &$\pm$& 7 &$\pm$& 17 &$\pm$& 38 \\
0.01 & $-$ & 0.02 & 1780 &$\pm$& 7 &$\pm$& 16 &$\pm$& 36 \\
0.02 & $-$ & 0.03 & 1609 &$\pm$& 6 &$\pm$& 14 &$\pm$& 32 \\
0.03 & $-$ & 0.05 & 1322 &$\pm$& 4 &$\pm$& 11 &$\pm$& 26 \\
0.05 & $-$ & 0.07 & 1005 &$\pm$& 4 &$\pm$& 8 &$\pm$& 20 \\
0.07 & $-$ & 0.10 & 724.9 &$\pm$& 2.5 &$\pm$& 6.0 &$\pm$& 14.5 \\
0.10 & $-$ & 0.15 & 462.8 &$\pm$& 1.5 &$\pm$& 4.0 &$\pm$& 9.3 \\
0.15 & $-$ & 0.20 & 284.4 &$\pm$& 1.2 &$\pm$& 2.5 &$\pm$& 5.7 \\
0.20 & $-$ & 0.30 & 158.9 &$\pm$& 0.6 &$\pm$& 1.4 &$\pm$& 3.2 \\
0.30 & $-$ & 0.40 & 80.73 &$\pm$& 0.44 &$\pm$& 0.84 &$\pm$& 1.61 \\
0.40 & $-$ & 0.60 & 37.10 &$\pm$& 0.21 &$\pm$& 0.32 &$\pm$& 0.74 \\
0.60 & $-$ & 0.80 & 15.16 &$\pm$& 0.14 &$\pm$& 0.15 &$\pm$& 0.30 \\
0.80 & $-$ & 1.20 & 5.54 &$\pm$& 0.06 &$\pm$& 0.06 &$\pm$& 0.11 \\
1.20 & $-$ & 2.00 & 1.286 &$\pm$& 0.020 &$\pm$& 0.021 &$\pm$& 0.026 \\
2.00 & $-$ & 4.00 & 0.185 &$\pm$& 0.005 &$\pm$& 0.003 &$\pm$& 0.004 \\ 
\hline
\end{tabular}
\label{tab:cen_phi}
\end{center}
\end{table}

\clearpage

\begin{center}
\begin{longtable}{rclrcl|ccccccc}
  \caption{Measured double differential cross-sections in interval regions of \zy and \zpt.
The first uncertainty is statistical, the second systematic, and the third is due to the luminosity. }
\label{tab:cen_Zpty} \\ \hline
\multicolumn{3}{c}{\zy}  & \multicolumn{3}{c|}{\zpt [\gevc]} &  \multicolumn{7}{c}{$d^2\sigma({\Zmm})/d\zpt d\zy$~[$\pb$]} \\ \hline
2.0 & $-$ & 2.5 & 0.0 & $-$ & 2.2 & 1.237 &$\pm$& 0.018 &$\pm$& 0.021 &$\pm$& 0.025 \\ 
2.0 & $-$ & 2.5 & 2.2 & $-$ & 3.4 & 2.43 &$\pm$& 0.04 &$\pm$& 0.07 &$\pm$& 0.05 \\ 
2.0 & $-$ & 2.5 & 3.4 & $-$ & 4.6 & 2.68 &$\pm$& 0.03 &$\pm$& 0.04 &$\pm$& 0.05 \\ 
2.0 & $-$ & 2.5 & 4.6 & $-$ & 5.8 & 2.49 &$\pm$& 0.03 &$\pm$& 0.04 &$\pm$& 0.05 \\ 
2.0 & $-$ & 2.5 & 5.8 & $-$ & 7.2 & 2.32 &$\pm$& 0.03 &$\pm$& 0.03 &$\pm$& 0.05 \\ 
2.0 & $-$ & 2.5 & 7.2 & $-$ & 8.7 & 2.11 &$\pm$& 0.03 &$\pm$& 0.03 &$\pm$& 0.04 \\ 
2.0 & $-$ & 2.5 & 8.7 & $-$ & 10.5 & 1.849 &$\pm$& 0.022 &$\pm$& 0.030 &$\pm$& 0.037 \\ 
2.0 & $-$ & 2.5 & 10.5 & $-$ & 12.8 & 1.542 &$\pm$& 0.018 &$\pm$& 0.026 &$\pm$& 0.031 \\ 
2.0 & $-$ & 2.5 & 12.8 & $-$ & 15.4 & 1.321 &$\pm$& 0.016 &$\pm$& 0.016 &$\pm$& 0.026 \\ 
2.0 & $-$ & 2.5 & 15.4 & $-$ & 19.0 & 1.030 &$\pm$& 0.012 &$\pm$& 0.013 &$\pm$& 0.021 \\ 
2.0 & $-$ & 2.5 & 19.0 & $-$ & 24.5 & 0.782 &$\pm$& 0.008 &$\pm$& 0.009 &$\pm$& 0.016 \\ 
2.0 & $-$ & 2.5 & 24.5 & $-$ & 34.0 & 0.517 &$\pm$& 0.005 &$\pm$& 0.005 &$\pm$& 0.010 \\ 
2.0 & $-$ & 2.5 & 34.0 & $-$ & 63.0 & 0.228 &$\pm$& 0.002 &$\pm$& 0.003 &$\pm$& 0.005 \\ 
2.0 & $-$ & 2.5 & 63.0 & $-$ & 270.0 & 0.0167 &$\pm$& 0.0002 &$\pm$& 0.0002 &$\pm$& 0.0003 \\ 
2.5 & $-$ & 3.0 & 0.0 & $-$ & 2.2 & 3.37 &$\pm$& 0.03 &$\pm$& 0.05 &$\pm$& 0.07 \\ 
2.5 & $-$ & 3.0 & 2.2 & $-$ & 3.4 & 6.53 &$\pm$& 0.06 &$\pm$& 0.08 &$\pm$& 0.13 \\ 
2.5 & $-$ & 3.0 & 3.4 & $-$ & 4.6 & 6.88 &$\pm$& 0.05 &$\pm$& 0.09 &$\pm$& 0.14 \\ 
2.5 & $-$ & 3.0 & 4.6 & $-$ & 5.8 & 6.80 &$\pm$& 0.06 &$\pm$& 0.11 &$\pm$& 0.14 \\ 
2.5 & $-$ & 3.0 & 5.8 & $-$ & 7.2 & 6.09 &$\pm$& 0.05 &$\pm$& 0.07 &$\pm$& 0.12 \\ 
2.5 & $-$ & 3.0 & 7.2 & $-$ & 8.7 & 5.57 &$\pm$& 0.04 &$\pm$& 0.06 &$\pm$& 0.11 \\ 
2.5 & $-$ & 3.0 & 8.7 & $-$ & 10.5 & 4.89 &$\pm$& 0.04 &$\pm$& 0.06 &$\pm$& 0.10 \\ 
2.5 & $-$ & 3.0 & 10.5 & $-$ & 12.8 & 4.04 &$\pm$& 0.03 &$\pm$& 0.04 &$\pm$& 0.08 \\ 
2.5 & $-$ & 3.0 & 12.8 & $-$ & 15.4 & 3.26 &$\pm$& 0.02 &$\pm$& 0.03 &$\pm$& 0.07 \\ 
2.5 & $-$ & 3.0 & 15.4 & $-$ & 19.0 & 2.60 &$\pm$& 0.02 &$\pm$& 0.03 &$\pm$& 0.05 \\ 
2.5 & $-$ & 3.0 & 19.0 & $-$ & 24.5 & 1.911 &$\pm$& 0.013 &$\pm$& 0.017 &$\pm$& 0.038 \\ 
2.5 & $-$ & 3.0 & 24.5 & $-$ & 34.0 & 1.202 &$\pm$& 0.008 &$\pm$& 0.010 &$\pm$& 0.024 \\ 
2.5 & $-$ & 3.0 & 34.0 & $-$ & 63.0 & 0.481 &$\pm$& 0.003 &$\pm$& 0.004 &$\pm$& 0.010 \\ 
2.5 & $-$ & 3.0 & 63.0 & $-$ & 270.0 & 0.0289 &$\pm$& 0.0003 &$\pm$& 0.0002 &$\pm$& 0.0006 \\ 
3.0 & $-$ & 3.5 & 0.0 & $-$ & 2.2 & 4.16 &$\pm$& 0.04 &$\pm$& 0.06 &$\pm$& 0.08 \\ 
3.0 & $-$ & 3.5 & 2.2 & $-$ & 3.4 & 8.06 &$\pm$& 0.06 &$\pm$& 0.10 &$\pm$& 0.16 \\ 
3.0 & $-$ & 3.5 & 3.4 & $-$ & 4.6 & 8.25 &$\pm$& 0.06 &$\pm$& 0.13 &$\pm$& 0.17 \\ 
3.0 & $-$ & 3.5 & 4.6 & $-$ & 5.8 & 8.17 &$\pm$& 0.06 &$\pm$& 0.10 &$\pm$& 0.16 \\ 
3.0 & $-$ & 3.5 & 5.8 & $-$ & 7.2 & 7.25 &$\pm$& 0.05 &$\pm$& 0.07 &$\pm$& 0.14 \\ 
3.0 & $-$ & 3.5 & 7.2 & $-$ & 8.7 & 6.26 &$\pm$& 0.05 &$\pm$& 0.10 &$\pm$& 0.13 \\ 
3.0 & $-$ & 3.5 & 8.7 & $-$ & 10.5 & 5.56 &$\pm$& 0.04 &$\pm$& 0.06 &$\pm$& 0.11 \\ 
3.0 & $-$ & 3.5 & 10.5 & $-$ & 12.8 & 4.51 &$\pm$& 0.03 &$\pm$& 0.04 &$\pm$& 0.09 \\ 
3.0 & $-$ & 3.5 & 12.8 & $-$ & 15.4 & 3.63 &$\pm$& 0.03 &$\pm$& 0.03 &$\pm$& 0.07 \\ 
3.0 & $-$ & 3.5 & 15.4 & $-$ & 19.0 & 2.81 &$\pm$& 0.02 &$\pm$& 0.02 &$\pm$& 0.06 \\ 
3.0 & $-$ & 3.5 & 19.0 & $-$ & 24.5 & 1.966 &$\pm$& 0.013 &$\pm$& 0.033 &$\pm$& 0.039 \\ 
3.0 & $-$ & 3.5 & 24.5 & $-$ & 34.0 & 1.175 &$\pm$& 0.008 &$\pm$& 0.010 &$\pm$& 0.024 \\ 
3.0 & $-$ & 3.5 & 34.0 & $-$ & 63.0 & 0.429 &$\pm$& 0.003 &$\pm$& 0.005 &$\pm$& 0.009 \\ 
3.0 & $-$ & 3.5 & 63.0 & $-$ & 270.0 & 0.0202 &$\pm$& 0.0002 &$\pm$& 0.0002 &$\pm$& 0.0004 \\ 
3.5 & $-$ & 4.0 & 0.0 & $-$ & 2.2 & 2.25 &$\pm$& 0.03 &$\pm$& 0.04 &$\pm$& 0.04 \\ 
3.5 & $-$ & 4.0 & 2.2 & $-$ & 3.4 & 4.44 &$\pm$& 0.05 &$\pm$& 0.09 &$\pm$& 0.09 \\ 
3.5 & $-$ & 4.0 & 3.4 & $-$ & 4.6 & 4.37 &$\pm$& 0.05 &$\pm$& 0.09 &$\pm$& 0.09 \\ 
3.5 & $-$ & 4.0 & 4.6 & $-$ & 5.8 & 4.25 &$\pm$& 0.05 &$\pm$& 0.06 &$\pm$& 0.08 \\ 
3.5 & $-$ & 4.0 & 5.8 & $-$ & 7.2 & 3.60 &$\pm$& 0.04 &$\pm$& 0.06 &$\pm$& 0.07 \\ 
3.5 & $-$ & 4.0 & 7.2 & $-$ & 8.7 & 3.29 &$\pm$& 0.03 &$\pm$& 0.05 &$\pm$& 0.07 \\ 
3.5 & $-$ & 4.0 & 8.7 & $-$ & 10.5 & 2.78 &$\pm$& 0.03 &$\pm$& 0.06 &$\pm$& 0.06 \\ 
3.5 & $-$ & 4.0 & 10.5 & $-$ & 12.8 & 2.37 &$\pm$& 0.02 &$\pm$& 0.03 &$\pm$& 0.05 \\ 
3.5 & $-$ & 4.0 & 12.8 & $-$ & 15.4 & 1.849 &$\pm$& 0.019 &$\pm$& 0.026 &$\pm$& 0.037 \\ 
3.5 & $-$ & 4.0 & 15.4 & $-$ & 19.0 & 1.407 &$\pm$& 0.014 &$\pm$& 0.025 &$\pm$& 0.028 \\ 
3.5 & $-$ & 4.0 & 19.0 & $-$ & 24.5 & 0.952 &$\pm$& 0.009 &$\pm$& 0.011 &$\pm$& 0.019 \\ 
3.5 & $-$ & 4.0 & 24.5 & $-$ & 34.0 & 0.576 &$\pm$& 0.006 &$\pm$& 0.008 &$\pm$& 0.012 \\ 
3.5 & $-$ & 4.0 & 34.0 & $-$ & 63.0 & 0.188 &$\pm$& 0.002 &$\pm$& 0.003 &$\pm$& 0.004 \\ 
3.5 & $-$ & 4.0 & 63.0 & $-$ & 270.0 & 0.00594 &$\pm$& 0.00012 &$\pm$& 0.00021 &$\pm$& 0.00012 \\ 
4.0 & $-$ & 4.5 & 0.0 & $-$ & 2.2 & 0.377 &$\pm$& 0.012 &$\pm$& 0.023 &$\pm$& 0.008 \\ 
4.0 & $-$ & 4.5 & 2.2 & $-$ & 3.4 & 0.702 &$\pm$& 0.020 &$\pm$& 0.026 &$\pm$& 0.014 \\ 
4.0 & $-$ & 4.5 & 3.4 & $-$ & 4.6 & 0.760 &$\pm$& 0.020 &$\pm$& 0.024 &$\pm$& 0.015 \\ 
4.0 & $-$ & 4.5 & 4.6 & $-$ & 5.8 & 0.762 &$\pm$& 0.022 &$\pm$& 0.025 &$\pm$& 0.015 \\ 
4.0 & $-$ & 4.5 & 5.8 & $-$ & 7.2 & 0.610 &$\pm$& 0.017 &$\pm$& 0.020 &$\pm$& 0.012 \\ 
4.0 & $-$ & 4.5 & 7.2 & $-$ & 8.7 & 0.509 &$\pm$& 0.015 &$\pm$& 0.026 &$\pm$& 0.010 \\ 
4.0 & $-$ & 4.5 & 8.7 & $-$ & 10.5 & 0.432 &$\pm$& 0.014 &$\pm$& 0.016 &$\pm$& 0.009 \\ 
4.0 & $-$ & 4.5 & 10.5 & $-$ & 12.8 & 0.392 &$\pm$& 0.010 &$\pm$& 0.022 &$\pm$& 0.008 \\ 
4.0 & $-$ & 4.5 & 12.8 & $-$ & 15.4 & 0.260 &$\pm$& 0.007 &$\pm$& 0.009 &$\pm$& 0.005 \\ 
4.0 & $-$ & 4.5 & 15.4 & $-$ & 19.0 & 0.213 &$\pm$& 0.006 &$\pm$& 0.009 &$\pm$& 0.004 \\ 
4.0 & $-$ & 4.5 & 19.0 & $-$ & 24.5 & 0.141 &$\pm$& 0.004 &$\pm$& 0.007 &$\pm$& 0.003 \\ 
4.0 & $-$ & 4.5 & 24.5 & $-$ & 34.0 & 0.0768 &$\pm$& 0.0021 &$\pm$& 0.0015 &$\pm$& 0.0015 \\ 
4.0 & $-$ & 4.5 & 34.0 & $-$ & 63.0 & 0.0211 &$\pm$& 0.0006 &$\pm$& 0.0009 &$\pm$& 0.0004 \\ 
4.0 & $-$ & 4.5 & 63.0 & $-$ & 270.0 & 0.00033 &$\pm$& 0.00003 &$\pm$& 0.00001 &$\pm$& 0.00001 \\ 
\hline
\end{longtable}
\end{center}

\begin{center}
\begin{longtable}{rclrcl|ccccccc}
	\caption{Measured double differential cross-sections in interval regions of \zy and \phistar. 
The first uncertainty is statistical, the second systematic, and the third is due to the luminosity. }
	\label{tab:cen_ZY_ZPHI} \\ \hline
\multicolumn{3}{c}{\zy} & \multicolumn{3}{c|}{\phistar} &  \multicolumn{7}{c}{$d^2\sigma(\Zmm)/d\phistar d\zy$~[$\pb$]} \\ \hline
2.0 & $-$ & 2.5 & 0.00 & $-$ & 0.01 & 437.3 &$\pm$& 4.7 &$\pm$& 4.3 &$\pm$& 8.7 \\ 
2.0 & $-$ & 2.5 & 0.01 & $-$ & 0.02 & 415.4 &$\pm$& 4.5 &$\pm$& 4.8 &$\pm$& 8.3 \\ 
2.0 & $-$ & 2.5 & 0.02 & $-$ & 0.03 & 373.3 &$\pm$& 4.3 &$\pm$& 3.0 &$\pm$& 7.5 \\ 
2.0 & $-$ & 2.5 & 0.03 & $-$ & 0.05 & 321.4 &$\pm$& 2.8 &$\pm$& 3.1 &$\pm$& 6.4 \\ 
2.0 & $-$ & 2.5 & 0.05 & $-$ & 0.07 & 248.0 &$\pm$& 2.5 &$\pm$& 1.9 &$\pm$& 5.0 \\ 
2.0 & $-$ & 2.5 & 0.07 & $-$ & 0.10 & 180.9 &$\pm$& 1.7 &$\pm$& 1.4 &$\pm$& 3.6 \\ 
2.0 & $-$ & 2.5 & 0.10 & $-$ & 0.15 & 122.6 &$\pm$& 1.1 &$\pm$& 1.3 &$\pm$& 2.5 \\ 
2.0 & $-$ & 2.5 & 0.15 & $-$ & 0.20 & 78.78 &$\pm$& 0.87 &$\pm$& 0.83 &$\pm$& 1.58 \\ 
2.0 & $-$ & 2.5 & 0.20 & $-$ & 0.30 & 45.68 &$\pm$& 0.47 &$\pm$& 0.56 &$\pm$& 0.91 \\ 
2.0 & $-$ & 2.5 & 0.30 & $-$ & 0.40 & 25.61 &$\pm$& 0.35 &$\pm$& 0.26 &$\pm$& 0.51 \\ 
2.0 & $-$ & 2.5 & 0.40 & $-$ & 0.60 & 12.57 &$\pm$& 0.17 &$\pm$& 0.16 &$\pm$& 0.25 \\ 
2.0 & $-$ & 2.5 & 0.60 & $-$ & 0.80 & 5.68 &$\pm$& 0.11 &$\pm$& 0.08 &$\pm$& 0.11 \\ 
2.0 & $-$ & 2.5 & 0.80 & $-$ & 1.20 & 2.34 &$\pm$& 0.05 &$\pm$& 0.03 &$\pm$& 0.05 \\ 
2.0 & $-$ & 2.5 & 1.20 & $-$ & 2.00 & 0.554 &$\pm$& 0.018 &$\pm$& 0.010 &$\pm$& 0.011 \\ 
2.0 & $-$ & 2.5 & 2.00 & $-$ & 4.00 & 0.0645 &$\pm$& 0.0039 &$\pm$& 0.0015 &$\pm$& 0.0013 \\ 
2.5 & $-$ & 3.0 & 0.00 & $-$ & 0.01 & 1145 &$\pm$& 8 &$\pm$& 12 &$\pm$& 23 \\ 
2.5 & $-$ & 3.0 & 0.01 & $-$ & 0.02 & 1091 &$\pm$& 7 &$\pm$& 9 &$\pm$& 22 \\ 
2.5 & $-$ & 3.0 & 0.02 & $-$ & 0.03 & 984.1 &$\pm$& 7.0 &$\pm$& 10.7 &$\pm$& 19.7 \\ 
2.5 & $-$ & 3.0 & 0.03 & $-$ & 0.05 & 815.6 &$\pm$& 4.5 &$\pm$& 5.7 &$\pm$& 16.3 \\ 
2.5 & $-$ & 3.0 & 0.05 & $-$ & 0.07 & 631.9 &$\pm$& 3.9 &$\pm$& 4.4 &$\pm$& 12.6 \\ 
2.5 & $-$ & 3.0 & 0.07 & $-$ & 0.10 & 459.0 &$\pm$& 2.7 &$\pm$& 3.0 &$\pm$& 9.2 \\ 
2.5 & $-$ & 3.0 & 0.10 & $-$ & 0.15 & 297.7 &$\pm$& 1.7 &$\pm$& 2.1 &$\pm$& 6.0 \\ 
2.5 & $-$ & 3.0 & 0.15 & $-$ & 0.20 & 187.0 &$\pm$& 1.3 &$\pm$& 1.6 &$\pm$& 3.7 \\ 
2.5 & $-$ & 3.0 & 0.20 & $-$ & 0.30 & 107.5 &$\pm$& 0.7 &$\pm$& 0.8 &$\pm$& 2.2 \\ 
2.5 & $-$ & 3.0 & 0.30 & $-$ & 0.40 & 55.69 &$\pm$& 0.52 &$\pm$& 0.40 &$\pm$& 1.11 \\ 
2.5 & $-$ & 3.0 & 0.40 & $-$ & 0.60 & 26.82 &$\pm$& 0.25 &$\pm$& 0.20 &$\pm$& 0.54 \\ 
2.5 & $-$ & 3.0 & 0.60 & $-$ & 0.80 & 11.60 &$\pm$& 0.17 &$\pm$& 0.09 &$\pm$& 0.23 \\ 
2.5 & $-$ & 3.0 & 0.80 & $-$ & 1.20 & 4.32 &$\pm$& 0.07 &$\pm$& 0.04 &$\pm$& 0.09 \\ 
2.5 & $-$ & 3.0 & 1.20 & $-$ & 2.00 & 1.112 &$\pm$& 0.026 &$\pm$& 0.024 &$\pm$& 0.022 \\ 
2.5 & $-$ & 3.0 & 2.00 & $-$ & 4.00 & 0.166 &$\pm$& 0.006 &$\pm$& 0.002 &$\pm$& 0.003 \\ 
3.0 & $-$ & 3.5 & 0.00 & $-$ & 0.01 & 1334 &$\pm$& 8 &$\pm$& 10 &$\pm$& 27 \\ 
3.0 & $-$ & 3.5 & 0.01 & $-$ & 0.02 & 1264 &$\pm$& 8 &$\pm$& 10 &$\pm$& 25 \\ 
3.0 & $-$ & 3.5 & 0.02 & $-$ & 0.03 & 1149 &$\pm$& 8 &$\pm$& 8 &$\pm$& 23 \\ 
3.0 & $-$ & 3.5 & 0.03 & $-$ & 0.05 & 934.0 &$\pm$& 4.8 &$\pm$& 6.6 &$\pm$& 18.7 \\ 
3.0 & $-$ & 3.5 & 0.05 & $-$ & 0.07 & 708.2 &$\pm$& 4.2 &$\pm$& 5.1 &$\pm$& 14.2 \\ 
3.0 & $-$ & 3.5 & 0.07 & $-$ & 0.10 & 509.1 &$\pm$& 2.9 &$\pm$& 4.2 &$\pm$& 10.2 \\ 
3.0 & $-$ & 3.5 & 0.10 & $-$ & 0.15 & 319.2 &$\pm$& 1.8 &$\pm$& 2.3 &$\pm$& 6.4 \\ 
3.0 & $-$ & 3.5 & 0.15 & $-$ & 0.20 & 194.7 &$\pm$& 1.4 &$\pm$& 1.4 &$\pm$& 3.9 \\ 
3.0 & $-$ & 3.5 & 0.20 & $-$ & 0.30 & 107.2 &$\pm$& 0.7 &$\pm$& 1.0 &$\pm$& 2.1 \\ 
3.0 & $-$ & 3.5 & 0.30 & $-$ & 0.40 & 53.11 &$\pm$& 0.51 &$\pm$& 0.66 &$\pm$& 1.06 \\ 
3.0 & $-$ & 3.5 & 0.40 & $-$ & 0.60 & 24.01 &$\pm$& 0.24 &$\pm$& 0.20 &$\pm$& 0.48 \\ 
3.0 & $-$ & 3.5 & 0.60 & $-$ & 0.80 & 9.50 &$\pm$& 0.15 &$\pm$& 0.10 &$\pm$& 0.19 \\ 
3.0 & $-$ & 3.5 & 0.80 & $-$ & 1.20 & 3.45 &$\pm$& 0.06 &$\pm$& 0.04 &$\pm$& 0.07 \\ 
3.0 & $-$ & 3.5 & 1.20 & $-$ & 2.00 & 0.763 &$\pm$& 0.021 &$\pm$& 0.009 &$\pm$& 0.015 \\ 
3.0 & $-$ & 3.5 & 2.00 & $-$ & 4.00 & 0.119 &$\pm$& 0.005 &$\pm$& 0.003 &$\pm$& 0.002 \\ 
3.5 & $-$ & 4.0 & 0.00 & $-$ & 0.01 & 729.2 &$\pm$& 6.2 &$\pm$& 6.3 &$\pm$& 14.6 \\ 
3.5 & $-$ & 4.0 & 0.01 & $-$ & 0.02 & 676.0 &$\pm$& 5.9 &$\pm$& 6.2 &$\pm$& 13.5 \\ 
3.5 & $-$ & 4.0 & 0.02 & $-$ & 0.03 & 607.1 &$\pm$& 5.6 &$\pm$& 4.5 &$\pm$& 12.1 \\ 
3.5 & $-$ & 4.0 & 0.03 & $-$ & 0.05 & 492.5 &$\pm$& 3.6 &$\pm$& 3.5 &$\pm$& 9.8 \\ 
3.5 & $-$ & 4.0 & 0.05 & $-$ & 0.07 & 361.9 &$\pm$& 3.0 &$\pm$& 3.4 &$\pm$& 7.2 \\ 
3.5 & $-$ & 4.0 & 0.07 & $-$ & 0.10 & 261.1 &$\pm$& 2.1 &$\pm$& 2.4 &$\pm$& 5.2 \\ 
3.5 & $-$ & 4.0 & 0.10 & $-$ & 0.15 & 160.8 &$\pm$& 1.3 &$\pm$& 1.7 &$\pm$& 3.2 \\ 
3.5 & $-$ & 4.0 & 0.15 & $-$ & 0.20 & 94.33 &$\pm$& 0.98 &$\pm$& 1.25 &$\pm$& 1.89 \\ 
3.5 & $-$ & 4.0 & 0.20 & $-$ & 0.30 & 50.61 &$\pm$& 0.51 &$\pm$& 0.46 &$\pm$& 1.01 \\ 
3.5 & $-$ & 4.0 & 0.30 & $-$ & 0.40 & 23.99 &$\pm$& 0.35 &$\pm$& 0.46 &$\pm$& 0.48 \\ 
3.5 & $-$ & 4.0 & 0.40 & $-$ & 0.60 & 9.79 &$\pm$& 0.16 &$\pm$& 0.08 &$\pm$& 0.20 \\ 
3.5 & $-$ & 4.0 & 0.60 & $-$ & 0.80 & 3.27 &$\pm$& 0.09 &$\pm$& 0.05 &$\pm$& 0.07 \\ 
3.5 & $-$ & 4.0 & 0.80 & $-$ & 1.20 & 0.927 &$\pm$& 0.034 &$\pm$& 0.017 &$\pm$& 0.019 \\ 
3.5 & $-$ & 4.0 & 1.20 & $-$ & 2.00 & 0.141 &$\pm$& 0.009 &$\pm$& 0.011 &$\pm$& 0.003 \\ 
3.5 & $-$ & 4.0 & 2.00 & $-$ & 4.00 & 0.0210 &$\pm$& 0.0023 &$\pm$& 0.0014 &$\pm$& 0.0004 \\ 
4.0 & $-$ & 4.5 & 0.00 & $-$ & 0.01 & 121.1 &$\pm$& 2.6 &$\pm$& 2.1 &$\pm$& 2.4 \\ 
4.0 & $-$ & 4.5 & 0.01 & $-$ & 0.02 & 112.8 &$\pm$& 2.5 &$\pm$& 2.0 &$\pm$& 2.3 \\ 
4.0 & $-$ & 4.5 & 0.02 & $-$ & 0.03 & 98.76 &$\pm$& 2.35 &$\pm$& 1.12 &$\pm$& 1.98 \\ 
4.0 & $-$ & 4.5 & 0.03 & $-$ & 0.05 & 80.85 &$\pm$& 1.50 &$\pm$& 1.78 &$\pm$& 1.62 \\ 
4.0 & $-$ & 4.5 & 0.05 & $-$ & 0.07 & 58.34 &$\pm$& 1.27 &$\pm$& 0.59 &$\pm$& 1.17 \\ 
4.0 & $-$ & 4.5 & 0.07 & $-$ & 0.10 & 39.44 &$\pm$& 0.85 &$\pm$& 0.62 &$\pm$& 0.79 \\ 
4.0 & $-$ & 4.5 & 0.10 & $-$ & 0.15 & 25.03 &$\pm$& 0.52 &$\pm$& 0.50 &$\pm$& 0.50 \\ 
4.0 & $-$ & 4.5 & 0.15 & $-$ & 0.20 & 13.26 &$\pm$& 0.38 &$\pm$& 0.28 &$\pm$& 0.27 \\ 
4.0 & $-$ & 4.5 & 0.20 & $-$ & 0.30 & 6.62 &$\pm$& 0.19 &$\pm$& 0.09 &$\pm$& 0.13 \\ 
4.0 & $-$ & 4.5 & 0.30 & $-$ & 0.40 & 2.85 &$\pm$& 0.13 &$\pm$& 0.19 &$\pm$& 0.06 \\ 
4.0 & $-$ & 4.5 & 0.40 & $-$ & 0.60 & 0.992 &$\pm$& 0.052 &$\pm$& 0.018 &$\pm$& 0.020 \\ 
4.0 & $-$ & 4.5 & 0.60 & $-$ & 0.80 & 0.238 &$\pm$& 0.025 &$\pm$& 0.013 &$\pm$& 0.005 \\ 
4.0 & $-$ & 4.5 & 0.80 & $-$ & 1.20 & 0.0248 &$\pm$& 0.0060 &$\pm$& 0.0026 &$\pm$& 0.0005 \\ 
4.0 & $-$ & 4.5 & 1.20 & $-$ & 2.00 & 0.00241 &$\pm$& 0.00140 &$\pm$& 0.00092 &$\pm$& 0.00005 \\ 
4.0 & $-$ & 4.5 & 2.00 & $-$ & 4.00 & 0.00060 &$\pm$& 0.00060 &$\pm$& 0.00055 &$\pm$& 0.00001 \\ 
\hline
\end{longtable}
\end{center}

\clearpage

\begin{table}[h]
\caption{Systematic uncertainties in the single differential cross-sections in interval regions of \zy, presented in percentage. The contributions from efficiency (Eff), background (BKG), final state radiation (FSR), closure test (Closure), and alignment and calibration (Alignment) are shown.}
\begin{center}
\begin{tabular}{rcl|ccccc} \hline
  \multicolumn{3}{c|}{\zy}  & Eff & BKG & FSR & Closure & Alignment \\
\hline
2.000 & $-$ & 2.125 & 0.70 & 0.36 & 0.12 & 0.91 & 0.27 \\
2.125 & $-$ & 2.250 & 0.68 & 0.28 & 0.20 & 0.71 & 0.22 \\
2.250 & $-$ & 2.375 & 0.66 & 0.28 & 0.03 & 0.58 & 0.14 \\
2.375 & $-$ & 2.500 & 0.65 & 0.27 & 0.08 & 0.21 & 0.07 \\
2.500 & $-$ & 2.625 & 0.65 & 0.22 & 0.10 & 0.14 & 0.06 \\
2.625 & $-$ & 2.750 & 0.65 & 0.19 & 0.01 & 0.19 & 0.08 \\
2.750 & $-$ & 2.875 & 0.65 & 0.18 & 0.15 & 0.06 & 0.13 \\
2.875 & $-$ & 3.000 & 0.64 & 0.14 & 0.03 & 0.01 & 0.10 \\
3.000 & $-$ & 3.125 & 0.64 & 0.15 & 0.03 & 0.01 & 0.06 \\
3.125 & $-$ & 3.250 & 0.65 & 0.11 & 0.04 & 0.06 & 0.07 \\
3.250 & $-$ & 3.375 & 0.65 & 0.11 & 0.08 & 0.39 & 0.10 \\
3.375 & $-$ & 3.500 & 0.65 & 0.07 & 0.16 & 0.30 & 0.09 \\
3.500 & $-$ & 3.625 & 0.65 & 0.07 & 0.05 & 0.29 & 0.10 \\
3.625 & $-$ & 3.750 & 0.66 & 0.10 & 0.18 & 0.75 & 0.17 \\
3.750 & $-$ & 3.875 & 0.67 & 0.09 & 0.33 & 0.27 & 0.13 \\
3.875 & $-$ & 4.000 & 0.68 & 0.09 & 0.04 & 0.75 & 0.13 \\
4.000 & $-$ & 4.250 & 0.70 & 0.09 & 0.14 & 0.81 & 0.07 \\
4.250 & $-$ & 4.500 & 0.78 & 0.25 & 0.28 & 2.85 & 0.49 \\ 
 
\hline
\end{tabular}
\label{tab:err_y}
\end{center}
\end{table}

\begin{table}[h]
\caption{Systematic uncertainties in the single differential cross-sections in interval regions of \zpt, presented in percentage. The contributions from efficiency (Eff), background (BKG), final state radiation (FSR), closure test (Closure), unfolding (Unfold), and alignment and calibration (Alignment) are shown.}
\begin{center}
\begin{tabular}{rcl|cccccc} \hline
 \multicolumn{3}{c|}{\zpt~[\gevc]}  & Eff & BKG & FSR & Closure & Alignment & Unfold \\
\hline
0.0 & $-$ & 2.2 & 0.81 & 0.05 & 0.24 & 0.50 & 1.10 & 0.21 \\
2.2 & $-$ & 3.4 & 0.83 & 0.08 & 0.67 & 0.18 & 0.82 & 0.33 \\
3.4 & $-$ & 4.6 & 0.77 & 0.07 & 0.51 & 0.45 & 0.77 & 0.51 \\
4.6 & $-$ & 5.8 & 0.80 & 0.08 & 0.18 & 0.31 & 0.86 & 0.46 \\
5.8 & $-$ & 7.2 & 0.78 & 0.07 & 0.24 & 0.44 & 0.61 & 0.20 \\
7.2 & $-$ & 8.7 & 0.78 & 0.10 & 0.66 & 0.33 & 0.73 & 0.27 \\
8.7 & $-$ & 10.5 & 0.79 & 0.09 & 0.51 & 0.24 & 0.59 & 0.32 \\
10.5 & $-$ & 12.8 & 0.77 & 0.10 & 0.43 & 0.34 & 0.39 & 0.30 \\
12.8 & $-$ & 15.4 & 0.77 & 0.11 & 0.36 & 0.22 & 0.43 & 0.20 \\
15.4 & $-$ & 19.0 & 0.77 & 0.12 & 0.52 & 0.43 & 0.36 & 0.09 \\
19.0 & $-$ & 24.5 & 0.77 & 0.11 & 0.62 & 0.46 & 0.32 & 0.15 \\
24.5 & $-$ & 34.0 & 0.77 & 0.08 & 0.14 & 0.42 & 0.24 & 0.10 \\
34.0 & $-$ & 63.0 & 0.76 & 0.07 & 0.13 & 0.72 & 0.14 & 0.03 \\
63.0 & $-$ & 270.0 & 0.77 & 0.11 & 0.37 & 0.54 & 0.27 & 0.06 \\ 
 
\hline
\end{tabular}
\label{tab:err_pt}
\end{center}
\end{table}

\begin{table}[h]
\caption{Systematic uncertainties in the single differential cross-sections in interval regions of \phistar, presented in percentage. The contributions from efficiency (Eff), background (BKG), final state radiation (FSR), closure test (Closure), and alignment and calibration (Alignment) are shown.}
\begin{center}
\begin{tabular}{rcl|ccccc} \hline
  \multicolumn{3}{c|}{\phistar}  & Eff & BKG & FSR & Closure & Alignment \\
\hline
0.00 & $-$ & 0.01 & 0.77 & 0.06 & 0.20 & 0.43 & 0.04 \\
0.01 & $-$ & 0.02 & 0.77 & 0.06 & 0.30 & 0.32 & 0.06 \\
0.02 & $-$ & 0.03 & 0.77 & 0.07 & 0.17 & 0.32 & 0.06 \\
0.03 & $-$ & 0.05 & 0.77 & 0.07 & 0.14 & 0.17 & 0.07 \\
0.05 & $-$ & 0.07 & 0.77 & 0.07 & 0.19 & 0.14 & 0.07 \\
0.07 & $-$ & 0.10 & 0.77 & 0.08 & 0.22 & 0.21 & 0.06 \\
0.10 & $-$ & 0.15 & 0.77 & 0.07 & 0.07 & 0.38 & 0.04 \\
0.15 & $-$ & 0.20 & 0.77 & 0.08 & 0.26 & 0.36 & 0.10 \\
0.20 & $-$ & 0.30 & 0.77 & 0.06 & 0.28 & 0.40 & 0.06 \\
0.30 & $-$ & 0.40 & 0.77 & 0.10 & 0.17 & 0.68 & 0.08 \\
0.40 & $-$ & 0.60 & 0.77 & 0.10 & 0.24 & 0.25 & 0.11 \\
0.60 & $-$ & 0.80 & 0.78 & 0.17 & 0.34 & 0.44 & 0.14 \\
0.80 & $-$ & 1.20 & 0.78 & 0.15 & 0.26 & 0.60 & 0.19 \\
1.20 & $-$ & 2.00 & 0.80 & 0.33 & 1.18 & 0.66 & 0.19 \\
2.00 & $-$ & 4.00 & 0.82 & 0.40 & 1.02 & 0.91 & 0.38 \\

\hline
\end{tabular}
\label{tab:err_phi}
\end{center}
\end{table}

\clearpage
\begin{center}
\begin{longtable}{rclrcl|ccccccc}
	\caption{Systematic uncertainties in the double differential cross-sections in interval regions of \zy and \zpt, presented in percentage. The contributions from efficiency (Eff), background (BKG), final state radiation (FSR), closure test (Closure), unfolding (Unfold), and alignment and calibration (Alignment) are shown.}
	\label{tab:err_Zpty} \\ \hline
\multicolumn{3}{c}{\zy} & \multicolumn{3}{c|}{\zpt~[\gevc]} & Eff & BKG & FSR & Closure & Alignment & Unfold \\ \hline
2.0 & $-$ & 2.5 & 0.0 & $-$ & 2.2 & 0.73 & 0.22 & 0.65 & 0.85 & 1.04 & 0.20 \\ 
2.0 & $-$ & 2.5 & 2.2 & $-$ & 3.4 & 0.73 & 0.18 & 2.11 & 0.48 & 1.38 & 0.40 \\ 
2.0 & $-$ & 2.5 & 3.4 & $-$ & 4.6 & 0.69 & 0.17 & 0.80 & 0.61 & 0.66 & 0.41 \\ 
2.0 & $-$ & 2.5 & 4.6 & $-$ & 5.8 & 0.74 & 0.22 & 0.04 & 0.06 & 1.18 & 0.47 \\ 
2.0 & $-$ & 2.5 & 5.8 & $-$ & 7.2 & 0.68 & 0.19 & 0.39 & 0.36 & 0.88 & 0.24 \\ 
2.0 & $-$ & 2.5 & 7.2 & $-$ & 8.7 & 0.68 & 0.19 & 0.31 & 0.54 & 0.77 & 0.29 \\ 
2.0 & $-$ & 2.5 & 8.7 & $-$ & 10.5 & 0.67 & 0.41 & 0.35 & 1.18 & 0.65 & 0.12 \\ 
2.0 & $-$ & 2.5 & 10.5 & $-$ & 12.8 & 0.68 & 0.17 & 1.36 & 0.32 & 0.62 & 0.33 \\ 
2.0 & $-$ & 2.5 & 12.8 & $-$ & 15.4 & 0.67 & 0.27 & 0.71 & 0.33 & 0.51 & 0.30 \\ 
2.0 & $-$ & 2.5 & 15.4 & $-$ & 19.0 & 0.69 & 0.19 & 0.47 & 0.80 & 0.48 & 0.18 \\ 
2.0 & $-$ & 2.5 & 19.0 & $-$ & 24.5 & 0.67 & 0.16 & 0.26 & 0.85 & 0.33 & 0.09 \\ 
2.0 & $-$ & 2.5 & 24.5 & $-$ & 34.0 & 0.67 & 0.24 & 0.14 & 0.55 & 0.33 & 0.08 \\ 
2.0 & $-$ & 2.5 & 34.0 & $-$ & 63.0 & 0.67 & 0.15 & 0.28 & 0.87 & 0.38 & 0.02 \\ 
2.0 & $-$ & 2.5 & 63.0 & $-$ & 270.0 & 0.67 & 0.25 & 0.65 & 1.09 & 0.24 & 0.09 \\ 
2.5 & $-$ & 3.0 & 0.0 & $-$ & 2.2 & 0.71 & 0.04 & 0.11 & 0.62 & 1.02 & 0.17 \\ 
2.5 & $-$ & 3.0 & 2.2 & $-$ & 3.4 & 0.78 & 0.08 & 0.59 & 0.04 & 0.66 & 0.41 \\ 
2.5 & $-$ & 3.0 & 3.4 & $-$ & 4.6 & 0.66 & 0.13 & 0.22 & 0.36 & 0.87 & 0.63 \\ 
2.5 & $-$ & 3.0 & 4.6 & $-$ & 5.8 & 0.68 & 0.13 & 0.34 & 0.88 & 0.99 & 0.46 \\ 
2.5 & $-$ & 3.0 & 5.8 & $-$ & 7.2 & 0.67 & 0.13 & 0.05 & 0.29 & 0.92 & 0.28 \\ 
2.5 & $-$ & 3.0 & 7.2 & $-$ & 8.7 & 0.66 & 0.16 & 0.30 & 0.62 & 0.44 & 0.32 \\ 
2.5 & $-$ & 3.0 & 8.7 & $-$ & 10.5 & 0.66 & 0.14 & 0.91 & 0.03 & 0.43 & 0.16 \\ 
2.5 & $-$ & 3.0 & 10.5 & $-$ & 12.8 & 0.66 & 0.22 & 0.10 & 0.44 & 0.43 & 0.18 \\ 
2.5 & $-$ & 3.0 & 12.8 & $-$ & 15.4 & 0.66 & 0.21 & 0.27 & 0.20 & 0.58 & 0.24 \\ 
2.5 & $-$ & 3.0 & 15.4 & $-$ & 19.0 & 0.65 & 0.24 & 0.87 & 0.42 & 0.41 & 0.04 \\ 
2.5 & $-$ & 3.0 & 19.0 & $-$ & 24.5 & 0.65 & 0.18 & 0.53 & 0.07 & 0.29 & 0.05 \\ 
2.5 & $-$ & 3.0 & 24.5 & $-$ & 34.0 & 0.65 & 0.08 & 0.13 & 0.36 & 0.30 & 0.07 \\ 
2.5 & $-$ & 3.0 & 34.0 & $-$ & 63.0 & 0.65 & 0.11 & 0.16 & 0.59 & 0.10 & 0.04 \\ 
2.5 & $-$ & 3.0 & 63.0 & $-$ & 270.0 & 0.65 & 0.15 & 0.09 & 0.14 & 0.32 & 0.02 \\ 
3.0 & $-$ & 3.5 & 0.0 & $-$ & 2.2 & 0.72 & 0.09 & 0.22 & 0.29 & 1.14 & 0.12 \\ 
3.0 & $-$ & 3.5 & 2.2 & $-$ & 3.4 & 0.66 & 0.14 & 0.54 & 0.07 & 0.76 & 0.33 \\ 
3.0 & $-$ & 3.5 & 3.4 & $-$ & 4.6 & 0.66 & 0.15 & 0.92 & 0.77 & 0.60 & 0.57 \\ 
3.0 & $-$ & 3.5 & 4.6 & $-$ & 5.8 & 0.68 & 0.19 & 0.19 & 0.03 & 0.84 & 0.48 \\ 
3.0 & $-$ & 3.5 & 5.8 & $-$ & 7.2 & 0.67 & 0.10 & 0.19 & 0.47 & 0.48 & 0.08 \\ 
3.0 & $-$ & 3.5 & 7.2 & $-$ & 8.7 & 0.73 & 0.17 & 1.16 & 0.13 & 0.69 & 0.30 \\ 
3.0 & $-$ & 3.5 & 8.7 & $-$ & 10.5 & 0.66 & 0.16 & 0.32 & 0.23 & 0.40 & 0.56 \\ 
3.0 & $-$ & 3.5 & 10.5 & $-$ & 12.8 & 0.65 & 0.16 & 0.48 & 0.12 & 0.21 & 0.36 \\ 
3.0 & $-$ & 3.5 & 12.8 & $-$ & 15.4 & 0.66 & 0.19 & 0.30 & 0.10 & 0.38 & 0.09 \\ 
3.0 & $-$ & 3.5 & 15.4 & $-$ & 19.0 & 0.65 & 0.23 & 0.24 & 0.13 & 0.24 & 0.07 \\ 
3.0 & $-$ & 3.5 & 19.0 & $-$ & 24.5 & 0.65 & 0.20 & 1.23 & 0.84 & 0.36 & 0.12 \\ 
3.0 & $-$ & 3.5 & 24.5 & $-$ & 34.0 & 0.65 & 0.16 & 0.17 & 0.44 & 0.15 & 0.17 \\ 
3.0 & $-$ & 3.5 & 34.0 & $-$ & 63.0 & 0.65 & 0.12 & 0.09 & 1.05 & 0.15 & 0.02 \\ 
3.0 & $-$ & 3.5 & 63.0 & $-$ & 270.0 & 0.65 & 0.20 & 0.73 & 0.26 & 0.33 & 0.13 \\ 
3.5 & $-$ & 4.0 & 0.0 & $-$ & 2.2 & 0.80 & 0.05 & 0.37 & 0.37 & 1.55 & 0.29 \\ 
3.5 & $-$ & 4.0 & 2.2 & $-$ & 3.4 & 0.73 & 0.20 & 1.12 & 0.33 & 1.31 & 0.50 \\ 
3.5 & $-$ & 4.0 & 3.4 & $-$ & 4.6 & 0.69 & 0.09 & 0.43 & 0.39 & 1.58 & 0.83 \\ 
3.5 & $-$ & 4.0 & 4.6 & $-$ & 5.8 & 0.73 & 0.09 & 0.12 & 0.08 & 1.07 & 0.67 \\ 
3.5 & $-$ & 4.0 & 5.8 & $-$ & 7.2 & 0.68 & 0.10 & 0.78 & 0.98 & 0.69 & 0.38 \\ 
3.5 & $-$ & 4.0 & 7.2 & $-$ & 8.7 & 0.67 & 0.29 & 0.86 & 0.25 & 1.07 & 0.54 \\ 
3.5 & $-$ & 4.0 & 8.7 & $-$ & 10.5 & 0.70 & 0.10 & 0.94 & 0.01 & 1.67 & 0.47 \\ 
3.5 & $-$ & 4.0 & 10.5 & $-$ & 12.8 & 0.71 & 0.09 & 0.86 & 0.53 & 0.44 & 0.54 \\ 
3.5 & $-$ & 4.0 & 12.8 & $-$ & 15.4 & 0.66 & 0.23 & 0.88 & 0.62 & 0.49 & 0.40 \\ 
3.5 & $-$ & 4.0 & 15.4 & $-$ & 19.0 & 0.67 & 0.15 & 0.72 & 1.37 & 0.55 & 0.23 \\ 
3.5 & $-$ & 4.0 & 19.0 & $-$ & 24.5 & 0.68 & 0.15 & 0.31 & 0.30 & 0.67 & 0.55 \\ 
3.5 & $-$ & 4.0 & 24.5 & $-$ & 34.0 & 0.66 & 0.11 & 0.27 & 1.00 & 0.54 & 0.12 \\ 
3.5 & $-$ & 4.0 & 34.0 & $-$ & 63.0 & 0.66 & 0.15 & 0.17 & 1.13 & 0.11 & 0.11 \\ 
3.5 & $-$ & 4.0 & 63.0 & $-$ & 270.0 & 0.66 & 0.30 & 0.73 & 3.36 & 0.67 & 0.24 \\ 
4.0 & $-$ & 4.5 & 0.0 & $-$ & 2.2 & 0.75 & 0.13 & 1.43 & 5.50 & 1.71 & 0.71 \\ 
4.0 & $-$ & 4.5 & 2.2 & $-$ & 3.4 & 0.72 & 0.47 & 0.92 & 2.30 & 2.35 & 1.21 \\ 
4.0 & $-$ & 4.5 & 3.4 & $-$ & 4.6 & 0.73 & 0.50 & 1.24 & 0.67 & 2.57 & 0.49 \\ 
4.0 & $-$ & 4.5 & 4.6 & $-$ & 5.8 & 0.81 & 0.43 & 0.71 & 1.29 & 2.36 & 1.33 \\ 
4.0 & $-$ & 4.5 & 5.8 & $-$ & 7.2 & 0.79 & 0.39 & 0.63 & 2.54 & 1.38 & 1.32 \\ 
4.0 & $-$ & 4.5 & 7.2 & $-$ & 8.7 & 0.85 & 0.20 & 3.74 & 1.99 & 1.86 & 1.78 \\ 
4.0 & $-$ & 4.5 & 8.7 & $-$ & 10.5 & 0.83 & 0.15 & 0.42 & 2.19 & 2.39 & 1.33 \\ 
4.0 & $-$ & 4.5 & 10.5 & $-$ & 12.8 & 0.72 & 0.47 & 0.36 & 3.73 & 3.65 & 2.06 \\ 
4.0 & $-$ & 4.5 & 12.8 & $-$ & 15.4 & 0.74 & 0.15 & 0.96 & 1.08 & 2.21 & 1.91 \\ 
4.0 & $-$ & 4.5 & 15.4 & $-$ & 19.0 & 0.74 & 0.22 & 3.72 & 0.07 & 2.06 & 0.98 \\ 
4.0 & $-$ & 4.5 & 19.0 & $-$ & 24.5 & 0.77 & 0.68 & 2.58 & 3.51 & 1.29 & 0.44 \\ 
4.0 & $-$ & 4.5 & 24.5 & $-$ & 34.0 & 0.71 & 0.24 & 0.94 & 0.69 & 0.96 & 1.01 \\ 
4.0 & $-$ & 4.5 & 34.0 & $-$ & 63.0 & 0.71 & 0.21 & 0.57 & 4.16 & 0.63 & 0.32 \\ 
4.0 & $-$ & 4.5 & 63.0 & $-$ & 270.0 & 0.69 & 0.45 & 1.40 & 0.93 & 2.68 & 1.06 \\ 
\hline
\end{longtable}
\end{center}

\begin{center}
\begin{longtable}{rclrcl|ccccc}
	\caption{Systematic uncertainties in the double differential cross-sections in interval regions of \zy and \phistar, presented in percentage. The contributions from efficiency (Eff), background (BKG), final state radiation (FSR), closure test (Closure), and alignment and calibration (Alignment) are shown.}
	\label{tab:err_ZY_ZPHI} \\ \hline
\multicolumn{3}{c}{\zy} & \multicolumn{3}{c|}{\phistar} &   Eff & BKG & FSR & Closure & Alignment \\ \hline
2.0 & $-$ & 2.5 & 0.00 & $-$ & 0.01 & 0.67 & 0.15 & 0.03 & 0.72 & 0.06 \\ 
2.0 & $-$ & 2.5 & 0.01 & $-$ & 0.02 & 0.67 & 0.19 & 0.50 & 0.77 & 0.15 \\ 
2.0 & $-$ & 2.5 & 0.02 & $-$ & 0.03 & 0.67 & 0.22 & 0.38 & 0.10 & 0.08 \\ 
2.0 & $-$ & 2.5 & 0.03 & $-$ & 0.05 & 0.67 & 0.18 & 0.22 & 0.61 & 0.06 \\ 
2.0 & $-$ & 2.5 & 0.05 & $-$ & 0.07 & 0.67 & 0.13 & 0.33 & 0.13 & 0.08 \\ 
2.0 & $-$ & 2.5 & 0.07 & $-$ & 0.10 & 0.67 & 0.14 & 0.26 & 0.15 & 0.11 \\ 
2.0 & $-$ & 2.5 & 0.10 & $-$ & 0.15 & 0.67 & 0.21 & 0.01 & 0.82 & 0.05 \\ 
2.0 & $-$ & 2.5 & 0.15 & $-$ & 0.20 & 0.67 & 0.17 & 0.45 & 0.67 & 0.06 \\ 
2.0 & $-$ & 2.5 & 0.20 & $-$ & 0.30 & 0.67 & 0.18 & 0.34 & 0.96 & 0.05 \\ 
2.0 & $-$ & 2.5 & 0.30 & $-$ & 0.40 & 0.67 & 0.28 & 0.35 & 0.60 & 0.10 \\ 
2.0 & $-$ & 2.5 & 0.40 & $-$ & 0.60 & 0.67 & 0.25 & 0.19 & 1.00 & 0.12 \\ 
2.0 & $-$ & 2.5 & 0.60 & $-$ & 0.80 & 0.67 & 0.31 & 0.86 & 0.76 & 0.30 \\ 
2.0 & $-$ & 2.5 & 0.80 & $-$ & 1.20 & 0.67 & 0.36 & 0.22 & 1.05 & 0.26 \\ 
2.0 & $-$ & 2.5 & 1.20 & $-$ & 2.00 & 0.67 & 0.69 & 1.42 & 0.72 & 0.18 \\ 
2.0 & $-$ & 2.5 & 2.00 & $-$ & 4.00 & 0.67 & 0.66 & 0.50 & 1.92 & 0.78 \\ 
2.5 & $-$ & 3.0 & 0.00 & $-$ & 0.01 & 0.65 & 0.13 & 0.38 & 0.70 & 0.05 \\ 
2.5 & $-$ & 3.0 & 0.01 & $-$ & 0.02 & 0.65 & 0.13 & 0.50 & 0.25 & 0.07 \\ 
2.5 & $-$ & 3.0 & 0.02 & $-$ & 0.03 & 0.65 & 0.16 & 0.33 & 0.78 & 0.09 \\ 
2.5 & $-$ & 3.0 & 0.03 & $-$ & 0.05 & 0.65 & 0.09 & 0.22 & 0.00 & 0.07 \\ 
2.5 & $-$ & 3.0 & 0.05 & $-$ & 0.07 & 0.65 & 0.13 & 0.17 & 0.02 & 0.10 \\ 
2.5 & $-$ & 3.0 & 0.07 & $-$ & 0.10 & 0.65 & 0.11 & 0.08 & 0.05 & 0.04 \\ 
2.5 & $-$ & 3.0 & 0.10 & $-$ & 0.15 & 0.65 & 0.11 & 0.05 & 0.27 & 0.05 \\ 
2.5 & $-$ & 3.0 & 0.15 & $-$ & 0.20 & 0.65 & 0.15 & 0.36 & 0.39 & 0.07 \\ 
2.5 & $-$ & 3.0 & 0.20 & $-$ & 0.30 & 0.65 & 0.08 & 0.09 & 0.40 & 0.11 \\ 
2.5 & $-$ & 3.0 & 0.30 & $-$ & 0.40 & 0.65 & 0.16 & 0.23 & 0.08 & 0.10 \\ 
2.5 & $-$ & 3.0 & 0.40 & $-$ & 0.60 & 0.65 & 0.10 & 0.31 & 0.01 & 0.13 \\ 
2.5 & $-$ & 3.0 & 0.60 & $-$ & 0.80 & 0.65 & 0.25 & 0.02 & 0.31 & 0.13 \\ 
2.5 & $-$ & 3.0 & 0.80 & $-$ & 1.20 & 0.65 & 0.28 & 0.47 & 0.19 & 0.23 \\ 
2.5 & $-$ & 3.0 & 1.20 & $-$ & 2.00 & 0.65 & 0.34 & 1.98 & 0.27 & 0.18 \\ 
2.5 & $-$ & 3.0 & 2.00 & $-$ & 4.00 & 0.66 & 0.28 & 0.35 & 0.60 & 0.27 \\ 
3.0 & $-$ & 3.5 & 0.00 & $-$ & 0.01 & 0.65 & 0.09 & 0.11 & 0.39 & 0.03 \\ 
3.0 & $-$ & 3.5 & 0.01 & $-$ & 0.02 & 0.65 & 0.08 & 0.14 & 0.36 & 0.06 \\ 
3.0 & $-$ & 3.5 & 0.02 & $-$ & 0.03 & 0.65 & 0.11 & 0.11 & 0.15 & 0.05 \\ 
3.0 & $-$ & 3.5 & 0.03 & $-$ & 0.05 & 0.65 & 0.16 & 0.15 & 0.14 & 0.06 \\ 
3.0 & $-$ & 3.5 & 0.05 & $-$ & 0.07 & 0.66 & 0.11 & 0.25 & 0.08 & 0.06 \\ 
3.0 & $-$ & 3.5 & 0.07 & $-$ & 0.10 & 0.65 & 0.17 & 0.35 & 0.32 & 0.07 \\ 
3.0 & $-$ & 3.5 & 0.10 & $-$ & 0.15 & 0.65 & 0.14 & 0.01 & 0.31 & 0.04 \\ 
3.0 & $-$ & 3.5 & 0.15 & $-$ & 0.20 & 0.65 & 0.13 & 0.22 & 0.05 & 0.15 \\ 
3.0 & $-$ & 3.5 & 0.20 & $-$ & 0.30 & 0.65 & 0.09 & 0.48 & 0.50 & 0.03 \\ 
3.0 & $-$ & 3.5 & 0.30 & $-$ & 0.40 & 0.65 & 0.18 & 0.12 & 1.03 & 0.09 \\ 
3.0 & $-$ & 3.5 & 0.40 & $-$ & 0.60 & 0.65 & 0.25 & 0.38 & 0.26 & 0.12 \\ 
3.0 & $-$ & 3.5 & 0.60 & $-$ & 0.80 & 0.66 & 0.30 & 0.53 & 0.58 & 0.08 \\ 
3.0 & $-$ & 3.5 & 0.80 & $-$ & 1.20 & 0.65 & 0.20 & 0.09 & 0.96 & 0.16 \\ 
3.0 & $-$ & 3.5 & 1.20 & $-$ & 2.00 & 0.66 & 0.77 & 0.37 & 0.45 & 0.31 \\ 
3.0 & $-$ & 3.5 & 2.00 & $-$ & 4.00 & 0.66 & 0.40 & 2.56 & 0.67 & 0.35 \\ 
3.5 & $-$ & 4.0 & 0.00 & $-$ & 0.01 & 0.67 & 0.11 & 0.43 & 0.32 & 0.09 \\ 
3.5 & $-$ & 4.0 & 0.01 & $-$ & 0.02 & 0.67 & 0.09 & 0.37 & 0.51 & 0.06 \\ 
3.5 & $-$ & 4.0 & 0.02 & $-$ & 0.03 & 0.67 & 0.13 & 0.12 & 0.24 & 0.10 \\ 
3.5 & $-$ & 4.0 & 0.03 & $-$ & 0.05 & 0.66 & 0.09 & 0.10 & 0.10 & 0.15 \\ 
3.5 & $-$ & 4.0 & 0.05 & $-$ & 0.07 & 0.67 & 0.14 & 0.22 & 0.62 & 0.10 \\ 
3.5 & $-$ & 4.0 & 0.07 & $-$ & 0.10 & 0.67 & 0.08 & 0.39 & 0.46 & 0.13 \\ 
3.5 & $-$ & 4.0 & 0.10 & $-$ & 0.15 & 0.66 & 0.10 & 0.29 & 0.78 & 0.07 \\ 
3.5 & $-$ & 4.0 & 0.15 & $-$ & 0.20 & 0.66 & 0.14 & 0.19 & 1.10 & 0.25 \\ 
3.5 & $-$ & 4.0 & 0.20 & $-$ & 0.30 & 0.66 & 0.14 & 0.52 & 0.28 & 0.07 \\ 
3.5 & $-$ & 4.0 & 0.30 & $-$ & 0.40 & 0.67 & 0.15 & 0.13 & 1.77 & 0.11 \\ 
3.5 & $-$ & 4.0 & 0.40 & $-$ & 0.60 & 0.66 & 0.25 & 0.17 & 0.16 & 0.33 \\ 
3.5 & $-$ & 4.0 & 0.60 & $-$ & 0.80 & 0.66 & 0.74 & 0.69 & 0.71 & 0.34 \\ 
3.5 & $-$ & 4.0 & 0.80 & $-$ & 1.20 & 0.66 & 0.14 & 0.28 & 1.55 & 0.58 \\ 
3.5 & $-$ & 4.0 & 1.20 & $-$ & 2.00 & 0.68 & 0.42 & 1.95 & 7.26 & 0.71 \\ 
3.5 & $-$ & 4.0 & 2.00 & $-$ & 4.00 & 0.68 & 0.50 & 1.78 & 5.58 & 2.56 \\ 
4.0 & $-$ & 4.5 & 0.00 & $-$ & 0.01 & 0.71 & 0.26 & 0.21 & 1.50 & 0.30 \\ 
4.0 & $-$ & 4.5 & 0.01 & $-$ & 0.02 & 0.71 & 0.05 & 1.57 & 0.31 & 0.16 \\ 
4.0 & $-$ & 4.5 & 0.02 & $-$ & 0.03 & 0.71 & 0.40 & 0.20 & 0.74 & 0.19 \\ 
4.0 & $-$ & 4.5 & 0.03 & $-$ & 0.05 & 0.71 & 0.25 & 0.38 & 2.03 & 0.10 \\ 
4.0 & $-$ & 4.5 & 0.05 & $-$ & 0.07 & 0.71 & 0.35 & 0.57 & 0.09 & 0.21 \\ 
4.0 & $-$ & 4.5 & 0.07 & $-$ & 0.10 & 0.71 & 0.27 & 0.68 & 1.20 & 0.18 \\ 
4.0 & $-$ & 4.5 & 0.10 & $-$ & 0.15 & 0.71 & 0.04 & 0.44 & 1.79 & 0.35 \\ 
4.0 & $-$ & 4.5 & 0.15 & $-$ & 0.20 & 0.71 & 0.18 & 1.64 & 0.98 & 0.36 \\ 
4.0 & $-$ & 4.5 & 0.20 & $-$ & 0.30 & 0.71 & 0.03 & 0.86 & 0.49 & 0.41 \\ 
4.0 & $-$ & 4.5 & 0.30 & $-$ & 0.40 & 0.75 & 0.38 & 1.09 & 6.43 & 0.65 \\ 
4.0 & $-$ & 4.5 & 0.40 & $-$ & 0.60 & 0.70 & 0.08 & 0.81 & 1.10 & 0.97 \\ 
4.0 & $-$ & 4.5 & 0.60 & $-$ & 0.80 & 0.70 & 0.25 & 0.07 & 5.31 & 1.62 \\ 
4.0 & $-$ & 4.5 & 0.80 & $-$ & 1.20 & 0.71 & 0.00 & 7.46 & 6.39 & 3.97 \\ 
4.0 & $-$ & 4.5 & 1.20 & $-$ & 2.00 & 12.06 & 18.37 & 24.14 & 0.23 & 19.98 \\ 
4.0 & $-$ & 4.5 & 2.00 & $-$ & 4.00 & 38.09 & 61.98 & 4.01 & 20.47 & 51.95 \\ 
\hline
\end{longtable}
\end{center}


\addcontentsline{toc}{section}{References}
\bibliographystyle{LHCb}
\bibliography{main,standard,LHCb-PAPER,LHCb-CONF,LHCb-DP,LHCb-TDR}
 
\newpage
\centerline
{\large\bf LHCb collaboration}
\begin
{flushleft}
\small
R.~Aaij$^{32}$,
A.S.W.~Abdelmotteleb$^{56}$,
C.~Abell{\'a}n~Beteta$^{50}$,
F.~Abudin{\'e}n$^{56}$,
T.~Ackernley$^{60}$,
B.~Adeva$^{46}$,
M.~Adinolfi$^{54}$,
H.~Afsharnia$^{9}$,
C.~Agapopoulou$^{13}$,
C.A.~Aidala$^{87}$
% ...

{\footnotesize \it

% ...
$^{9}$Universit{\'e} Clermont Auvergne, CNRS/IN2P3, LPC, Clermont-Ferrand, France\\
% ...
$^{13}$LPNHE, Sorbonne Universit{\'e}, Paris Diderot Sorbonne Paris Cit{\'e}, CNRS/IN2P3, Paris, France\\
% ...
$^{32}$Nikhef National Institute for Subatomic Physics, Amsterdam, Netherlands\\
% ...
$^{46}$Instituto Galego de F{\'\i}sica de Altas Enerx{\'\i}as (IGFAE), Universidade de Santiago de Compostela, Santiago de Compostela, Spain\\
% ...
$^{50}$Physik-Institut, Universit{\"a}t Z{\"u}rich, Z{\"u}rich, Switzerland\\
% ...
$^{54}$H.H. Wills Physics Laboratory, University of Bristol, Bristol, United Kingdom\\
% ...
$^{56}$Department of Physics, University of Warwick, Coventry, United Kingdom\\
% ...
$^{60}$Oliver Lodge Laboratory, University of Liverpool, Liverpool, United Kingdom\\
% ...
$^{68}$Syracuse University, Syracuse, NY, United States\\
% ...
$^{87}$University of Michigan, Ann Arbor, United States, associated to $^{68}$\\
\bigskip
% ...
}
\end{flushleft}

\end{document}